# Extirpation of Atlantic Cod from a Northwest Atlantic ecosystem in the absence of predator control: inference from an ecosystem model of intermediate complexity

Steven P. Rossi[a]*, Sean P. Cox[a], Hugues P. Benoît[b]

[a] School of Resource and Environmental Management Simon Fraser University, 8888 University Drive, Burnaby, British Columbia, V5A 1S6, Canada

[b]Maurice Lamontagne Institute, Fisheries and Oceans Canada, Mont-Joli, Quebec, G5H 3Z4, Canada

*Corresponding author. Email: rossi@burnsideanalytics.com

## Abstract

Atlantic cod (*Gadus morhua*) in the southern Gulf of St. Lawrence (sGSL) declined to low abundance in the early 1990s and have since failed to recover due to high natural mortality, which has been linked to grey seal (*Halichoerus grypus*) predation. Increased grey seal harvests have been suggested to improve cod survival, however, predicting the response of cod to seal abundance changes in the sGSL is complicated by a hypothesized triangular food web involving seals, cod, and small pelagic fishes, wherein the pelagic fishes are prey for cod and grey seals, but may also prey on young cod. Grey seals may therefore have an indirect positive effect on prerecruit cod survival via predation on pelagic fish. Using a multispecies model of intermediate complexity fitted to various scientific and fisheries data, we found that seal predation accounted for the majority of recent cod mortality and that cod will likely be extirpated without a strong and rapid reduction in grey seal abundance. We did not find evidence that reducing grey seal abundance will result in large increases to herring biomass that could impair cod recovery.

## 1 Introduction

Managing mammalian predator abundance has been a common tactic in human societies for at least two millennia (Reynolds and Tapper, 1996). Reducing predator abundance has sometimes been a means to improve human safety, whereas, in other cases, it has been aimed at improving the survival of species of interest (or, "target" species). In this latter case, predator reduction initiatives often implicitly assume that ecosystems can be reduced to simple predator-prey systems in which reducing predator abundance results in sustained increases in prey survival; however, reducing predator abundance may yield a wide range of unintended consequences (Bax, 1998; Yodzis, 2001; Bowen & Lidgard, 2013). For instance, removing a top predator from an ecosystem may improve the short-term survival of not only the target species, but all species consumed by that predator. If one or more of these other species are predators of the target species, then the long-term, total predation mortality for the target species may increase in the absence of the top predator due to the improved survival of other predators. It is therefore important to consider wider ecosystem implications when forecasting the impact of predator reductions on target species.

Marine mammal predation on commercially valuable fish stocks is increasingly used to justify predator control programs, especially where marine mammal populations (particularly pinnipeds) are growing in response to reduced exploitation. For example, predation by recovering pinnipeds on several sockeye (*Oncorhynchus nerka*) and chinook salmon (*O. tshawytscha*) stocks in the northeast Pacific likely now accounts for more mortality than all sources of fishing combined (Magera et al., 2013; Wargo Rub et al., 2019; Walters et al., 2020). Similarly, in the northwest Atlantic, where grey seal (*Halichoerus grypus*) abundance has grown especially large, natural mortality (*M*) for numerous fish stocks has increased to the point where some stocks, particularly in the southern Gulf of St. Lawrence (sGSL), are being evaluated by the Committee on the Status of Endangered Wildlife in Canada for enhanced risk of extirpation (e.g., Swain et al., 2015). While targeted pinniped reductions have been suggested as a possible means of recovering these threatened fish stocks (e.g., FRCC, 2011; SSCFO, 2012), the wider implications of such interventions for the sGSL ecosystem are unknown.

Developing quantitative models of interactions between predators and target species is an important step toward understanding target species responses to changes in predator abundance. The specific approach used to model species interactions should be linked to scientific goals (i.e., understanding seal predation impacts) and management objectives (i.e., recovering threatened fish populations). "Food web" or "whole ecosystem" models, such as ECOPATH with ECOSIM (Christensen and Pauly, 1992; Polovina, 1984; Walters et al., 1997; 2000) and ATLANTIS (Fulton et al., 2004), consider populations across all trophic levels of an ecosystem. These models are often complex and are intended for strategic use (i.e., broad scale, long-term planning). In contrast, Models of Intermediate Complexity for Ecosystem assessments (MICE; Plagányi et al., 2014) include only species considered to have important interactions with the species of interest. MICE are tactically focused, with outputs that can be used in short-term decision making (Plagányi et al., 2014), and are well suited for investigating interactions among higher trophic level species, as these species require relatively fewer linkages than lower trophic level species. In particular, MICE are useful for understanding the effects of pinniped predation on fish *M* and forecasting responses to pinniped abundance changes while accounting for indirect effects (e.g., Punt and Butterworth, 1995).

### 1.1 Cod and species interactions in the southern Gulf of St. Lawrence

In this paper, we investigate the extirpation risk for Atlantic cod (*Gadus morhua*) in the southern Gulf of St. Lawrence (sGSL; Atlantic Canada, Northwest Atlantic Fisheries Organization (NAFO) Division 4T) via a MICE consisting of Atlantic cod, Atlantic herring (*Clupea harengus*), and Canadian-origin grey seals (hereafter referred to as "cod", "herring", and "seals", respectively). The sGSL encompasses the Magdalen Shallows, with depths mostly less than 100 m, and the Laurentian Channel, with depths up to 500 m (Fig. 1). Cod and herring reside in the sGSL from the spring to fall, where they spawn and feed, while adult herring and all stages of cod overwinter in the Cabot Strait (NAFO Subdivision 4Vn). Cod in the sGSL form a single spawning stock, while herring consist of genetically distinct spring and fall spawning components (Lamichhaney et al., 2017), the latter of which is further disaggregated by region within the sGSL (North, Middle, and South) for management purposes to account for strong spawning-site fidelity (DFO, 2018). Grey seals in the Northwest Atlantic form a single population but are subdivided in Canadian waters into Scotian Shelf and Gulf of St. Lawrence herds (hereafter referred to as "Shelf" and "Gulf" herds, respectively) for management purposes. The Shelf herd consists of seals from Sable Island, the largest grey seal colony in the world, as well as seals from smaller whelping grounds along coastal Nova Scotia. Gulf herd seals pup on pack-ice and small islands in the Gulf of St. Lawrence. Grey seals forage widely within their range, including within the sGSL and on cod/herring overwintering grounds (Breed et al., 2006; Harvey et al., 2011; Swain et al., 2015).

Cod were fished to low abundance in the early 1990s and have failed to recover due to an increase in *M* among older cod (Swain and Benoît, 2015), which is concurrent with a rapid increase in seal abundance (Hammill et al., 2017a). Early maturation, environmental conditions, parasites and unreported catch were investigated as possible causes of elevated *M* among older cod from the mid-1990s to present, but none were found to be important contributing factors (Swain et al., 2011). In contrast, the hypothesis that seal predation has driven increases in cod *M* has been supported by bioenergetic models, shifts in cod distribution to areas with lower seal abundance, and population models linking cod *M* to seal abundance (Benoît et al., 2011; Swain et al., 2015). Specifically, increased cod *M* appears be the result of a predation-driven Allee effect, suggesting that the extirpation of cod is likely without large reductions in seal predation (Neuenhoff et al., 2019).

Predicting ecosystem responses to grey seal abundance reductions in the sGSL is complicated by a hypothesized triangular food web involving seals, cod, and pelagic fishes such as herring and mackerel (*Scomber scombrus*) that are important prey for seals (Bowen et al., 1993; Hammill et al., 2007; 2014) and cod (Hanson and Chouinard, 2002; Hanson, 2011). Cod recruitment success in the sGSL has a strongly negative relationship with pelagic fish biomass, potentially resulting from predation or competition by pelagics with early life history stages of cod (Swain and Sinclair, 2000). Thus, seals may have an indirect positive effect on prerecruit cod survival via predation on pelagics, suggesting that reducing seal abundance could result in reduced recruitment success for cod. For instance, Punt and Butterworth (1995) analyzed a pinniped cull in a three-species food web in the Benguela ecosystem and found that reducing pinniped abundance had a neutral or negative effect on the target species due to the resulting increased abundance of an intermediate species that consumes the target species. Alternatively, pelagics released from seal predation may instead be consumed by recovering cod, resulting in improved cod survival through cultivation effects (Walters and Kitchell, 2001). Management actions aimed at reducing cod mortality may be the only means of recovering cod in the sGSL, so it is critical to better understand the importance of these processes and their relative impacts on the efficacy of potential management actions.

We develop a MICE to evaluate the effects of changes in seal and/or herring abundance on cod survival and recruitment in the sGSL. Compared to existing cod-seal modelling approaches for northwest Atlantic ecosystems (Fu et al., 2001; Mohn and Bowen, 1996; Neuenhoff et al., 2019; Trzcinski et al., 2006), MICE allow for indirect effects of seal reductions on cod survival via other species to be explicitly evaluated. Our model linked cod mortality to local seal abundance and herring mortality to the local abundance of both cod and seals (Fig. 2). Additionally, we modelled cod recruitment as a function of herring biomass to account for herring effects (predation, competition) on prerecruit cod. Under this model formulation, seal abundance reductions have a positive direct effect on adult cod survival and an uncertain, potentially negative indirect effect on cod recruitment. The net effect of seal reductions on cod productivity is therefore unknown and may be negative if reduced seal predation on adult cod is sufficiently offset by increased predation on and/or competition with young cod. This model may therefore yield a range of outcomes for cod depending on the net effect of reduced seal abundance. Our results over a wide range of model assumptions and sensitivity tests suggest that cod recovery in the sGSL is highly unlikely without rapid reductions in seal abundance.

## 2 Methods

### 2.1 Data

Annual fishery landings of cod (1971-2018) in the sGSL included directed fishing from fixed and mobile gears, as well as cod bycatch from other groundfish fisheries but not from invertebrate fisheries, as landings of commercial-size cod in these fisheries were negligible (Swain et al., 2011). Herring landings included catches from fixed (gillnet) gear fisheries on spawning grounds in Div. 4T and mobile (purse seine) gear fisheries in Div. 4T and Subdiv. 4Vn.

Fisheries and Oceans Canada (DFO) has monitored relative cod abundance in the sGSL using stratified-random bottom-trawl research vessel (RV) surveys each September since 1971. Changes in fishing efficiency by different RVs were accounted for by applying conversion factors estimated using results of comparative fishing experiments. We also obtained cod abundance indices from a sentinel longline (LL) survey, conducted each summer and fall (1995-2017) at fixed sites, and a mobile sentinel (MS) bottom-trawl survey, conducted each August since 2003 using the same stratified-random design as the RV survey (Swain et al., 2019).

We used spring herring relative abundance data (1994-2017) from DFO acoustic surveys in the western portion of Div. 4T each fall. Herring catch per unit effort (CPUE) time-series (1986-2018) were constructed for each subpopulation from commercial gillnet catch and effort data. Herring data prior to 1978 were unavailable, so we used regionally-aggregated biomass estimates as a biomass index for 1971-1978 (Cleary 1982).

Counts of newly weaned seals at the main breeding colonies in Canadian waters (1971-2016; den Heyer et al. 2017; Hammill et al. 2017b), corrected for pups that were unseen or died prior to surveying, were used to infer seal abundance. We used samples of age and reproductive status from seals in the Gulf of St. Lawrence between late May and November in intermittent years (1982-2015; Hammill and Gosselin 1995). We also used estimates of seal removals from the Canadian commercial harvest, nuisance license kills, bounty kills/culls, and scientific sampling (Hammill et al. 2017a). Removals were aggregated into young of year (YOY) and age-1+ bins.

We used seal and cod diet samples to estimate consumption rates and size-selectivity of herring. The seal diet was inferred from prey hard parts found in the digestive tracts of grey seals collected (i) in coastal areas of the sGSL between late spring and August (1985-2004; Hammill et al., 2007), (ii) from the west coast of Cape Breton Island between September and January (1996-2011; Hammill et al., 2014)., and (iii) in the Cabot Strait, mostly between October and December (2010-2011; Hammill et al., 2014). Seals were sampled on or near shore and the inferred diets likely reflect feeding that occurred near (~30 km) the sampling site (Benoît et al., 2011a). The available diet information was assumed to be spatially representative. Cod in the sGSL have been sampled for diet information since 1959 (Hanson and Chouinard, 2002; Benoît and Rail, 2016). Samples were taken throughout the sGSL and at different times of year when cod were either aggregated or dispersed. Herring consumed by grey seals ranged between 9-39 cm, with 50% of consumed herring between 25-30 cm. Cod tended to consume smaller herring than seals (50% of consumed herring between 13-23 cm, range: 7-29 cm). Length frequencies of herring in predator diets were converted to age-frequencies using annual, subpopulation-specific herring age-length keys, which were based on fishery catch-at-age (gillnet and purse seine combined; François Turcotte, Fisheries and Oceans Canada, unpubl. data).

The movement of seals has been tracked using satellite telemetry since the mid-1990s (Benoît and Rail, 2016; Breed et al., 2006; Harvey et al., 2008). We used this data to infer the monthly seal presence in areas occupied by cod and herring (defined as Div. 4T from May to October and both Div. 4T and Subdiv. 4Vn from November to April), which we then averaged into annual foraging rates (Table 2).

**2.2 Model of Intermediate Complexity for Ecosystem assessment (MICE)**

Our analysis had three steps. First, the MICE was fitted to commercial fishery and survey catch-at-age data for cod and herring, as well as pup production and reproductive data for grey seals, to estimate intraspecific parameters for each species. Intraspecific parameters estimated for cod and herring included recruitment, initial abundance, fishery/survey selectivity, and catchability, while those for seals were maturity rates, initial abundance, the scale and shape of the density dependence relationship, and the maximum reproductive rate. The MICE was also simultaneously fitted to bioenergetically-derived consumption rates and observed age-composition of herring in seal and cod diets to estimate parameters for per-capita consumption and age/size preference by seals and cod. Second, functional responses and the stock-recruitment relationships were post-fitted to MICE estimates of prey consumption and abundance/biomass. Finally, the fitted model, functional responses, and stock-recruitment relationships were projected in stochastic simulations for 50 years under a range of seal and herring harvest levels. These simulations accounted for parameter uncertainty (via random parameter draws from Bayes joint posterior distributions) as well as uncertainty about future variability in natural processes (i.e., recruitment, non-predation mortality, etc.).

Model equations and notation are listed in Appendix A.

***2.2.1 Population modeling***

The MICE consisted of an age-structured model for each species (indexed by $i$; 1=seal, 2=cod, 3=herring) in which abundance $N$ decayed exponentially according to an annual instantaneous total mortality rate $Z$ (yr$^{-1}$), i.e.,

$$N_{i,x,a,t} = N_{i,x,a-1,t-1} \exp\left(-Z_{i,x,a-1,t-1}\right) \quad (1)$$

where $x$, $a$, and $t$ index subpopulation, age, and year, respectively. Each combination of herd and sex was considered a separate grey seal subpopulation. Herring had separate subpopulations for spring spawners (Spring) and fall spawners in each region: north (Fall-N), middle (Fall-M), and south (Fall-S).

Seal pup production was the product of female abundance and pregnancy rates, which were a logistic function of age. Seal recruitment at age-1 was then pup production from the previous year adjusted for mortality arising from weaning, poor ice condition, harvest, and density dependence. Fish ($i$>1) recruitment at age-2 for each subpopulation was modelled as a temporal random walk:

$$N_{i,x,2,t} = N_{i,x,2,t-1} \exp\left(\varepsilon_{i,x,t}^{(\mathrm{R})}\right) \quad (2)$$

We defined $Z$ as the sum of fishing/hunting mortality $F$ and natural mortality $M$, i.e.,

$$Z_{i,x,a,t} = F_{i,x,a,t} + M_{i,x,a,t} \quad (3)$$

where $M$ may arise from predation by one of the predators in our model ($M^{(P)}$) or from another source ($M^{(O)}$):

$$M_{i,x,a,t} = M_{i,a,t}^{(P)} + M_{i,x,a,t}^{(O)} \quad (4)$$

Cod $M^{(O)}$ for ages 2-4 was estimated as a single parameter and was assumed to be time-invariant, while $M^{(O)}$ for ages 5+ varied as a random walk after 1978 to account for non-predation factors (poor growth/condition, unreported catch) that led to elevated cod $M$ in this period (Bousquet et al., 2010; Swain et al., 2011). Herring $M^{(O)}$ was estimated as separate random walks for ages 2-6 and ages 7-11+. To aid model tractability, we assumed herring $M^{(O)}$ in the first time step at 0.4 yr$^{-1}$ for all ages.

We explicitly modelled three predation links in the MICE: (i) grey seal predation on cod aged 5-12+ yr (hereafter “5+ cod”), (ii) grey seal predation on all ages of herring, and (iii) cod predation on all ages of herring. We did not consider a reciprocal effect of prey on predators, such as increased predator $M$ when consumption rates were low, as cod and grey seals are largely generalist predators and thus decreased predation on one species may be compensated by increased predation on another. Additionally, the grey seal population has demonstrated an ability to continue expanding despite low abundance of cod, herring, and mackerel across their range.

For each prey species ($i$ > 1), the instantaneous predation mortality rate ($M^{(P)}$) was calculated as

$$M^{(P)}_{i,a,t} = \sum_j \sum_y \sum_b N_{j,y,b,t}\, m^{(P)}_{j,y,b,i,a,t} \quad (5)$$

where $j$, $y$ and $b$ index predator species, predator subpopulation and predator age, respectively, while $m^{(P)}_{j,y,b,i,a,t}$ represents the annual instantaneous mortality rate imposed on prey $i$ of age $a$ by an individual predator $j$,$y$ at age $b$. We calculated $m^{(P)}$ as

$$m^{(P)}_{j,y,b,i,a,t} = S^{(\mathrm{P})}_{j,i,a}\, f_{j,y}\, \varphi_{j,y,i,t}\, \rho_{j,b} \quad (6)$$

where $S^{(\mathrm{P})}_{j,i,a}$ is the age-selectivity of the prey $i$ to predator $j$ ($0 \le S^{(\mathrm{P})} \le 1$), $f$ is the proportion of each year that the range of predator $j,y$ overlaps with the range of prey, $\varphi$ is the maximum per-capita rate (across predator ages) at which predator $j,y$ consumes prey $i$, and $\rho$ is the relative consumption-at-age for each predator ($0 < \rho \le 1$). We included $\rho$ to account for the different rates at which predators of different ages consume prey (i.e., older/larger predators consume more prey than younger/smaller predators). All 5+ cod were assumed to be fully vulnerable to seal predation (i.e., $S^{(\mathrm{P})}_{j=1,i=2,a} = 1$ for $a \ge 5$). The predation rate on younger cod appears to be small and thus for the model these fish were considered invulnerable (i.e., $S^{(\mathrm{P})}_{j=1,i=2,a} = 0$ for $a < 5$). The age-selectivity of herring to seal predation was assumed to be a logistic function of herring age, i.e.,

$$S^{(\mathrm{P})}_{j=1,i=3,a} = \left(1 + \exp\left[\frac{-\log(19)(a - b^{50\%})}{b^{95\%} - b^{50\%}}\right]\right)^{-1} \quad (7)$$

where $b^{50\%}$ and $b^{95\%}$ were estimated parameters representing the herring ages at which selectivity to herring predation was 0.50 and 0.95, respectively. The age-selectivity of herring to cod predation was assumed to be proportional to a gamma distribution, i.e.,

$$S^{(\mathrm{P})}_{j=2,i=3,a} \propto \frac{1}{\Gamma(k)\theta^k}\, \theta^{k-1} \exp\left(-\frac{a}{\theta}\right) \quad (8)$$

where $k$ and $\theta$ were estimated parameters representing the shape and scale, respectively, of the selectivity function. We chose these distributions after initial trials using a range of selectivity functions suggested that herring selectivity to seal predation was a monotonically increasing function of age whereas selectivity to cod predation was dome-shaped. We assumed that cod and herring had complete spatial overlap throughout the year, while spatial overlap between seals and both fish species was set equal to the mean of the monthly proportion of time that satellite-tracked seals in each herd and for each sex spent near cod and herring (Table 2). Per-capita predation rates were assumed to vary as a temporal random walk with shared annual deviations between predator subpopulations (e.g. Cook & Trijoulet, 2016), i.e.,

$$\varphi_{j,y,i,t} = \varphi_{j,y,i,t-1} \exp\left(\varepsilon_{j,i,t}^{(\varphi)}\right) \quad (9)$$

Relative consumption ($\rho$) was modelled as a logistic function of predator age, i.e.,

$$\rho_{j,y,b} = \left(1 + \exp\left[\frac{-\log(19)\left(b - \rho_{j,y}^{50\%}\right)}{\rho_{j,y}^{95\%} - \rho_{j,y}^{50\%}}\right]\right)^{-1} \quad (10)$$

where $\rho_{j,y}^{50\%}$ and $\rho_{j,y}^{95\%}$ are estimated parameters representing the ages at which consumption reaches 50% and 95% of peak levels, respectively.

Annual per-capita consumption (in numbers) by predator $j$ (of subpopulation $y$ and age $b$) of prey $i$ (of age $a$) was calculated from the Baranov catch equation, i.e.,

$$\hat{c}_{j,y,b,i,x,a,t}^{(N)} = \frac{m_{j,y,b,i,a,t}^{(P)}}{Z_{i,x,a,t}} N_{i,x,a,t}\left(1 - \exp\left(-Z_{i,x,a,t}\right)\right) \quad (11)$$

We scaled per-capita consumption by prey weight-at-age and summed over prey subpopulations to obtain the total consumption in weight per-predator for each predator/prey pair:

$$\hat{c}_{j,y,b,i,a,t} = \sum_x \hat{c}_{j,y,b,i,x,a,t}^{(N)} w_{i,x,a,t} \quad (12)$$

where $w_{i,x,a,t}$ was the annual weight-at-age for each prey subpopulation.

The age-composition of consumed prey was calculated by converting total consumed numbers-at-age to proportions-at-age:

$$\hat{u}_{j,a,t}^{(\mathrm{P})} = \frac{\sum_y \sum_b \sum_x \hat{\boldsymbol{c}}_{\boldsymbol{j,y,b,i,x,a,t}}^{(N)} N_{j,y,b,t}}{\sum_a \sum_y \sum_b \sum_x \hat{\boldsymbol{c}}_{\boldsymbol{j,y,b,i,x,a,t}}^{(N)} N_{j,y,b,t}} \quad (13)$$

The abundance of cod or herring that was vulnerable to the commercial fishery ($g = 1$) or surveys ($g > 1$) was calculated as

$$V_{g,i,x,a,t} = N_{i,x,a,t} S_{g,i,x,a,t}^{(\mathrm{F})} \exp\left(-d_{g,i,x} Z_{i,x,a,t}\right) \quad (14)$$

where $S^{(F)}$ was age-selectivity to the fishery and $d$ was the timing of the fishery (expressed as the approximate ordinal date of the fishery divided by 365). We estimated $F$ for each fish species and subpopulation by iteratively solving the Baranov catch equation:

$$C^{(F)}_{i,x,t} = \sum_a \frac{F_{i,x,t}}{Z_{i,x,a,t}} V_{g=1,i,x,a,t} w^{(\mathrm{F})}_{g=1,i,x,a,t} \left(1 - \exp\left(-Z_{i,x,a,t}\right)\right) \quad (15)$$

Similar equations were used to solve for $F$ for YOY and age-1+ seals.

We calculated model-predicted cod and herring biomass indices ($g$>1) as the product of vulnerable biomass and fishery/survey catchability ($q$)

$$\hat{I}_{g,i,t} = q_{g,i,x,t} \sum_a V_{g,i,x,a,t} w^{(\mathrm{F})}_{g,i,x,a,t} \quad (16)$$

We assumed that $q$ was time-invariant for all cod surveys. For the herring gillnet CPUE indices, $q$ varied as a random walk to allow for changes in $q$ that are expected to arise from stock and fishery changes. For instance, $q$ is expected to increase as a stock declines and occupies smaller areas (Winters and Wheeler 1985), though time/area closures implemented since 2010 are expected to decrease $q$ (Swain 2016). Cod and herring age-composition was calculated by converting vulnerable numbers-at-age to proportions-at-age, i.e.,

$$\hat{u}^{(\mathrm{F})}_{g,ix,a,t} = \frac{V_{g,i,x,a,t}}{\sum_a V_{g,i,x,a,t}} \quad (17)$$

Grey seal foraging effort (seal-yrs) near sGSL cod and herring was the product of seal abundance and foraging rates $f$, i.e.,

$$E_{y,t} = f_{j=1,y} \sum_a N_{i=1,y,a,t} \quad (18)$$

### 2.2.2 *Model fitting*

Cod and herring biomass indices were assumed to arise from lognormal distributions, while age-proportions of cod and herring in fishery and survey catches were assumed to arise from logistic-normal distributions (Schnute and Haigh 2007, Francis 2014). Process errors in fish recruitment, prey consumption and herring catchability were assumed to arise from zero-mean lognormal distributions. Grey seal pup production observations were also assumed to arise from lognormal distributions, while the number of pregnancies were assumed to arise from age-specific binomial distributions.

Stock assessment models with predation may be fit to total estimates of prey consumption as a means of bounding the model consumption within a plausible range (e.g., Cook et al. 2015). Without priors or external information about consumption, predation mortality may absorb statistical noise and produce estimates of prey consumption that are considered biologically impossible. Total prey consumption has previously been estimated for the species under consideration by first calculating per-capita consumption then scaling by predator abundance as estimated by single-species population models (e.g., Benoît and Rail 2016). We could not incorporate these estimates into the model, as the predator abundance estimates were based on the same data that are used in our analysis. Instead, we fit the model to per-capita consumption-at-age ($\hat{c}$) based on bioenergetic consumption estimates, the spatiotemporal overlap between predator and prey species, and the proportional contribution of prey to predator diets (Appendix B). For each predator subpopulation and each associated prey species, we assumed a lognormal distribution for prey consumption at predator age $b$, centered on $c$, i.e.,

$$\ln \sum_a \hat{c}_{j,y,b,i,a,t} \sim \mathrm{N}\left(\ln c_{j,y,b,i,t}, \sigma^2_{\mathrm{c},j}\right) \quad (19)$$

We chose $\sigma_{\mathrm{c},j}$ = 1 as the standard deviation for seal predation on cod and herring to allow the model to deviate from the prior mean while preventing model-predicted consumption from becoming implausibly small or large (i.e., the consumption by a predator of particular species should not be larger than the total bioenergetic-predicted consumption of all species by that predator). In contrast, initial trials revealed that using $\sigma_{\mathrm{c},j}$ = 1 for cod predation on herring allowed the model to estimate consumption levels that were inconsistent with previous analyses of cod consumption, so we used a smaller value of $\sigma_{\mathrm{c},j}$ = 0.1 for this predation link. Sensitivity analyses are provided to show how model outputs respond to alternative values of $\sigma_{\mathrm{c,j}}$.

We also fit the model to the observed age-composition of herring in seal and cod diets from diet studies. The subpopulation-of-origin of herring in the diet composition analyses were unknown so we converted the entire set of length frequencies from these studies into a separate set of age frequencies for each subpopulation using age-length keys. We then aggregated age frequencies across subpopulations to create a single set of age frequencies for each predator. We assumed a logistic-normal likelihood for the age-composition of herring in predator diets.

The model was implemented using the Template Model Builder (TMB; Kristensen *et al.* 2016) package within R version 3.6.3 (R Core Team 2019). Model estimates and uncertainty were based on Hamiltonian Monte Carlo (HMC) samples from the joint posterior distribution. We ran 4 HMC chains for 2000 iterations each, discarding the first half of each chain as a warm-up. Starting values for each chain were randomly sampled based on maximum likelihood estimates and associated standard errors. We monitored convergence using the potential scale reduction factor on rank-normalized split chains ($\hat{R}$) and the effective sample size of the rank-normalized draws (Vehtari 2020).

### *2.2.3 Extirpation risk projections*

To evaluate the effects of seal and herring abundance changes on sGSL ecosystem dynamics, we projected the model forward in time (50 years) under a range of seal and herring harvest strategies. For each posterior sample, we post-fitted functional responses and stock-recruitment functions, then projected the model based on the posterior sample, the post-fitted structural forms, and a harvest plan for seals and herring. We post-fitted functional response and stock-recruitment relationships to model estimates rather than estimating them within the model to avoid imposing structure on relationships for which the data are typically uninformative (Cook and Trijoulet, 2016). For cod and each seal subpopulation, we fitted the relationship between prey density and the consumption rate by predators with a functional response (dropping predator species and subpopulation indices for simplification):

$$\varphi_{i,t} = \frac{\eta_i B_{i,t}^{\lambda_i - 1}}{1 + \sum_l \left(\eta_l h_l B_{l,t}^{\lambda_l}\right)} \qquad (20)$$

where $\varphi_{i,t}$ is the annual consumption rate of prey species $i$ per unit of prey $i$ biomass, $B_{i,t}$ represents the biomass of prey $i$ (summed across subpopulations) available to the predator, $\eta_i$ represents the rate at which the predator encounters prey $i$, $h_i$ represents the time the predator spends consuming prey $i$ (or the "handling time"), and $\lambda_i$ determines the shape of the functional response. This equation describes a multispecies functional response for seals and a single-species response for cod. The biomass of prey species not explicitly included in our analysis is assumed to be constant over time.

We characterized the cod stock-recruitment relationship in the projections using an extended Ricker function that accounted for a herring effect on prerecruit cod (Minto and Worm 2012):

$$N_{i=2,x=1,a=2,t} = \hat{R}_t = \beta_0 C_{t-2} \exp\left(-\beta_C C_{t-2} - \beta_H \sum_x H_{x,t-2}\right) \qquad (21)$$

where $C_t$ and $H_{x,t}$ are the spawning biomasses of cod and herring, respectively, in year $t$, $\beta_0$ and $\beta_C$ are standard Ricker parameters representing fecundity and density-dependence, respectively, and $\beta_H$ represents the strength of herring predation on or competition with prerecruit cod. For each posterior sample, we fitted the full extended Ricker model (ER), as well as a standard Ricker (SR) version of ER in which the herring effect ($\beta_H$) was fixed at 0. Ricker models were fitted to posterior estimates of cod recruitment ($R$) assuming lognormal errors, i.e.,

$$\ln R_t = \ln \hat{R}_t + \varepsilon_t \qquad (22)$$

where the residuals $\varepsilon$ were either *iid* ($\varepsilon_t \sim \mathrm{N}(0, \sigma_S^2)$) or AR(1) autocorrelated ($\varepsilon_t \sim \mathrm{N}(\phi \varepsilon_{t-1}, \sigma_S^2)$). We used Durbin-Watson tests to determine whether projections should be based on either *iid* or AR(1) models. We similarly projected herring recruitment for each subpopulation using a standard Ricker function, though we considered Beverton-Holt recruitment as a sensitivity analysis (Appendix C).

We projected the model under all combinations of the following four controls:

1) Annual seal quota (thousands of seals): (0, 1, 2, …, 20)
2) Proportion of quota allocated to YOY seals: (0.5, 0.75)
3) Length of seal harvest (yrs): (5, 10)
4) Constant herring $F$ (yr$^{-1}$): (0, 0.1, 0.2, …, 1.0)

To simulate the effect of a short-term increase in seal harvest, the seal quota was taken only in the initial projection years (i.e., quota from a 5-year harvest was taken only in the first five years of the projection). Additionally, the quota was only taken from the Gulf herd as these seals spend significantly more time foraging near sGSL cod and herring than Shelf herd seals. The additional quota in initial years and the targeting of seals in the sGSL are consistent with an adaptive management approach proposed a decade ago as part of a review of grey seal impacts in Eastern Canada (DFO 2011; Hammill and Swain 2011). Annual Shelf harvest levels in the projections, as well as annual Gulf harvest levels after the initial harvest period, were randomly sampled from historical harvest observations. Seal harvests were applied to sex- and age-classes in the projections, as in the historical model, in proportion to the relative abundance of those classes. We assumed no commercial fishery landings of cod in the projections consistent with the moratorium on directed fishing in place for the stock since 2009. The random walks in recruitment and consumption in the model were replaced by the post-fitted recruitment and consumption relationships. Specifically, Eq (9) was replaced by Eq (20), while Eq (2) was replaced by Eq (21) for cod and Eq (23) for herring.

To evaluate cod recovery potential, we compared cod SSB at the end of the projection period with a limit reference point (LRP) of 80 kt, which was based on the lowest spawning stock biomass (SSB) from which the stock has recovered (Chouinard et al., 2003).

## 3 Results

Overall, the model fitted the seal pup production and reproductive data, as well as cod and herring fishery and survey catch-at-age, reasonably well, and we did not detect issues with convergence (Fig. 3; Appendix D). Estimated seal abundance increased from approximately 15,000 in 1971 to 428,000 in 2018, corresponding to a rise in foraging effort in the sGSL from approximately 6,000 seal-yrs to nearly 43,000 seal-yrs over that time frame (Fig. 4a). Estimated predation $M$ for 5+ cod grew from 0.02 $yr^{-1}$ in the early 1980s to 0.83 $yr^{-1}$ in 2018 (Fig. 4b). Other $M$ estimates for 5+ cod rose from 0.2 $yr^{-1}$ in the 1970s to 0.40 $yr^{-1}$ in the early 1990s and has since declined to 0.25 $yr^{-1}$ (Fig. 4b). Estimated cod fishing mortality rose to 0.55 $yr^{-1}$ in the mid-1970s and again early 1990s as the cod population was collapsing (Fig. 4b). Fishing was a minor source of cod mortality from 1993 to 2008 and was negligible thereafter (Fig. 4b). Estimated annual cod consumption by seals rose from less than 3 kt in the early 1970s to 16 kt in the mid-2000s and has since declined to 10 kt (Fig. 4c). Estimated annual cod consumption by seals has exceeded cod fishery landings since 1993 (Fig 4c.). Estimated cod SSB grew rapidly in the late 1970s and remained high until the mid-1980s before declining to low levels (Fig. 4d). Cod SSB declined precipitously in recent years, falling from 32 kt in 2016 to 17 kt in 2018. Estimated cod recruitment has steadily declined from nearly 1.5 billion in the early 1980s to less than 60 million in recent years (Fig 4e). The estimated cod recruitment rate (age-2 abundance divided by SSB two years earlier) rose dramatically in the late 1970s to more than 8,000 recruits per tonne of SSB, contributing to the rapid recovery of cod during this time (Fig. 4f). The cod recruitment rate doubled from the early 1990s to the early 2010s but has since declined (Fig. 4f).

Seal and cod predation represented moderate sources of mortality for herring at various periods. Seals preferentially selected for larger herring, while cod selected for moderately-sized (age 3-7) herring (Fig. 5). Seal predation mortality for herring rose steadily from less than 0.04 $yr^{-1}$ the 1970s to nearly 0.18 $yr^{-1}$ in recent years (Fig. 6a). Cod predation was an appreciable source of herring mortality in the mid- to late-1980s but has since declined to negligible levels (Fig. 6a). Other herring $M$ for ages 2-6 increased steadily from 0.42 $yr^{-1}$ in the mid-1980s to more than 0.70 $yr^{-1}$ in recent years, but was relatively stable around 0.22 $yr^{-1}$ for ages 7-11+ over this period (Fig. 6b). Fishery landings of herring were larger than estimated seal and cod consumption in all years except for the mid-1980s during the peak years of cod predation (Fig. 6c). Herring SSB rapidly increased from 122 kt in 1977 to nearly 400 kt in 2006 before gradually declining to less than 200 kt in recent years (Fig. 6d). Rapid cod population growth in the late 1970s overlapped with a period of low and declining herring SSB.

The estimated predation rate imposed on age 5+ cod per seal was largely steady until the 2000s, at which point it increased sharply (e.g., Fig. 7). The predation rate imposed on herring per seal declined steeply in the early 1980s and has increased steadily since then (Fig. 7). Functional responses fit the model estimates of prey consumption by seals closely and exhibited strongly hyperbolic patterns consistent with a Type II functional response (Fig. 7). The relationship between cod predation and herring biomass was less clear; the shape parameter was greater than 1 (indicating a sigmoidal or Type III functional response) for 85% of posterior samples, while a hyperbolic functional response emerged in other samples (Appendix D)

Durbin-Watson tests for the *iid* stock-recruitment models indicated the presence of lag-1 autocorrelation in residuals, so inference was based on the AR(1) models. Cod recruitment in the extended Ricker AR(1) model was negatively associated with herring biomass. For instance, an increase in herring SSB from the mean estimated level (274 kt) to the maximum estimated level 393 kt) corresponded to a 32% decrease in cod recruitment in the extended Ricker AR(1) model, while a decrease in herring SSB from the mean level to the minimum level (122 kt) corresponded to a 40% increase in cod recruitment (Fig. 8).

Cod failed to recover in nearly all projection scenarios in which annual seal quotas were less than 5,000 seals (Fig. 9). Five-year harvests were less effective than ten-year harvests at facilitating cod recovery, as five-year harvests required extremely high quotas (e.g., >15,000 seals) and required a sufficiently high proportion of age 1+ seals to be targeted, to adequately reduce grey seal predation on cod (Fig. 9). There was a positive relationship between herring *F* and cod recovery; cod were generally unlikely to recover when herring *F* was less than 0.2 $yr^{-1}$ (Fig. 9). Targeting higher proportions of YOY seals was relatively ineffective in decreasing seal predation on cod as quotas for YOY seals quickly exceeded the number of pups being born, leaving much of the quota unfilled.

Even in the most optimistic cod projections, cod SSB continued to decline for 10 years and did not surpass the LRP for more than 20 years. We consider harvest strategies that set an annual quota of 12,000 seals targeting 50% YOY for either 5 or 10 years, with either low (0.1 $yr^{-1}$) or high (0.6 $yr^{-1}$) herring *F* as this set of strategies produces a range of outcomes for each species that can be generalized (Fig. 10). Cod recovery was only likely under longer (10 year) seal harvests (Fig. 10c) that rapidly removed the entire Gulf herd (Fig. 9a). The Gulf herd quickly recovered after shorter (5 year) seal harvests in which it was not driven to zero abundance (Fig. 10a), resulting in predation mortality rates that the still-collapsed cod stock was unable to withstand (Fig. 10d). Following five-year seal harvests, seals still imposed high predation mortality rates despite reduced abundance due to the hyperbolic functional response, which specifies exponentially increasing mortality with declining prey abundance. Seal foraging effort increased in years following 10-year seal harvests due to the growing Scotian Shelf herd; however, it took more than 50 years for seal foraging effort in the sGSL to recover to pre-harvest levels (Fig. 10b), by which point cod had sufficiently recovered to levels that could sustain seal predation (Fig. 10c). Seal harvests that enabled cod recovery were ultimately negative for herring, as increased predation from recovered cod outweighed reduced predation from seals (Fig. 9e).

Model estimates were relatively insensitive to the choice of consumption prior for seal predation (Appendix E). Additionally, using Beverton-Holt functions to project herring recruitment instead of Ricker functions had negligible impacts on projections (Appendix C).

## 4 Discussion

Many studies have linked reduced survival of cod in the sGSL to increases in seal predation (e.g., Benoît et al., 2011; Swain et al., 2015; Neuenhoff et al., 2019). In this paper, we analyzed a triangular food web of grey seals, cod and herring in the sGSL and found that cod failed to recover in the absence of grey seal abundance reductions. The seal quotas required to sufficiently reduce predation mortality on cod to allow for the chance of cod survival were significantly higher than current removal levels and would likely collapse the grey seal herd in the Gulf of St. Lawrence. Herring biomass was found to have a negative relationship with cod recruitment; however, seal predation accounted for a smaller proportion of herring mortality than fishing or other sources of natural mortality, so herring did not increase greatly in abundance in response to grey seal reductions. Additionally, while our analysis suggests that cod may rebound in response to sufficiently severe grey seal reductions, cod spawning biomass continued to decline for about a decade in these scenarios to less than 1,500 t and was not projected to be at the LRP for several decades. During this period, cod would be vulnerable to extirpation by environmental stochasticities and/or processes not captured by our model.

Grey seal impacts on sGSL cod were last evaluated by Neuenhoff et al. (2019), who found that the extirpation of sGSL cod was likely without a strong and rapid reduction in seal abundance ($E$<27,000 seal-years by 2024). In comparison to Neuenhoff et al. (2019), our analysis (i) added eight years of cod removals in fisheries directed to other stocks and survey catch-at-age data that indicated further population decline (e.g., the cod RV biomass index fell by 60% between 2010 and 2018), (ii) incorporated an updated grey seal population dynamics model with new pup production observations in 2016 and several years of new reproductive data (Rossi et al., 2021), rather than simply fitting to output from the seal assessment, (iii) incorporated herring dynamics, and (iv) estimated parameters for all species under consideration in a single integrated framework. Adding the 2016 Gulf grey seal pup production observation to the model changed the perception of the Gulf population from one that was growing to one that had reached its carrying capacity and stabilized. The change in Gulf estimates was not a result of changes in assumptions between our model and the previous grey seal assessment model, as similar estimates were obtained when the 2016 pup production observations were added to that assessment model (Hammill et al., 2017a). Our estimates of seal foraging effort varied from those used in Neuenhoff et al. (2019), particularly in the last two decades, for which our model predicted that seal foraging effort levelled off around 41,000 seal-years, while the foraging effort estimates used in Neuenhoff et al. (2019) continued to increase rapidly to nearly 78,000 seal-years in 2014. Despite these differences, as well as differences in modelling approaches, our estimates of cod abundance and predation mortality broadly match those of Neuenhoff et al. (2019).

Extending the cod-seal analysis of Neuenhoff et al. (2019) to a MICE allowed us to test hypotheses about the effect of management actions aimed at grey seals or herring on cod recovery in a single framework, avoiding ad hoc procedures for propagating uncertainty from one species to another, and without modelling the entire ecosystem. However, despite these benefits, the MICE was time-consuming to construct and tune, and still required many simplifying assumptions. In particular, the inclusion of herring in the model was challenging given that herring (i) consisted of multiple subpopulations in the sGSL, (ii) had more volatile dynamics than seals or cod, and (iii) had noisy and/or conflicting data. Accounting for nonstationarities in herring catchability, selectivity, and natural mortality increased model complexity and runtime. Multiple biomass indices existed for each herring subpopulation and these indices were difficult to reconcile with one another, leading to the exclusion of some fishery-independent herring data. Moreover, the herring data we included for 1971-1977, which is excluded from assessment models (DFO 2018), was less reliable and coarser than the 1978-2018 data; however we needed to include herring data prior to 1978 since it was important to capture this period when cod rebounded from low abundance. The model estimated that a very large herring population could weaken cod recruitment, but it is difficult to imagine that the herring population would grow to such levels given that $M$ is currently elevated for young herring and the population is actively fished. We also note that our analysis was applied to an ecosystem that has been closely studied for decades, and for which an array of data exists for each species under consideration, including relative abundance-at-age, diet composition, and spatiotemporal predator foraging behaviours. While gaps exist in our data, these types of data may be entirely unavailable for species of concern in other, less studied ecosystems, making our approach difficult to apply to other systems.

Hyperbolic (Type II) functional responses for seal consumption of both cod and herring emerged from MICE estimates. This functional response states that per-capita consumption of prey increases exponentially as prey abundance declines, even when prey are at low abundance. These dynamics lead to the extirpation of the target prey species if the abundance of that species falls below some critical threshold. Sigmoidal (Type III) functional responses could emerge at lower prey densities if seals switch to alternative prey. In this case, the primary prey species may be trapped at low abundance in a "predator pit", i.e., the prey species would not be extirpated by predation, since predation mortality decreases at very low prey abundance, but the prey species also fails to recover, since increases in abundance are countered by increased predation mortality. Evidence for prey switching by seals in the literature is mixed; fine spatial and temporal scale analysis of North Sea grey seals diet showed evidence of prey switching at low prey density (Smout et al. 2014), whereas population modelling of grey seal predation on West of Scotland cod suggested that the functional response was hyperbolic (Cook et al. 2015). A hyperbolic functional response was also observed in harbour seal (*Phoca vitulina*) predation on salmon (Middlemas et al. 2006). While it is difficult to predict if prey switching will occur for grey seals preying on cod or herring in the sGSL, we note that cod are already at extremely low abundance and their aggregative behaviour while overwintering suggests that seals may continue to target cod even as cod abundance further declines. A hyperbolic functional response is also plausible for herring, given that herring aggregate during the winter (Chouinard and Hurlbut 2011), and also while spawning when they are particularly energy dense. Additionally, herring occupy coastal waters where seals are more likely to occur.

We did not model an effect of prey on seal mortality or reproductive success. This assumption is appropriate as grey seals are highly mobile generalist predators and are therefore less sensitive to changes in the biomass of a specific prey compared to predators that are more stationary and/or have more specialized diets. On the other hand, the seal model in our analysis assumes that no exchange between herds and therefore that no recolonization takes place once the Gulf herd reaches zero abundance. This assumption is more tenuous, given the large number of grey seals in adjacent ecosystems and the slowing of growth on Sable Island, suggesting that seals from Sable Island may be seeking less crowded whelping grounds. However, the data needed to characterize grey seal colonization is unavailable.

We aimed for the herring component of the MICE to match the accepted assessment model for herring as closely as a possible, though our inclusion of data for 1971-1977 and our exclusion of some survey data (e.g., catch-at-age from experimental gillnets, fall acoustic surveys, and the RV survey) suggest that notable differences between the two models are likely. Indeed, the MICE estimated smaller overall biomass than the assessment model throughout the time series. Additionally, the MICE estimated an increasing trend in herring $M$ for ages 2-6, with terminal values greater than 0.70 $yr^{-1}$, whereas the assessment model estimated a fairly stable trend in $M$ for the spring subpopulation and decreasing trends in the fall subpopulations, terminating at 0.05 $yr^{-1}$. It is not surprising that $M$ estimates for young fish could diverge between the two models, as juvenile $M$ is correlated with recruitment and is therefore difficult to reliably estimate. Given that $M$ of 0.05 $yr^{-1}$ is anomalously low for small fish that are eaten by numerous predators (e.g., Benoît and Rail.2016), we considered the MICE estimates to be more plausible. Furthermore, we note that the increasing trend for other $M$ for juvenile herring since the early 1980s estimated by the MICE is consistent with the increase in predators in the ecosystem that feed on herring of these sizes, notably Northern gannets (*Morus bassanus*), cormorants (*Phalacrocorax auratus* and *P. carbo*) and bluefin tuna (*Thunnus thynnus*) (Benoît and Rail, 2016).

A shortcoming of MICE is that the appropriate degree of complexity (e.g., number of species or interactions) is unknown *a priori*. Yodzis (1998) demonstrated that ignoring feeding links accounting for more than 10% of consumption by or for a species in a Benguela ecosystem food web model led to unreliable model estimates. In our model, seal predation accounts for about 80% of cod mortality and there is no evidence that other suspected cod predators are important drivers of the remaining mortality. Studies of pinniped diets in Atlantic Canada found no evidence of piscivory in the sGSL among harp seals (*Phoca groenlandica*) and hooded seals (*Cystophora cristata*), and found harbour seals (*Phoca vitulina*) to be a minor predator of sGSL cod (Hammill and Stenson, 2000). Expanding the model to include harbour seals would be difficult as these seals are not regularly surveyed in Atlantic Canada and their abundance is unknown (e.g., Mosnier et al. 2023). Juvenile white sharks (*Carcharodon carcharias*) use the sGSL and are piscivorous; however the population is generally considered depleted. Atlantic Bluefin Tuna (*Thunnus thynnus*) use the sGSL (Block et al., 2019) but cod are a minor prey (Pleizner et al. 2012; Varela et al., 2020).

We chose herring to represent pelagic fish in the hypothesized triangular food web for sGSL cod, as age-structured herring data were available; however, indirect effects of seal reductions on cod survival could also be expressed through mackerel, the other common small pelagic species in the system (Swain and Sinclair, 2000). We tested cod stock-recruitment models that included mackerel spawning biomass as a covariate using estimates from DFO (2023), but found a negligible effect compared to the effect of herring.

Evaluating the logistics of grey seal predator control was outside the scope of this analysis; however, large-scale reductions of the grey seal population may be infeasible. The difficulty of hunting and recovering grey seals renders many potential commercial hunts unprofitable, so scaling up commercial harvests would likely require government subsidies. Indeed, recent commercial quotas have gone unfilled and most hunting licenses are inactive. Additionally, the status of Brion Island, which is a provincial sanctuary and largest grey seal colony in the Gulf of St Lawrence, and Sable Island National Park, precludes hunting at these locations where predator control programs would likely be most effective. Similar barriers exist for managing the growing grey seal colonies in US waters. The US Marine Mammal Protection Act generally prohibits the use of hunting in managing seal populations in US waters, though the lethal removal of California sea lions (*Zalophus californianus*) predating upon Pacific salmon and steelhead (*Oncorhynchus spp.*) below the Bonneville Dam has recently been authorized (Marine Mammals; Pinniped Removal Authority; Revised Authorization, 2019). More generally, current reference points under the precautionary approach for grey seals result in a narrow scope for removals that might facilitate rebuilding in sGSL cod (Rossi et al. 2021).

When ecosystems have been altered and one species of interest is at risk of extinction due at least partly to a newly-abundant predator and/or competitor, natural resource managers are confronted with a choice between active control of the predator/competitor, passive controls such as reducing anthropogenic impacts, or "letting nature run its course" (Lessard et al., 2005). There may also be calls for further research to better understand predation impacts and indirect impacts of active controls before making decisions. In the case of cod in the southern Gulf of St. Lawrence, there are few options for passive controls, as directed fishing has already been curtailed and unreported catch is thought to be low. Letting nature run its course will likely lead to the extirpation of cod from this ecosystem (Swain and Chouinard, 2008; Neuenhoff et al., 2019; this study). Our study demonstrates that there are opportunity costs associated with further research, as was recommended as part of a large review of grey seal predation impacts in Atlantic Canada (DFO 2011; Hammill and Swain 2011), In the time it has taken collect additional data on seal diets, establish predation links to cod, and analyze the indirect effects of a seal cull via pelagics, the cod population has declined to critically low levels with little potential for recovery. In other ecosystems where cod mortality and seal abundance are both elevated (O'Boyle and Sinclair 2012; Rossi et al. 2019), such as the Scotian Shelf and Georges Bank, significantly less is known about seal foraging behaviours and diets (e.g., Rossi et al. 2023), so developing MICE to analyze potential responses seal population management in these ecosystems would be more difficult than for the sGSL. Decisions between active, passive, or no ecosystem controls for these ecosystems will therefore need to made without a full understanding of the possible effects.

**Acknowledgments**

We thank Doug Swain for providing feedback at each stage of the analysis. We also thank Michael Bradford for helpful discussion. Finally, we thank François Turcotte for providing herring age-length keys.

## Tables

Table 1. Summary of data used in population model. Indices represent pup abundance for grey seals and age-aggregated biomass for the fish species. Abbreviations: *RS* reproductive status, *BE* bioenergetic estimate, *AC* age-composition, *FI* fishery independent, *FD* fishery dependent, *g* fishing gear.

| Source | Data type | Ages | Years | *g* | Comment |
|---|---|---|---|---|---|
| ***Grey Seal*** | | | | | |
| Pup production survey | Index | 0 | 1971-2016 | 1,2 | Herd-specific (1=Shelf, 2=Gulf) |
| Reproductive survey | RS | 4-8+ | 1982-2015 | - | Gulf samples only |
| Per-capita consumption of cod and herring | BE | 0-30 | 1971-2018 | - | Year-invariant (see Appendix B) |
| Herring-at-age in diet | AC | - | 1986-2011 | - | Year-, herd- and age-invariant |
| ***Atlantic Cod*** | | | | | |
| Commercial fishery | Catch, AC | 2-12 | 1971-2018 | 1 | FD |
| Bottom trawl (RV) survey | Index, AC | 2-11 | 1971-2018* | 2 | FI |
| Mobile sentinel (MS) survey | Index, AC | 2-11 | 2003-2018 | 3 | FI |
| Longline (LL) sentinel survey | Index, AC | 5-11 | 1995-2017 | 4 | FI |
| Per-capita consumption of herring | BE | 2-12 | 1978-2018 | - | See Appendix B |
| Herring-at-age in diet | AC | - | 1990-2013 | - | |
| ***Atlantic Herring*** | | | | | |
| Commercial fishery | Catch, AC | 2-11 | 1971-2018 | 1 | FD, stock-specific |
| Gillnet CPUE | Index, AC | 4-10 | 1986-2018† | 2 | FD, stock-specific |
| Acoustic survey | Index, AC | 4-8 | 1994-2017 | 3 | FI |
| Assessment biomass series | Index | 4-11 | 1971-1977 | 6 | Aggregated across all subpopulations |

* No data in 2003 † Spring data starts in 1990

Table 2. Proportion of time spent by satellite-tracked grey seals in areas occupied by sGSL cod or herring (from Benoît and Rail, 2016).

| Month | Area | Gulf Male | Gulf Female | Shelf Male | Shelf Female |
|---|---|---|---|---|---|
| Jan | 4T + 4Vn | 0.647 | 0.534 | 0.020 | 0.091 |
| Feb | 4T + 4Vn | 0.577 | 0.417 | 0.059 | 0.017 |
| Mar | 4T + 4Vn | 0.441 | 0.208 | 0.000 | 0.000 |
| Apr | 4T + 4Vn | 0.553 | 0.302 | 0.000 | 0.000 |
| May | 4T | 0.522 | 0.226 | 0.000 | 0.000 |
| Jun | 4T | 0.876 | 0.606 | 0.000 | 0.035 |
| Jul | 4T | 0.798 | 0.615 | 0.031 | 0.065 |
| Aug | 4T | 0.801 | 0.605 | 0.067 | 0.055 |
| Sep | 4T | 0.842 | 0.549 | 0.056 | 0.045 |
| Oct | 4T | 0.908 | 0.562 | 0.046 | 0.055 |
| Nov | 4T + 4Vn | 0.946 | 0.582 | 0.012 | 0.045 |
| Dec | 4T + 4Vn | 0.841 | 0.707 | 0.003 | 0.045 |
| Mean | | 0.729 | 0.493 | 0.024 | 0.038 |

**Figures**

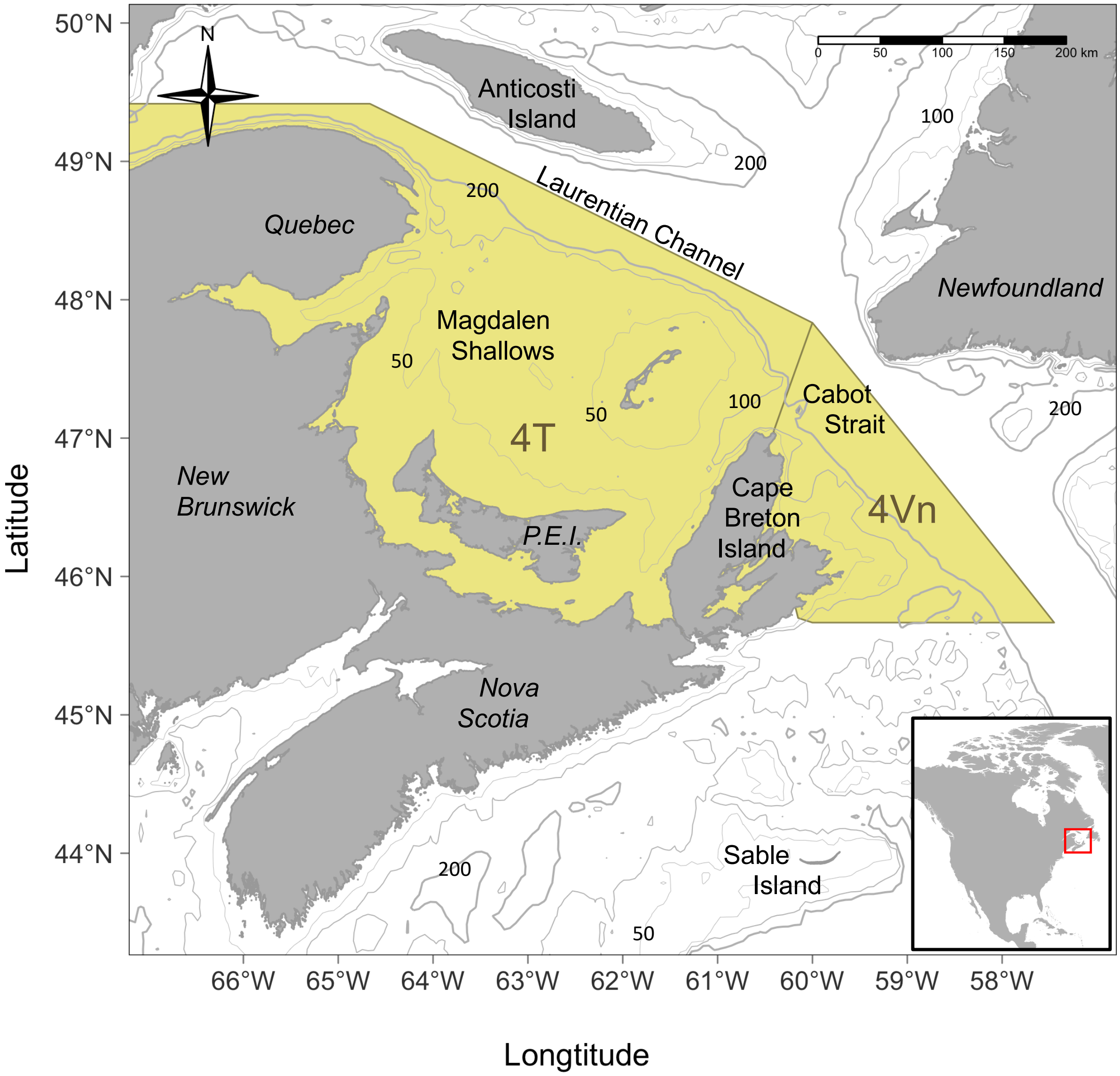

Figure 1. The southern Gulf of St. Lawrence (sGSL) and place names mentioned in the paper. Grey lines indicate the 50, 100, and 200 m depth contours; shaded regions indicate NAFO Divisions occupied by the sGSL stocks of cod and herring (4T and 4Vn). Abbreviations: *P.E.I.* - Prince Edward Island.

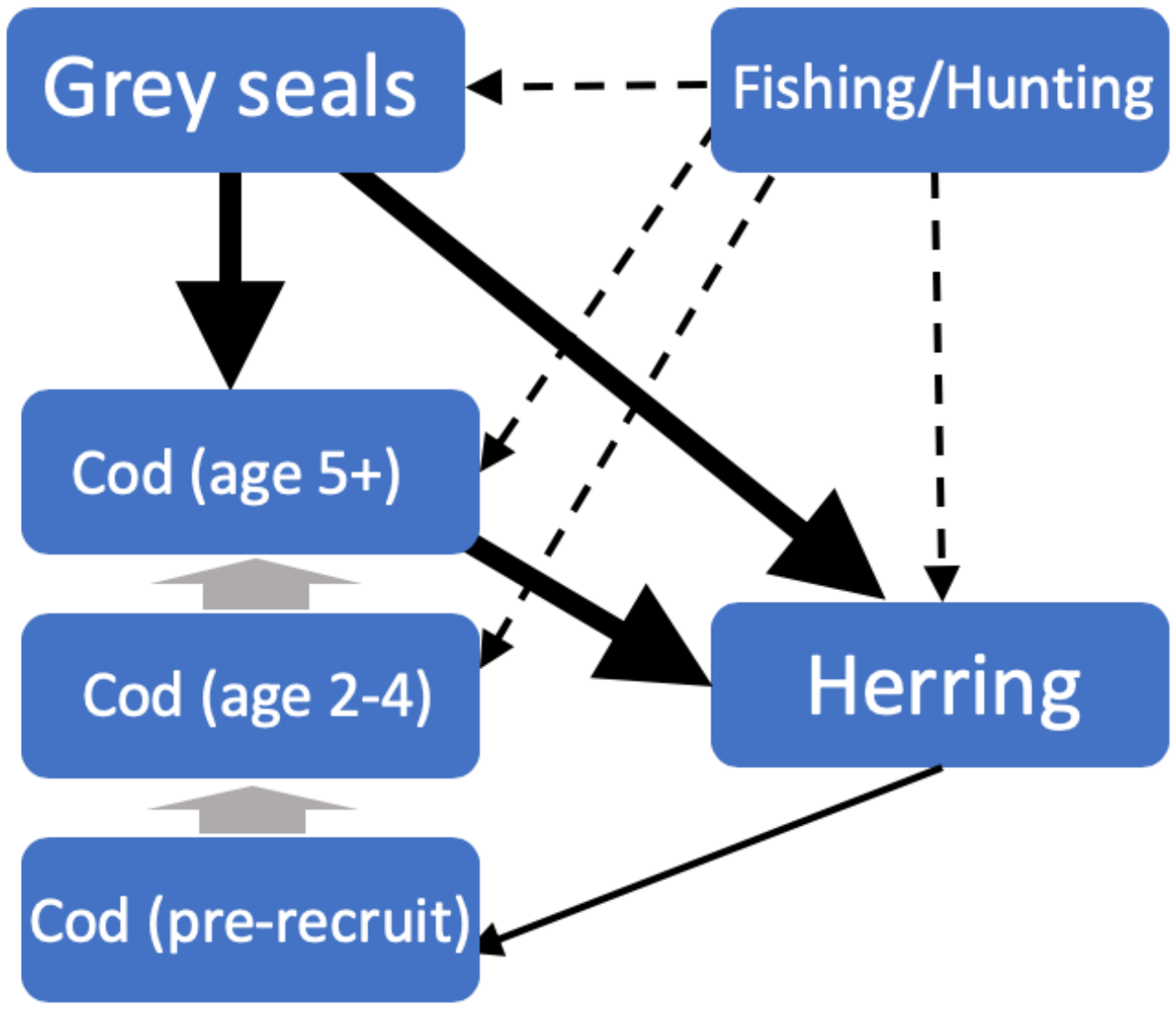

Figure 2. Schematic of seal-cod-herring triangular food web model. Thick black arrows indicate an explicitly modelled predation link (with arrows pointing towards prey), thin black arrows indicate an implicitly modelled link (i.e., via a stock-recruitment function), broken arrows indicate anthropogenic links, and grey arrows indicate growth through life stages.

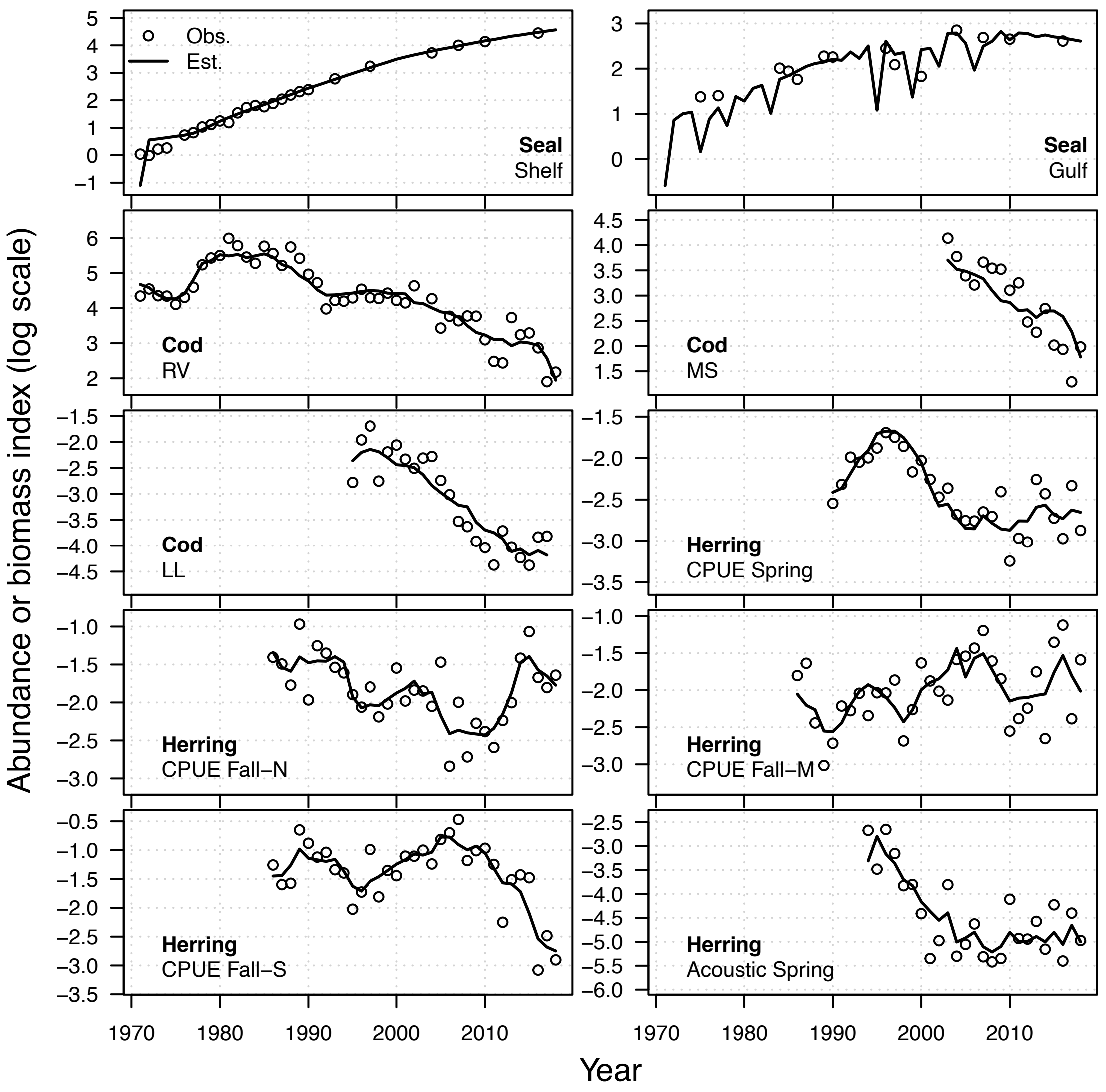

Figure 3. Model fits (posterior modes; lines) to observed indices (circles) for each modelled species. The abundance index represents pup production for grey seals and vulnerable biomass for cod and herring.

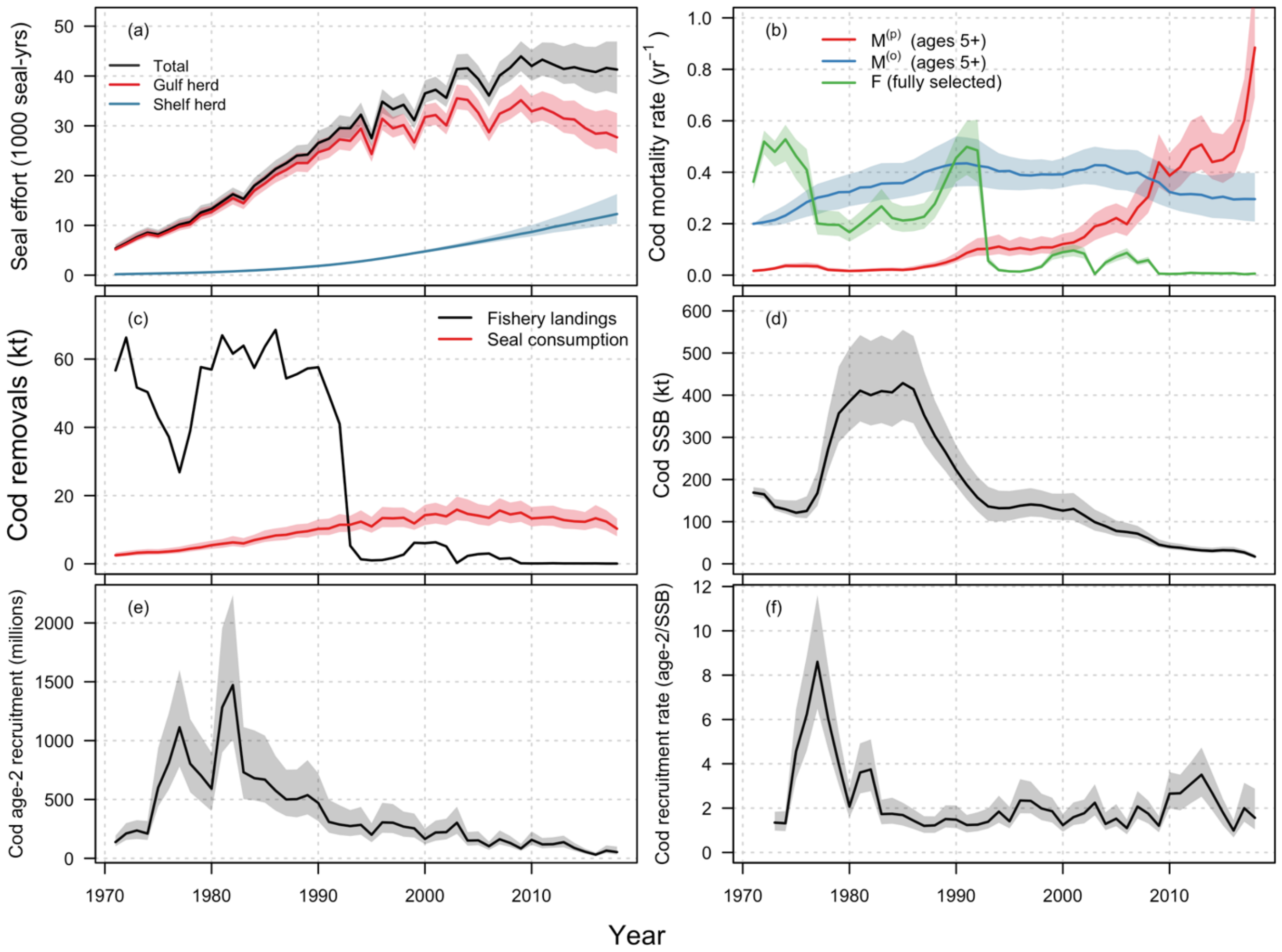

Figure 4. Model estimates of (a) grey seal foraging effort in the sGSL, (b) instantaneous cod mortality rates, (c) cod removals from fisheries (observed) and seal consumption (estimated), (d) cod spawning stock biomass on Jan 1, (e) cod recruitment, and (f) cod recruits-per-spawner. Lines represent posterior modes (except for fishery landings in (c) which were observed) and shaded regions represent central 95% uncertainty intervals.

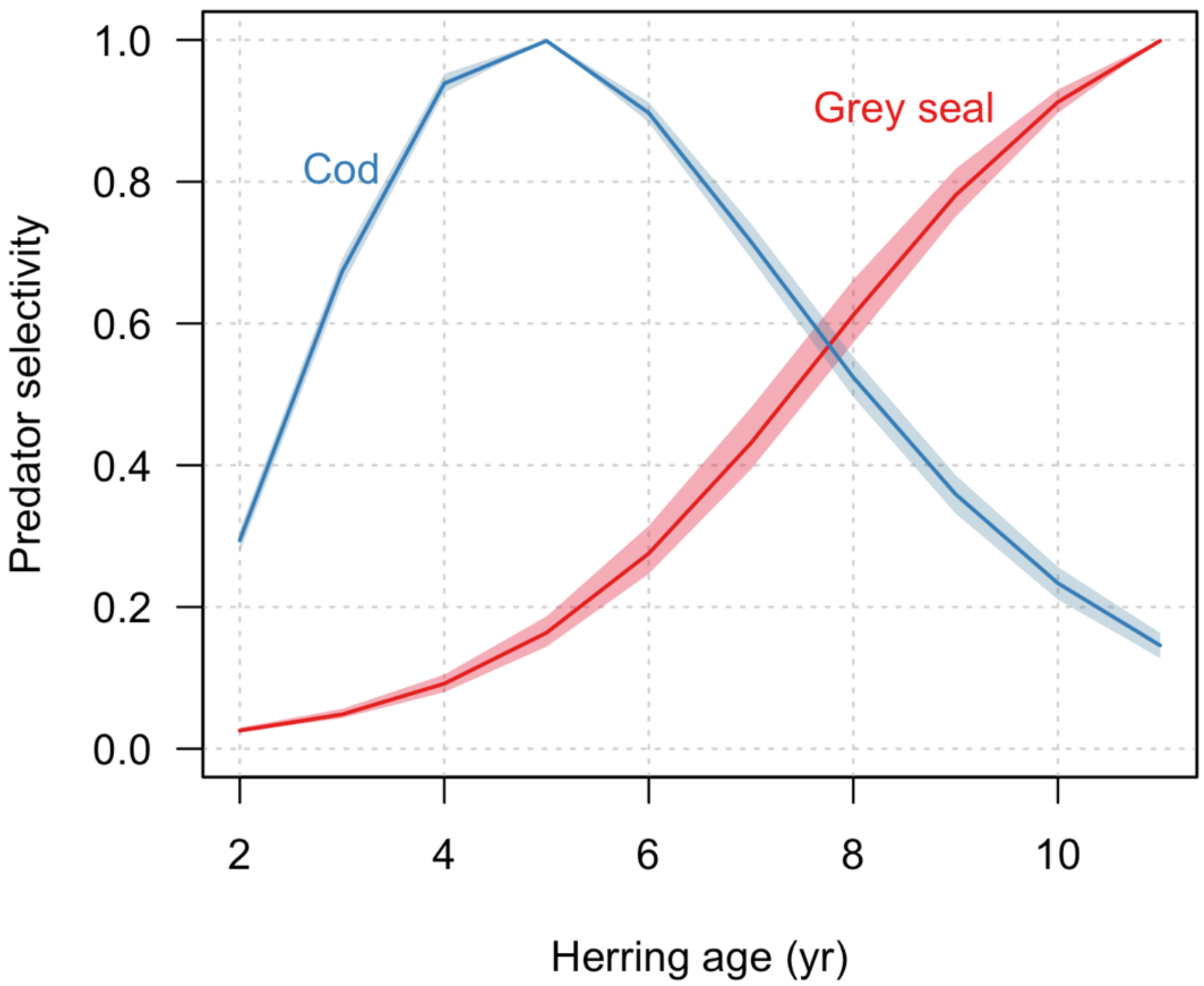

Figure 5. Estimated selectivity-at-age functions for grey seal (red) and cod (blue) predation on herring. Lines represent posterior modes, while shaded regions represent central 95% posterior intervals.

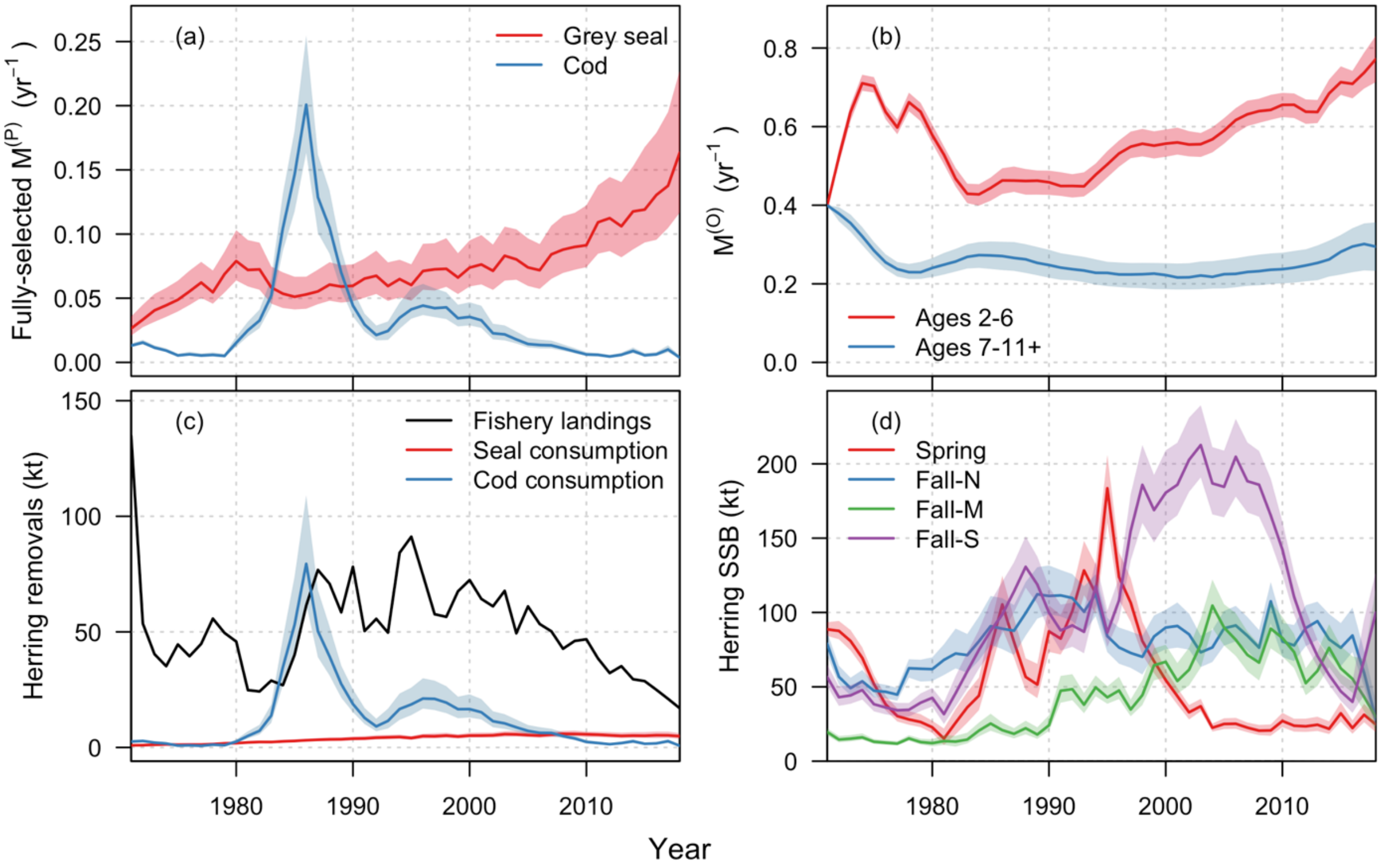

Figure 6. Model estimates of herring dynamics including (a) fully-selected predation mortality imposed by grey seals and cod, (b) natural mortality from sources other than predation, (c) removals by fisheries (observed) and predation (estimated), and (d) spawning stock biomass on Jan 1. Lines represent posterior modes (except for fishery landings in (c) which were observed) and shaded regions represent central 95% uncertainty intervals.

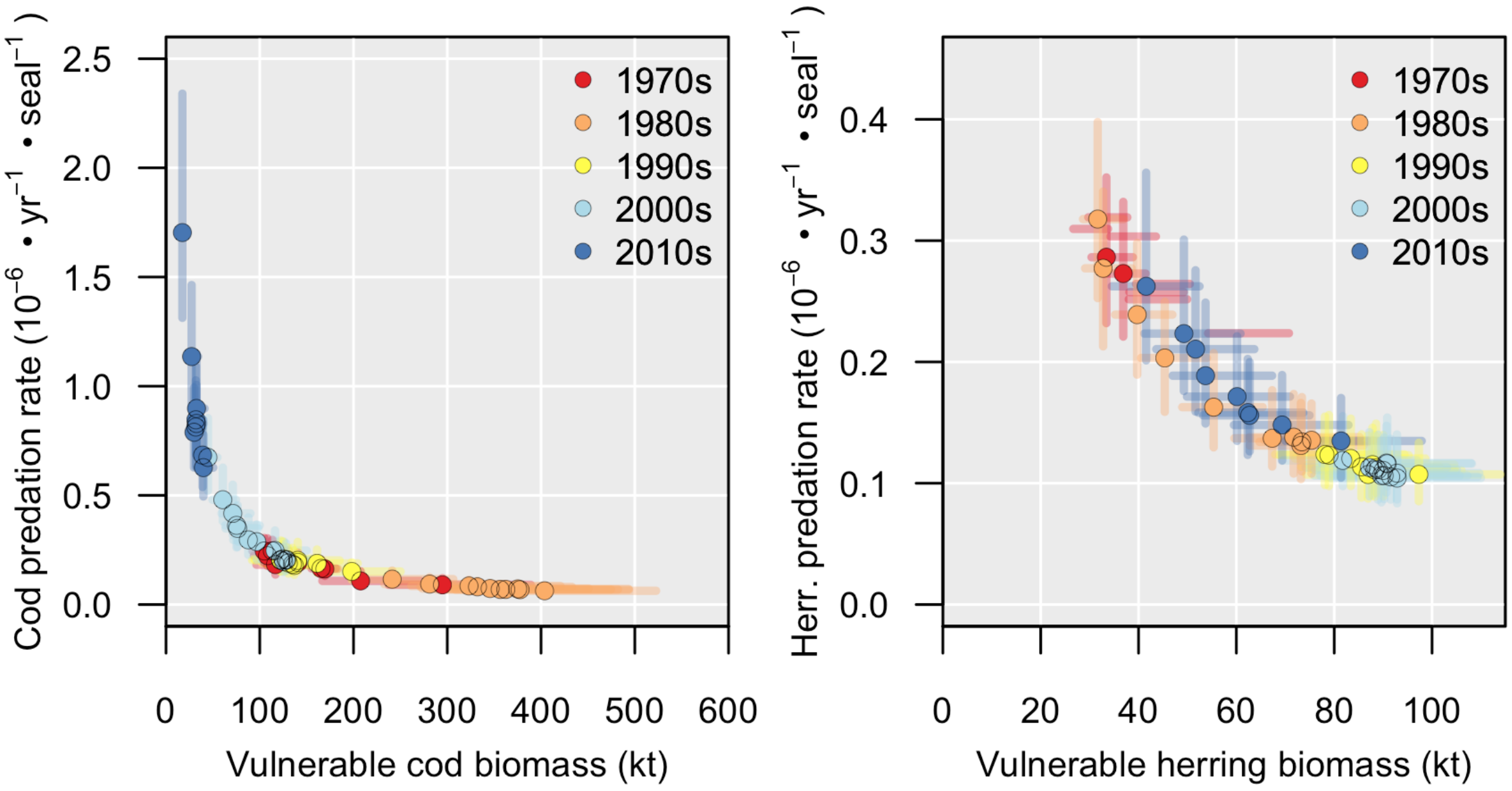

Figure 7. Model estimates of per-capita cod and herring consumption by Shelf Male grey seals (circles represent posterior modes while vertical and horizontal coloured lines represent central 95% uncertainty intervals).

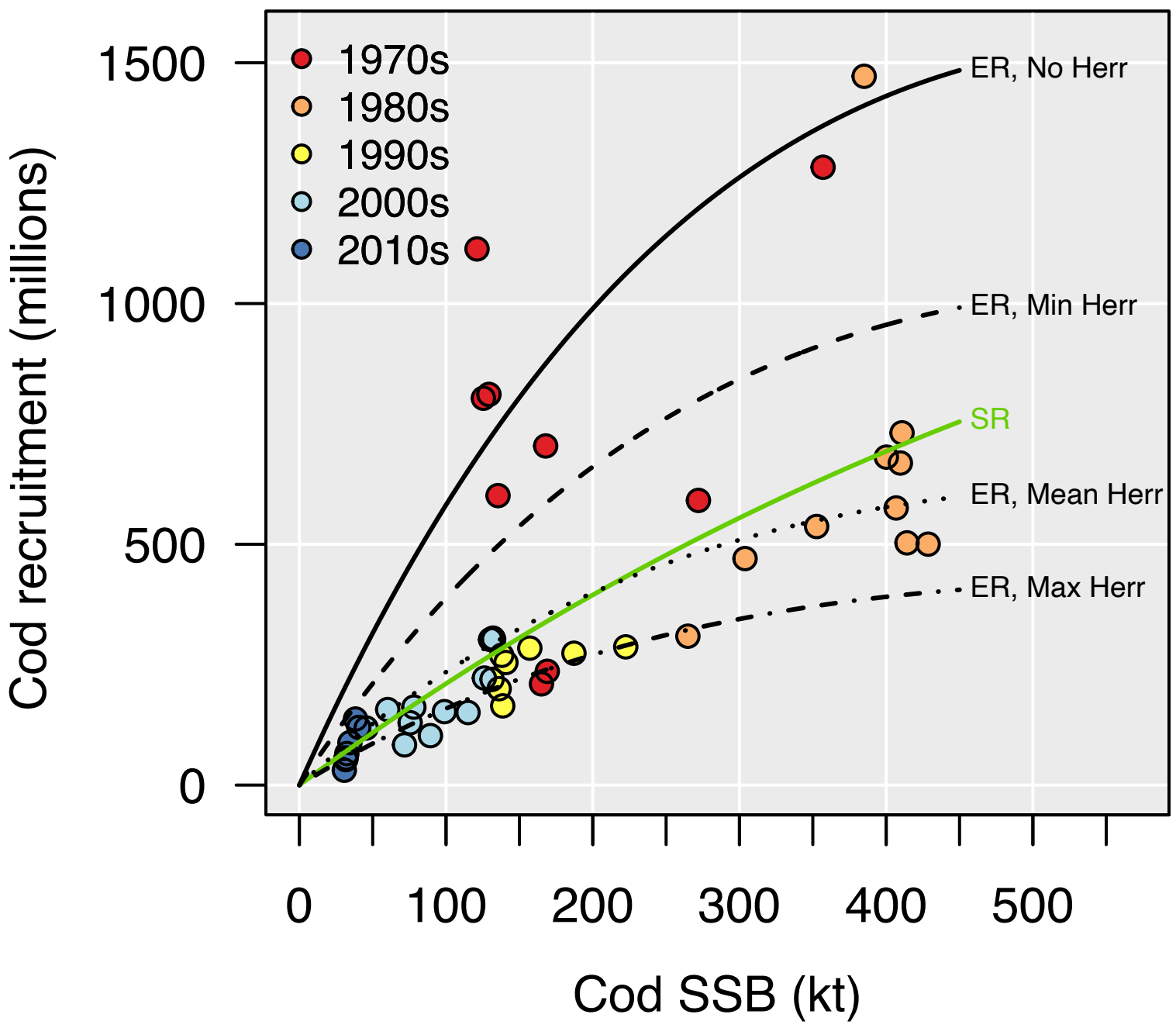

Figure 8. Posterior modes of cod recruitment and cod spawning biomass (circles), with fitted standard Ricker (green line) and extended Ricker stock-recruitment function under four levels of herring SSB (no herring (0 kt), and the minimum (122 kt), mean (274 kt) and maximum (393 kt) model-estimated levels for 1971-2018). Stock-recruitment models were fitted for illustrative purposes (in projections we fitted a separate stock-recruitment function to each posterior sample).

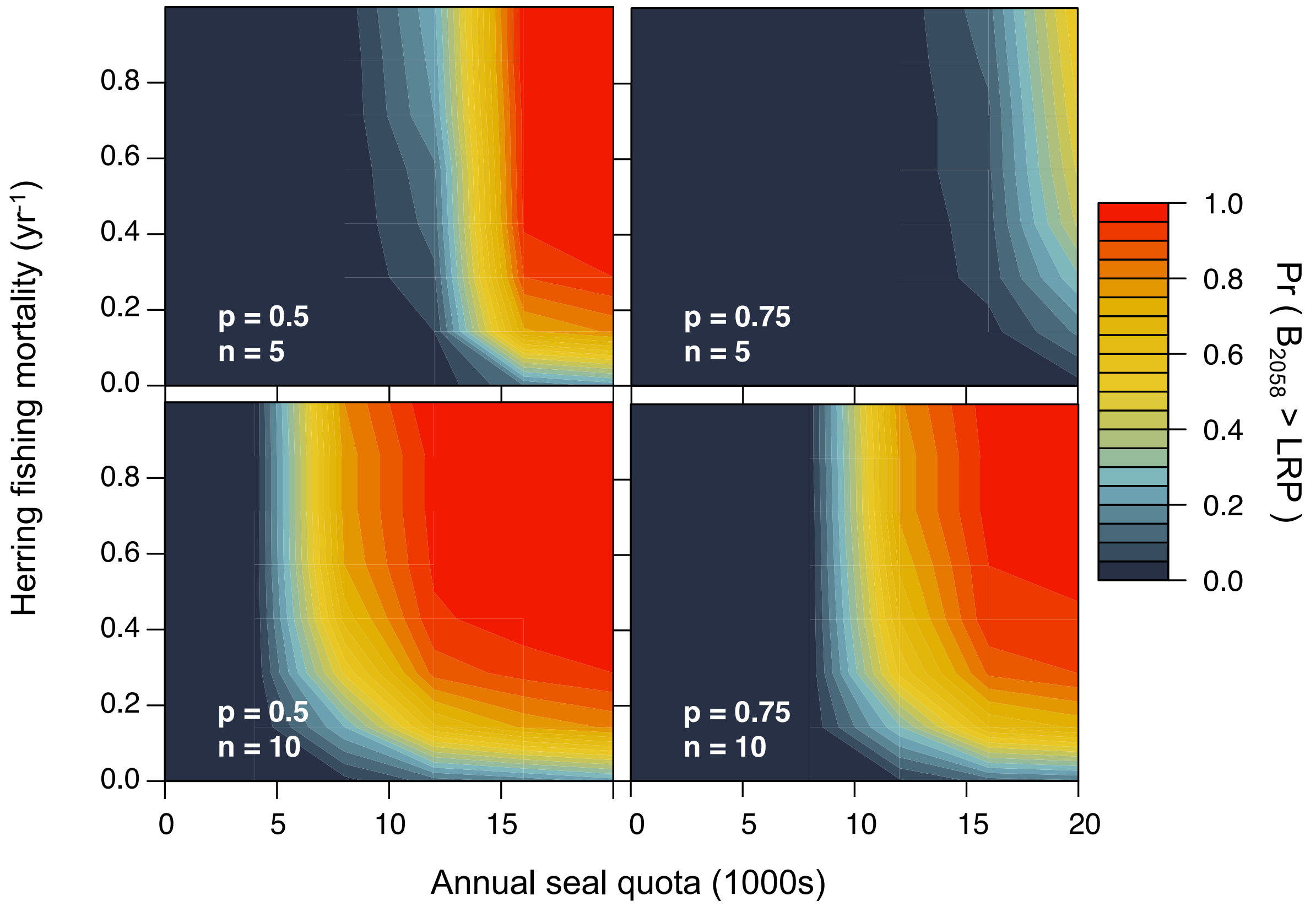

Figure 9. Probability of cod biomass in 2058 exceeding the limit reference point (LRP = 80 kt) given varying levels of seal quota (x-axis) and herring fishing mortality (y-axis), and four combinations of the proportion of YOY seals targeted for removal (p) and the length of the seal harvest period (n; number of years).

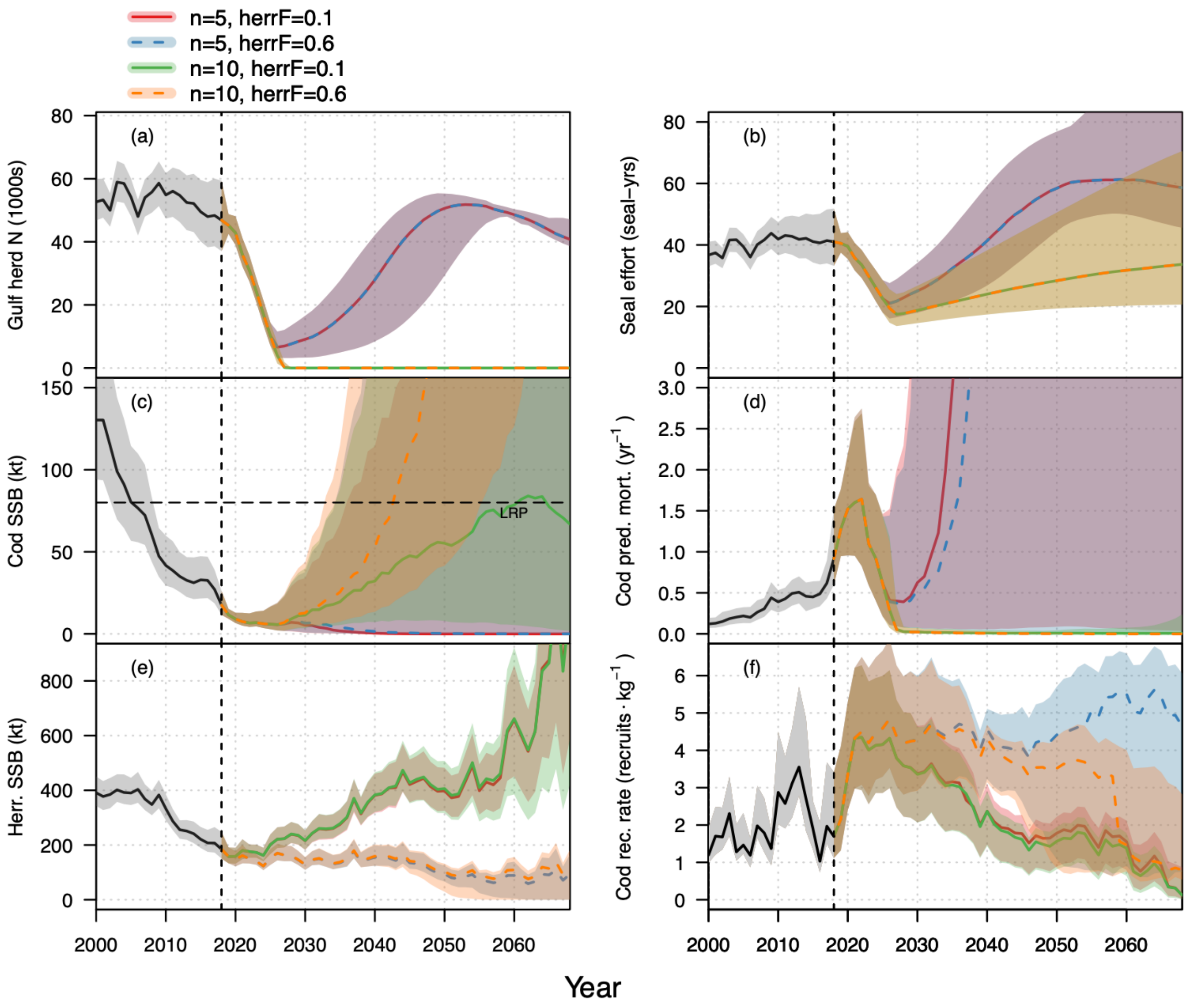

Figure 10. Model projections with annual quotas of 12,000 seals targeting 50% YOY under four combinations of seal harvest period (n; years) and herring fishing mortality (herrF; $yr^{-1}$). Lines represent posterior modes while shaded regions indicate the central 95% uncertainty interval. Black lines and grey shaded regions represent historical estimates while coloured lines and shaded regions represent projections.

## Appendix A. Model notation and equations.

*Table A1. Notation for Grey Seal/Cod/Herring model. * indicates parameters that were directly estimated, ** indicates model arrays for which some subset of values were directly estimated, *** indicates parameters that were conditionally estimated, and **** indicates parameters that were post-fitted (i.e., fitted to model estimates) for use in projections.*

| Symbol | Description | Value |
|---|---|---|
| **Indices** | | |
| $i, j$ | Species ($i$ = prey, $j$ = predator) | 1=Grey Seal, 2=Cod, 3=Herring |
| $x, y$ | Subpopulation ($x$ = prey, $y$ = predator) | |
| $a, b$ | Age (yr) ($a$ = prey, $b$ = predator) | $a_i^{(\mathrm{R})}, a_i^{(\mathrm{R})}+1, \ldots, A_i-1, A_i$ |
| $t$ | Year | 1,2,…,48 |
| $g$ | Survey index | Table 1 |
| **Data and inputs** | | |
| $a_i^{(\mathrm{R})}$ | Age-at-recruitment for species $i$ (yr) | 1,2,2 |
| $A_i$ | Maximum age-class for species $i$ (yr) | 30,12,11 |
| $I_{g,i,t}$ | Abundance or biomass index | |
| $u_{g,i,x,a,t}^{(\mathrm{F})}$ | Fishery/survey age-composition ($i \geq 2$) | |
| $u_{j,a,t}^{(\mathrm{P})}$ | Age-composition of Herring in predator diet ($j \leq 2$) | |
| $C_{j,y,b,i,t}^{(\mathrm{P})}$ | Bioenergetic-derived prey consumption by predator $jy$ at age $b$ of prey species $i$ at age $a$ (kt) | |
| $C_{x,t}^{(\mathrm{pup})}$ | Age-0 Grey Seal removals (1000s) | |
| $C_{x,t}^{(1+)}$ | Age-1+ Grey Seal removals (1000s) | |
| $C_{i,x,t}^{(\mathrm{F})}$ | Observed commercial fishery catch ($i \geq 2$) | |
| $p_{x,t}^{(\mathrm{ice})}$ | Proportion of Grey Seals born on pack-ice | |
| $s_{x,t}^{(\mathrm{ice})}$ | Survival rate of Grey Seals born on pack-ice | |
| $f_{j,y}$ | Proportion of time predator *j* in subpopulation *y* spends foraging near prey | $f_{j=1,y=1} = 0.024$<br>$f_{j=1,y=2} = 0.038$<br>$f_{j=1,y=3} = 0.729$<br>$f_{j=1,y=4} = 0.493$<br>$f_{j=2,y=1} = 1$ |
| $m_{i,a,t}$ | Proportion mature-at-age | |
| $w_{i,x,a,t}^{(\mathrm{S})}$ | Fish weight at start of year (kt) | |
| $w_{g,i,x,a,t}^{(\mathrm{F})}$ | Average weight-at-age of species *ix* caught by gear *g* (kt) ($i \geq 2$) | |
| $d_{g,i,x}$ | Survey date (expressed as Julian date / 365) | |
| $p_{x,a}^{(\mathrm{init})}$ | Initial proportion of Grey Seals by age and sex/subpopulation | |
| $\gamma_{a,t}^{(n)}$ | Number of seals sampled for reproductive status | |
| $\gamma_{a,t}^{(k)}$ | Observed number of pregnant seals | |
| $\zeta_{i,x}$ | Spawn timing (expressed as Julian date / 365) | $\zeta_{j=2,y=1} = 0.37$<br>$\zeta_{j=3,y=1} = 0.25$<br>$\zeta_{j=3,y=2} = 0.58$<br>$\zeta_{j=3,y=3} = 0.58$<br>$\zeta_{j=3,y=4} = 0.58$ |
| **Parameters** | | |
| $N_x^{(\mathrm{init})}$ | Initial Grey Seal abundance (‘000s) | * |
| $D_{x,1}, D_{x,2}$ | Seal juvenile density dependence half-saturation and shape | * |
| $\bar{\gamma}$ | Mature female seal reproductive rate | * |
| $a^{(50\%)}, a^{(95\%)}$ | Female seal age-at-50% and 95% maturity | * |
| $\bar{R}_x$ | Initial equilibrium Herring recruitment (millions) | * |
| $\varepsilon_{i,x,t}^{(\mathrm{R})}$ | Recruitment process errors ($i \geq 2$) | * |
| $M_{i,x,a,t}$ | Instantaneous natural mortality rate ($\mathrm{yr}^{-1}$) | ** |
| $\varepsilon_t^{(\mathrm{M,c})}$ | Cod other natural mortality process errors ($t > 1$) | * |
| $\varepsilon_t^{(\mathrm{M,h1})}$ | Herring other natural mortality process errors, ages 2-6 ($t > 1$) | * |
| $\varepsilon_t^{(\mathrm{M,h2})}$ | Herring other natural mortality process errors, ages 7-11+ ($t > 1$) | * |
| $\varphi_{j,y,i,t}$ | Per-capita capacity of predator *jy* to prey on species *i* ($i \geq 2, j < 3$) | ** |
| $\varepsilon_{j,i,t}^{(\varphi)}$ | Predation capacity process errors ($t > 1$) | * |
| $s_{g,i,x,t}^{50\%}, s_{g,i,x,t}^{95\%}$ | Age-at-50% and 95% selectivity for fishing gear *g* ($i \geq 2$) | * |
| $\rho_{j,y}^{50\%}, \rho_{j,y}^{95\%}$ | Age at which prey consumption by predator *jy* is 50% and 95% of its maximum consumption levels | * |
| $k_j$, $\theta_j$ | Shape and scale parameters for predator selectivity of Herring | * |
| $q_{g,i,x,t}$ | Catchability of species *ix* to gear *g* ($i \geq 2$, $g > 1$) | * |
| $\varepsilon_{g,i,x,t}^{(\mathrm{q})}$ | Catchability process errors ($t > 1$) | * |
| $\tau_{g,i,t}^{(I)}$ | Abundance or biomass index variance | ** (see Table 3) |
| $\tau_{g,i,\mathrm{x}}^{(u)}$ | Fishery/survey age-composition variance | ** |
| $\tau_{j,t}^{(P)}$ | Age-composition of herring in predator diet variance | ** |
| $F_{x,t}^{(1+)}$ | Grey seal exploitation rate ($\mathrm{yr}^{-1}$) | *** |
| $F_{i,x,t}$ | Fully-selected fishery exploitation rate ($\mathrm{yr}^{-1}$; $i > 1$) | *** |
| $\eta_{j,y,i}$ | Encounter rate between predator *jy* and prey *i* | **** |
| $h_{j,y,i}$ | Time predator *jy* spends consuming prey i | **** |
| $\beta_0, \beta_C, \beta_H$ | Parameters for fecundity, density-dependence and Herring effect in Ricker stock-recruitment function for Cod | **** |
| $\Psi_{1,x}, \Psi_{2,x}$ | Fecundity and density-dependence parameters of Herring stock-recruitment function in projections | **** |

**State variables**

| | | |
|---|---|---|
| $N_{i,x,a,t}$ | Abundance (thousands for Grey Seals, millions for Cod and Herring) | ** |
| $Z_{i,x,a,t}$ | Instantaneous total mortality rate (yr$^{-1}$) | |
| $M_{i,a,t}^{(P)}$ | Instantaneous predation mortality rate (yr$^{-1}$) | |
| $m_{j,y,b,i,a,t}^{(P)}$ | Fully-selected instantaneous predation mortality imposed on prey *i* by predator *jy* at age *b* (yr$^{-1}$) | |
| $M_{i,x,a,t}^{(O)}$ | Other (non-predation) natural mortality rate (yr$^{-1}$) | |
| $S_{j,i,a}^{(\mathrm{P})}$ | Selectivity of species *i* at age *a* by predator *j* | |
| $S_{g,i,x,a,t}^{(\mathrm{F})}$ | Selectivity of species *ix* at age *a* to fishing gear *g* | |
| $B_{i,x,t}^{(\mathrm{S})}$ | Spawner biomass ($i \geq 2$) | |
| $B_{j,i,x,a,t}^{(\mathrm{P})}$ | Biomass of prey *ix* at age *a* available to predator *j* ($i \geq 2, j < 3$) | |
| $V_{g,i,x,a,t}$ | Number of fish vulnerable to fishing gear *g* (millions) | |
| $G_{x,t}$ | Grey Seal pup production (thousands) | |
| $G_{x,t}^{*}$ | Grey Seal post-weaning pup abundance (thousands) | |
| $D_{x,t}$ | Density-dependent Grey Seal pup survival rate | |
| $E_{y,t}$ | Grey seal foraging effort in the sGSL (seal-yrs) | |
| $\rho_{j,y,b}$ | Relative prey consumption by predator *jy* at age *b* | |
| $\hat{c}_{j,y,b,i,\mathrm{x},a,t}^{(N)}$ | Per-capita catch by predator *jy* at age *b* of prey *ix* at age *a* (millions) | |
| $\hat{c}_{j,y,b,i,a,t}$ | Per-capita catch by predator *jy* at age *b* of prey *i* at age *a* (kt) | |
| $m_{i,a,t}$ | Proportion mature-at-age for species *i* in year *t* | |
| $\gamma_{a,t}$ | Grey Seal reproductive rate-at-age in year *t* | |
| $s_{x,a=1}^{(\mathrm{init})}$ | Initial Herring survivorship | |
| $\hat{I}_{g,i,t}$ | Predicted abundance/biomass index | |
| $\hat{u}_{g,i,x,a,t}^{(\mathrm{F})}$ | Predicted fishery/survey age-composition | |

$\hat{u}_{j,a,t}^{(\mathrm{P})}$ Predicted age-composition of Herring consumed by predator $j$

*Table A2. Population dynamics equations for Grey Seal/Cod/Herring model.*

| Equation | Formula |
|---|---|
| *Grey seal demographic rates* | |
| Proportion of females mature-at-age, $i = 1$, $a < 4$ | $m_{i=1,a,t} = 0$ |
| Proportion of females mature-at-age, $i = 1$, $4 \le a \le 11$ | $m_{i=1,a,t} = \left(1 + \exp\left[\frac{-\ln(19)\left(a - a^{(50\%)}\right)}{a^{(95\%)} - a^{(50\%)}}\right]\right)^{-1}$ |
| Proportion of females mature-at-age, $i = 1$, $a \ge 12$ | $m_{i=1,a,t} = 1$ |
| Annual reproductive rate-at-age | $\gamma_{a,t} = \bar{\gamma} m_{i=1,a,t}$ |
| *Selectivity* | |
| Fishery and survey selectivity (i > 1) | $S^{(\mathrm{F})}_{g,i,x,a,t} = \left(1 + \exp\left[\frac{-\ln(19)\left(a - s^{50\%}_{g,i,x,t}\right)}{s^{95\%}_{g,i,x,t} - s^{50\%}_{g,i,x,t}}\right]\right)^{-1}$ |
| Seal selectivity of cod | $S^{(\mathrm{P})}_{j=1,i=2,a} = \begin{cases} 0, & a \le 4 \\ 1, & a \ge 5 \end{cases}$ |
| Herring selectivity by seal | $S^{(\mathrm{P})}_{j=1,i=3,a} = \left(1 + \exp\left[\frac{-\log(19)\left(a - b^{50\%}\right)}{b^{95\%} - b^{50\%}}\right]\right)^{-1}$ |
| Herring selectivity by cod | $S^{(\mathrm{P})}_{j=2,i=3,a} \propto \frac{1}{\Gamma(k)\theta^{k}} \theta^{k-1} \exp\left(-\frac{a}{\theta}\right)$ |
| *Prey consumption* | |
| Relative prey consumption-at-age ($i \le 2$) | $\rho_{j,y,b} = \left(1 + \exp\left[\frac{-\log(19)\left(b - \rho^{50\%}_{j,y}\right)}{\rho^{95\%}_{j,y} - \rho^{50\%}_{j,y}}\right]\right)^{-1}$ |
| Per-capita predation capacity (Historical) ($t > 2$; $j \le 2$; $i > j$) | $\varphi_{j,y,i,t} = \varphi_{j,y,i,t-1} \exp\left(\varepsilon^{(\varphi)}_{j,i,t}\right)$ |
| Per-capita predation capacity (Projection) ($t > 2$; $j \le 2$; $i > j$) | $\varphi_{j,y,i,t} = \frac{\eta_{j,y,i}\left(\sum_x B_{i,x,t}\right)^{\lambda_{j,y,i}-1}}{1 + \sum_l \left(\eta_{j,y,l} h_{j,y,l} \left(\sum_x B_{l,x,t}\right)^{\lambda_{j,y,l}}\right)}$ |
| *Mortality rates* | |
| Seal exploitation | $F_{i,x,a,t} = F^{(1+)}_{x,t}$ |
| Predation mortality per predator | $m^{(P)}_{j,y,b,i,a,t} = S^{(\mathrm{P})}_{j,i,a} f_{j,y} \varphi_{j,y,i,t} \rho_{j,y,b}$ |
| Total predation mortality | $M_{i,a,t}^{(P)} = \sum_j \sum_y \sum_b N_{j,y,b,t}\, m_{j,y,b,i,a,t}^{(P)}$ |
| Other natural mortality ($t \geq 2$, $i$ = 2) | $M_{i=2,x=1,a,t}^{(O)} = M_{2,1,a,t-1}^{(O)} \exp\left(\varepsilon_t^{(\mathrm{M,c})}\right)$ |
| Other natural mortality ($t \geq 2$, $i$ = 3, $a \leq 7$) | $M_{i=3,x,a,t}^{(O)} = M_{3,x,a,t-1}^{(O)} \exp\left(\varepsilon_t^{(\mathrm{M,h1})}\right)$ |
| Other natural mortality ($t \geq 2$, $i$ = 3, $a \geq 8$) | $M_{i=3,x,a,t}^{(O)} = M_{3,x,a,t-1}^{(O)} \exp\left(\varepsilon_t^{(\mathrm{M,h2})}\right)$ |
| Total mortality rate | $Z_{i,x,a,t} = S_{g=1,i,x,a,t}^{(\mathrm{F})} F_{i,x,t} + M_{i,a,t}^{(P)} + M_{i,x,a,t}^{(O)}$ |
| *Initial abundance* | |
| Initial grey seal abundance | $N_{i=1,x,a,t=1} = N_x^{(\mathrm{init})} p_{x,a}^{(\mathrm{init})}$ |
| Herring survivorship, (a=1) | $s_{x,a=1}^{(\mathrm{init})} = 1$ |
| Herring survivorship, (a<11) | $s_{x,a}^{(\mathrm{init})} = s_{x,a-1}^{(\mathrm{init})} \exp\left(-M_{3,x,a-1,1} - S_{1,3,x,a-1,1}^{(\mathrm{F})} F_{3,x,1}\right)$ |
| Herring survivorship, (a=11) | $s_{x,a}^{(\mathrm{init})} = \dfrac{s_{x,a-1}^{(\mathrm{init})} \exp\left(-M_{3,x,a-1,1} - S_{1,3,x,a-1,1}^{(\mathrm{F})} F_{3,x,1}\right)}{\left(1 - \exp\left(-M_{3,x,a,1} - S_{1,3,x,a,1}^{(\mathrm{F})} F_{3,x,1}\right)\right)}$ |
| Herring abundance | $N_{i=3,x,a,t=1} = \bar{R}_x s_{x,a}^{(\mathrm{init})}$ |
| *State dynamics* | |
| Shelf seal pup production ($x \in \{1,2\}$) | $G_{x,t} = 0.5 \sum_a N_{1,2,a,t} \gamma_{a,t}$ |
| Gulf seal pup production ($x \in \{3,4\}$) | $G_{x,t} = 0.5 \sum_a N_{1,4,a,t} \gamma_{a,t}$ |
| Post-weaning pup abundance | $G_{x,t}^{*} = 0.95 \left(G_{x,t} p_{x,t}^{(\mathrm{ice})} s_{x,t}^{(\mathrm{ice})} + G_{x,t}\left[1 - p_{x,t}^{(\mathrm{ice})}\right]\right)$ |
| Post-harvest pup abundance | $Y_{x,t} = \left(G_{x,t}^{*} - C_{x,t}^{(\mathrm{pup})}\right)$ |
| Shelf seal recruitment ($x$={1,2}, $a$ = 1, $t \geq 2$) | $N_{1,x,1,t} = 0.9 Y_{x,t-1} \left(\dfrac{D_{x,1}^{D_{x,2}}}{D_{x,1}^{D_{x,2}} + \left[\sum_{x=1}^{2} Y_{x,t-1}\right]^{D_{x,2}}}\right)$ |
| Gulf seal recruitment ($x$={3,4}, $a$ = 1, $t \geq 2$) | $N_{1,x,1,t} = 0.9 Y_{x,t-1} \left(\dfrac{D_{x,1}^{D_{x,2}}}{D_{x,1}^{D_{x,2}} + \left[\sum_{x=3}^{4} Y_{x,t-1}\right]^{D_{x,2}}}\right)$ |
| Cod/herring recruitment (Historical) ($t \geq 2$; $i \geq 2$) | $N_{i,x,2,t} = N_{i,x,2,t-1} \exp\left(\varepsilon^{(\mathrm{R})}_{i,x,t}\right)$ |
| Cod/herring recruitment (Projection) ($i = 2$) | $N_{i,x,2,t} = \beta_0 B_{2,1,t-2} \exp\left(-\beta_C B_{2,1,t-2} - \beta_H \sum_x B_{3,x,t-2}\right)$ |
| Recruitment (Projection) ($i = 3$) | $N_{i,x,2,t} = \Psi_{x,1} B_{3,x,t-2} / \left(1 + \Psi_{x,2} B_{3,x,t-2}\right)$ |
| Abundance, $a_i^{(\mathrm{R})} < a < A_i$ | $N_{i,x,a,t} = N_{i,x,a-1,t-1} \exp\left(-Z_{i,x,a-1,t-1}\right)$ |
| Spawner biomass on Jan 1 ($i \geq 2$) | $B^{(1)}_{i,x,t} = \sum_a N_{i,x,a,t} m_{i,a,t} w^{(\mathrm{S})}_{i,a,t}$ |
| Spawner biomass at time of spawning ($i \geq 2$) | $B_{i,x,t} = \sum_a N_{i,x,a,t} m_{i,a,t} w^{(\mathrm{S})}_{i,a,t} \left(-\zeta_{i,x} Z_{i,x,a,t}\right)$ |
| Seal foraging effort | $E_{y,t} = f_{j=1,y} \sum_a N_{i=1,y,a,t}$ |
| Per-capita prey consumption (numbers) | $\hat{c}^{(N)}_{j,y,b,i,\mathrm{x},a,t} = \frac{m^{(P)}_{j,y,b,i,a,t}}{Z_{i,x,a,t}} N_{i,x,a,t}\left(1 - \exp\left(-Z_{i,x,a,t}\right)\right)$ |
| Per-capita prey consumption (weight) | $\hat{c}_{j,y,b,i,a,t} = \sum_x \hat{c}^{(N)}_{j,y,b,i,\mathrm{x},a,t} w_{i,x,a,t}$ |
| Abundance vulnerable to fishery/survey ($i \geq 2$) | $V_{g,i,x,a,t} = N_{i,x,a,t} S^{(\mathrm{F})}_{g,i,x,a,t} \exp\left(-d_{g,i,x} Z_{i,x,a,t}\right)$ |
| Age 1+ Shelf seal removals | $C_t^{(1+,\mathrm{S})} = \sum_{x=1}^{2} \sum_a \frac{F^{(1+)}_{x,t}}{Z_{1,x,a,t}} N_{1,x,a,t}\left(1 - \exp\left(-Z_{1,x,a,t}\right)\right)$ |
| Age 1+ Gulf seal removals | $C_t^{(1+,\mathrm{G})} = \sum_{x=3}^{4} \sum_a \frac{F^{(1+)}_{x,t}}{Z_{1,x,a,t}} N_{1,x,a,t}\left(1 - \exp\left(-Z_{1,x,a,t}\right)\right)$ |
| Fishery catch ($i \geq 2$) | $C^{(F)}_{i,x,t} = \sum_a \frac{F_{i,x,t}}{Z_{i,x,a,t}} V_{1,i,x,a,t} w^{(\mathrm{F})}_{1,i,x,a,t}\left(1 - \exp\left(-Z_{i,x,a,t}\right)\right)$ |
| *Predicted observations* | |
| Shelf seal pup production indices | $\hat{I}_{g=1,i=1,t} = \sum_{x=1}^{2} G^*_{x,t}$ |
| Gulf seal pup production indices | $\hat{I}_{g=2,i=1,t} = \sum_{x=3}^{4} G^*_{x,t}$ |
| Biomass indices ($i \geq 2$, $g > 1$) | $\hat{I}_{g,i,t} = q_{g,i,x,t} \sum_a V_{g,i,x,a,t} w^{(\mathrm{F})}_{g,i,x,a,t}$ |
| Fishery/survey age-composition ($i \geq 2$) | $\hat{u}^{(\mathrm{F})}_{g,i,x,a,t} = \frac{V_{g,i,x,a,t}}{\sum_a V_{g,i,x,a,t}}$ |
| Age-composition of herring in predator diet | $\hat{u}^{(\mathrm{P})}_{j,a,t} = \frac{\sum_y \sum_b \sum_x \hat{c}^{(N)}_{j,y,b,i=3,\mathrm{x},a,t} N_{j,y,b,t}}{\sum_a \sum_y \sum_b \sum_x \hat{c}^{(N)}_{j,y,b,i=3,\mathrm{x},a,t} N_{j,y,b,t}}$ |

*Table A3. Objective function components for Grey Seal/Cod/Herring model.*

| **Model variable** | **Distribution** | **Notes** |
|---|---|---|
| *Likelihood* | | |
| Abundance or biomass indices | $\ln I_{g,i,t} \sim N\left(\ln \hat{I}_{g,i,t}, \tau^{(I)}_{g,i,t}\right)$ | - $\tau^{(I)}$ for grey seals was fixed at herd- and year-specific values<br>- $\tau^{(I)}$for cod was conditionally estimated by gear<br>- $\tau^{(I)}$for herring was fixed at 0.1 for CPUE indices and 0.2 for all other indices |
| Fishery/survey age-composition | $\left\{u^{(\mathrm{F})}_{g,i,x,a,t}\right\}_{a=a^{(\min)}_{g,i}}^{a^{(\max)}_{g,i}} \sim P\left(N\left(\hat{u}^{(F)}_{g,i,x,a,t}, \tau^{(u)}_{g,i,x}\right)\right)$ | -$\tau^{(u)}$ was conditionally estimated<br>-$a^{(\min)}_{g,i}$ and $a^{(\max)}_{g,i}$ are the minimum and maximum ages respectively of sampled fish |
| Age-composition of herring in predator diets | $\left\{u^{(\mathrm{P})}_{j,a,t}\right\}_{a=a^{(\mathrm{R})}_{i=3}}^{A_{i=3}} \sim P\left(N\left(\hat{u}^{(\mathrm{P})}_{j,a}, \tau^{(\mathrm{P})}_{j,t}\right)\right)$ | -*P*(*N*($u,\tau$)) is the logistic-normal distribution<br>- $\tau^{(\mathrm{P})}$ was conditionally estimated |
| Per-capita consumption | $\ln \sum_a c_{j=1,y,b,i,a,t} \sim \mathrm{N}\left(\ln \hat{c}_{j=1,y,b,i,t}, 1\right)$<br>$\ln \sum_a c_{j=2,y,b,i,a,t} \sim \mathrm{N}\left(\ln \hat{c}_{j=2,y,b,i,t}, 0.1^2\right)$ | -For $i < j$ |
| Grey seal pregnancy rates | $\gamma^{(k)}_{a,t} \sim B\left(\gamma^{(n)}_{a,t}, \gamma_{a,t}\right)$ | |
| *Priors* | | |
| Recruitment deviations | $\varepsilon_{i,x,t}^{(\mathrm{R})} \sim N(0,1)$ | -High variance to account for potentially large interannual changes in recruitment |
| Consumption deviations | $\varepsilon_{j=1,i,t}^{(\varphi)} \sim N(0,0.25^2)$<br>$\varepsilon_{j=2,i,t}^{(\varphi)} \sim N(0,0.05^2)$ | -Variance chosen to allow for large decadal-scale shifts in consumption while preventing excessive interannual variation. Seal (*j*=1) was less prone to chasing noise so higher variance was allowed. |
| Catchability deviations | $\varepsilon_{x,t}^{(q)} \sim N(0,0.05^2)$ | -$\varepsilon_{x,t}^{(q)}$ was estimated between 1991-2018 for Spring (*x*=1) herring and between 1987-2018 for Fall (*x*>1) herring to match herring assessment (DFO 2018)<br>-Variance chosen to match herring assessment (DFO 2018) |
| Cod other natural mortality deviations | $\varepsilon_{t}^{(\mathrm{M,c})} \sim N(0,0.05^2)$ | -$\varepsilon_{t}^{(\mathrm{M})}$ was fixed at 0 until 1978 and was estimated thereafter |
| Herring other natural mortality deviations | $\varepsilon_{t}^{(\mathrm{M,h1})} \sim N(0,0.05^2)$<br>$\varepsilon_{t}^{(\mathrm{M,h2})} \sim N(0,0.05^2)$ | |
| Herring selectivity | $s_{g,i=3,x,t}^{50\%} \sim N(5,3^2)$<br>$\left(s_{g,i=3,x,t}^{95\%} - s_{g,i=3,x,t}^{50\%}\right) \sim N(1.5,2^2)$ | -Applied to all uniquely estimated Herring selectivity parameters |

| | |
|---|---|
| Herring pre-1978 biomass index catchability | $q_{g=3,i=3,x,t} \sim N(1,1)$ |
| Grey seal initial abundance | $N_h^{(\text{init})} \sim N(5,5^2)$ |
| Grey seal density dependence | $D_{h=1,1} \sim N(25,25^2)$<br>$D_{h=2,2} \sim N(5,10^2)$<br>$D_{h,2} \sim N(1,3^2)$ |
| Grey seal female age-at-maturity | $a^{(50\%)} \sim N(5,4^2)$<br>$\left(a^{(95\%)} - a^{(50\%)}\right) \sim N(2,2^2)$ |

## Appendix B. Estimating Grey Seal and Atlantic Cod prey consumption-at-age from bioenergetics

We used bioenergetic models to estimate age-specific, per-capita prey consumption by Northwest Atlantic Grey Seals (hereafter “seals”) and southern Gulf of St. Lawrence (sGSL) Atlantic Cod (hereafter “cod”). We estimated seal consumption of cod and sGSL Atlantic Herring (hereafter “herring”), and cod consumption of herring, based on total consumption estimates, the spatiotemporal overlap between predator and prey species, and the proportional contribution of prey to predator diets.

### Grey seals

Benoît et al. (2011b) estimated the daily gross energy intake (GEI) of individual seals based on a number of factors, including body mass, metabolism, assimilation efficiency and age-specific growth premiums. We converted daily GEI to monthly GEI, then divided by the average energy of prey (Trzcinski et al., 2006) to estimate the amount of biomass needed to maintain growth (Figure B1).

The proportion of each month seals spend foraging near cod and herring was estimated from the movements of satellite-tracked seals (Benoît and Rail, 2016; Breed et al., 2006; Harvey et al., 2008). Cod and herring were assumed to occupy the same areas each month (specifically, the sGSL year-round, plus NAFO Subdivision 4Vn from November to April).

The relative contribution of prey to the seal diet is highly uncertain due to spatiotemporal gaps in diet sampling as well heterogeneities in foraging behaviour across seasons, areas and individuals. The seal diet has previously been inferred from prey hard parts found in the digestive tracts of grey seals collected (i) in coastal areas of the sGSL between late spring and August (Hammill et al., 2007), (ii) from the west coast of Cape Breton Island between September and January, and (iii) in the Cabot Strait near St Paul Island, mostly between October and December (Hammill et al., 2014). Seals were sampled on or near shore and the inferred diets likely reflect feeding that occurred near (~30 km) the sampling site (Benoît et al., 2011a). We assume that the observed inshore diet also reflects the offshore diet, given the considerable spatial overlap among these species (Benoît and Rail, 2016).

For seal consumption of herring, we assumed that the Cape Breton samples represented the fall (Sept-Nov) seal diet, the Cabot Strait samples represented the winter seal diet (Dec-Mar), and the sGSL samples represented the seal diet from April to August (Figure B2). For seal consumption of cod, the Cabot Strait samples represented the seal diet while cod are migrating and overwintering (Oct-May) while the sGSL samples represented the summer (June-Sept) seal diet. The winter Cabot Strait samples were assumed to represent feeding on migrating cod, given that migrating cod occur at densities similar to that on the overwintering ground. This is reflected by the intense fisheries that used to target these migrating fish.

Seal consumption of individual prey species was calculated as the product of total prey consumption, monthly foraging behaviour, and diet composition, each of which were assumed to be year-invariant. Gulf herd seals consumed significantly more prey in the sGSL than Shelf herd seals (Figure B3).

**Atlantic Cod**

Benoît and Rail (2016) estimated monthly, size-specific prey consumption by cod $\left(C_{l,m}\right)$ from mean stomach content mass and gastric evacuation rates. We converted size-specific consumption to age-specific consumption using an age-length key and diet composition estimates, i.e.,

$$c_{a,t,m} = \sum_{l} p_{a,l,t} C_{l,m} d_{l,t} \quad (1)$$

where $c_{a,t,m}$ is the per-capita consumption-at-age for cod, $p_{a,l,t}$ is the proportion of age-$a$ cod at length *l* in year *t* and $d_{l,t}$ is the proportional contribution of herring to the of length-$l$ cod diet. Monthly consumption rates were converted to annual rates, accounting for mortality each month (Figure B4), i.e.,

$$c_{a,t} = \sum_{m} c_{a,t,m} \exp\left(-Z_{a,m} \frac{m}{12}\right) \quad (2)$$

where $Z_{a,m}$ is the annual instantaneous cod mortality rate (Swain et al., 2015).

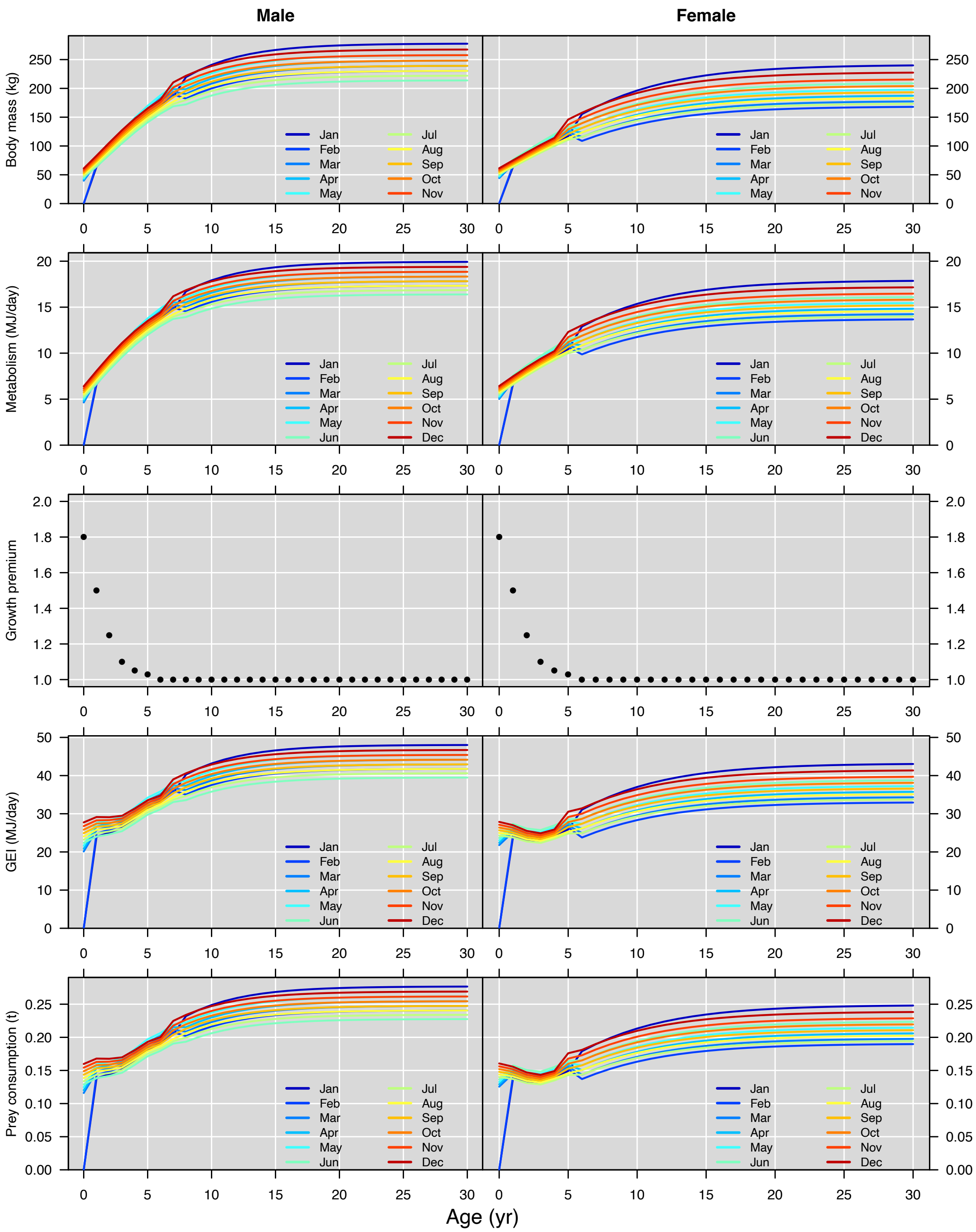

Figure B1. Grey seal body mass (top row, estimated from Gompertz growth model), daily metabolism (second row, assuming Kleiber allometric relationship between body mass and metabolism), growth premium for younger seals (third row), daily gross energy intake (fourth row, product of metabolism, growth premium, and conversion factors) and monthly prey consumption (bottom row, quotient of gross energy intake and average energy of prey).

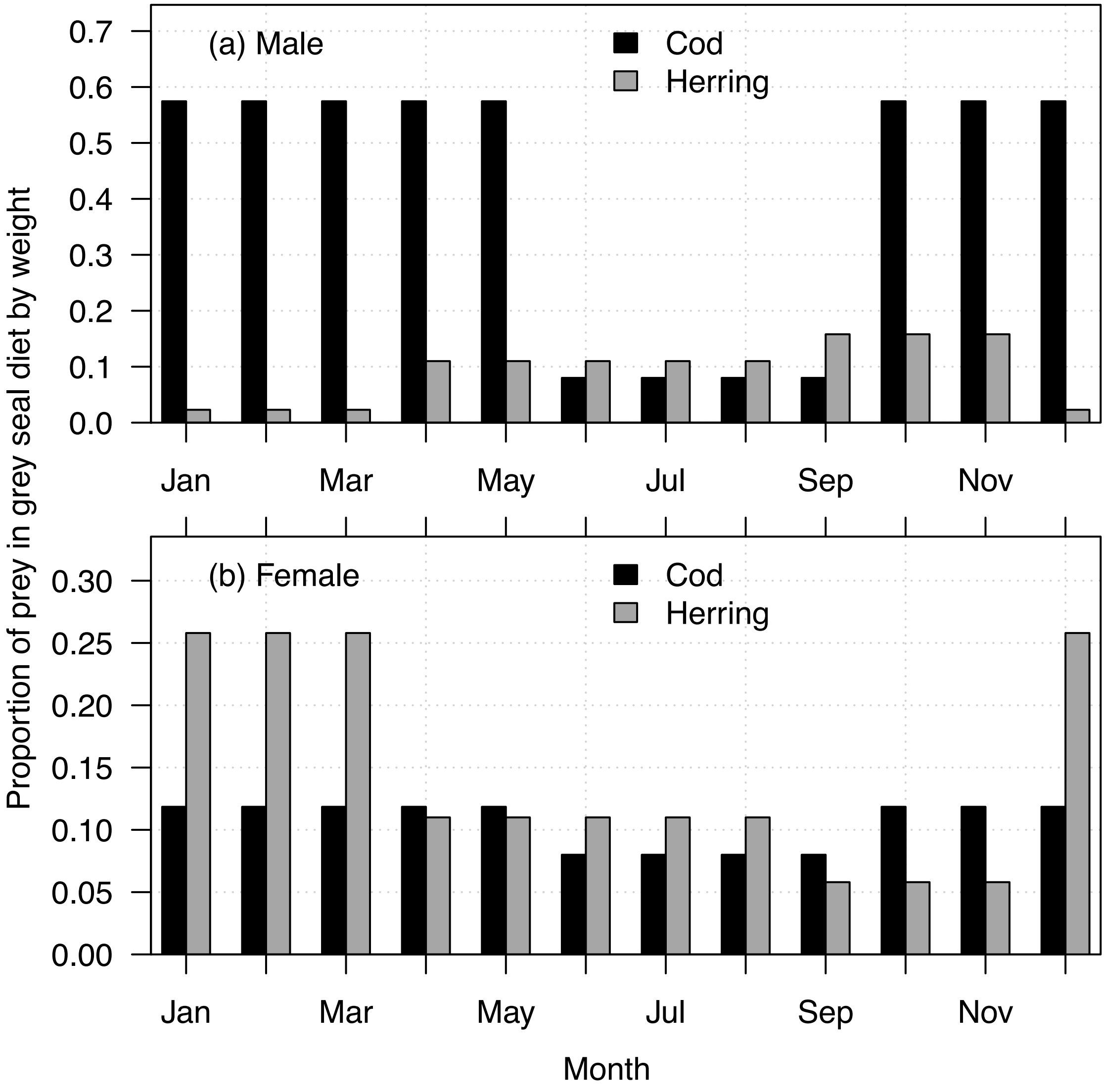

Figure B2. Monthly diet composition of (a) male and (b) female Grey Seals foraging near sGSL Cod and Herring.

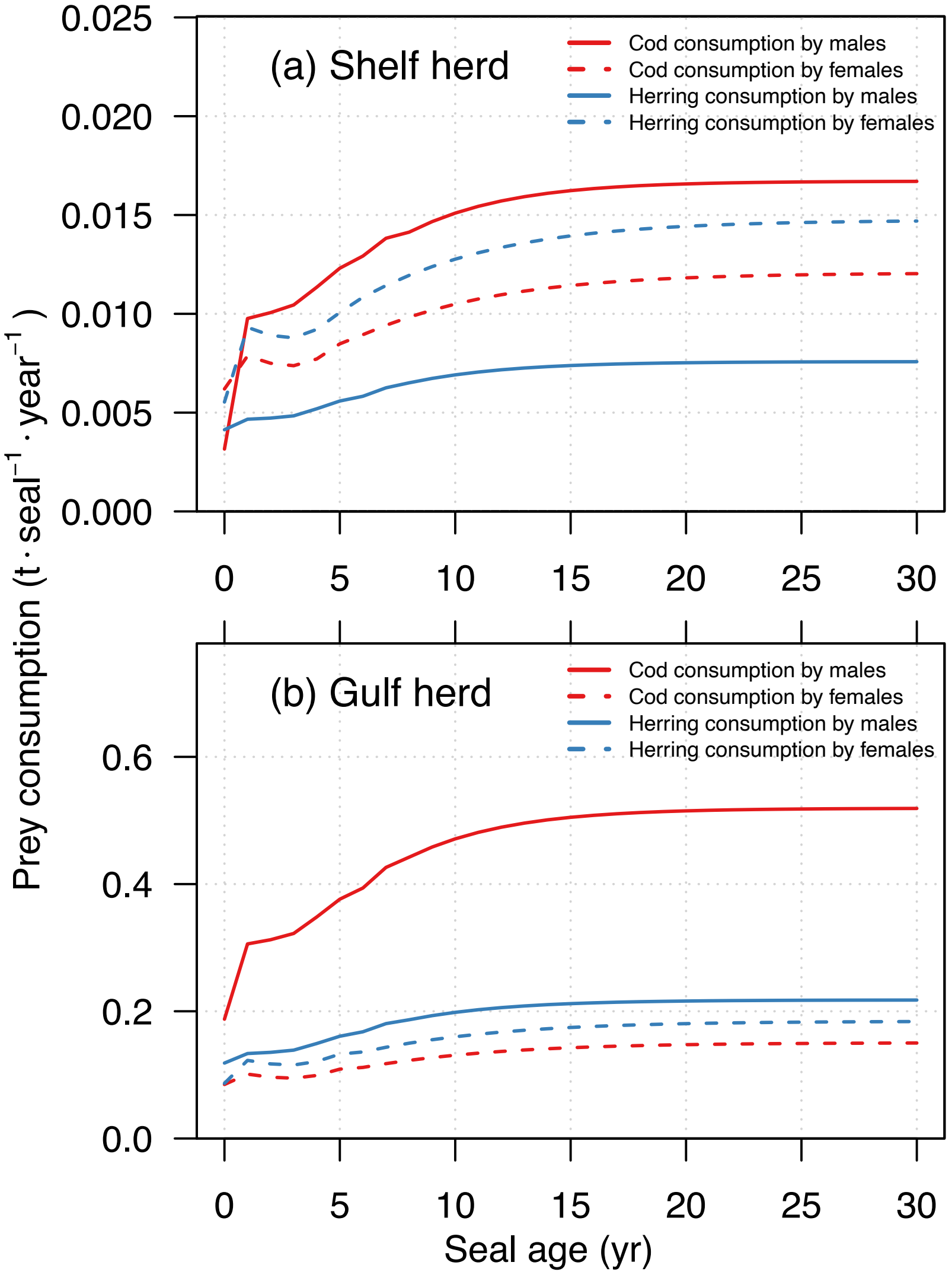

Figure B3. Annual per-capita consumption-at-age in the sGSL (plus NAFO Subdivision 4Vn from November to April) by Grey Seals from the (a) Shelf herd and (b) Gulf herd.

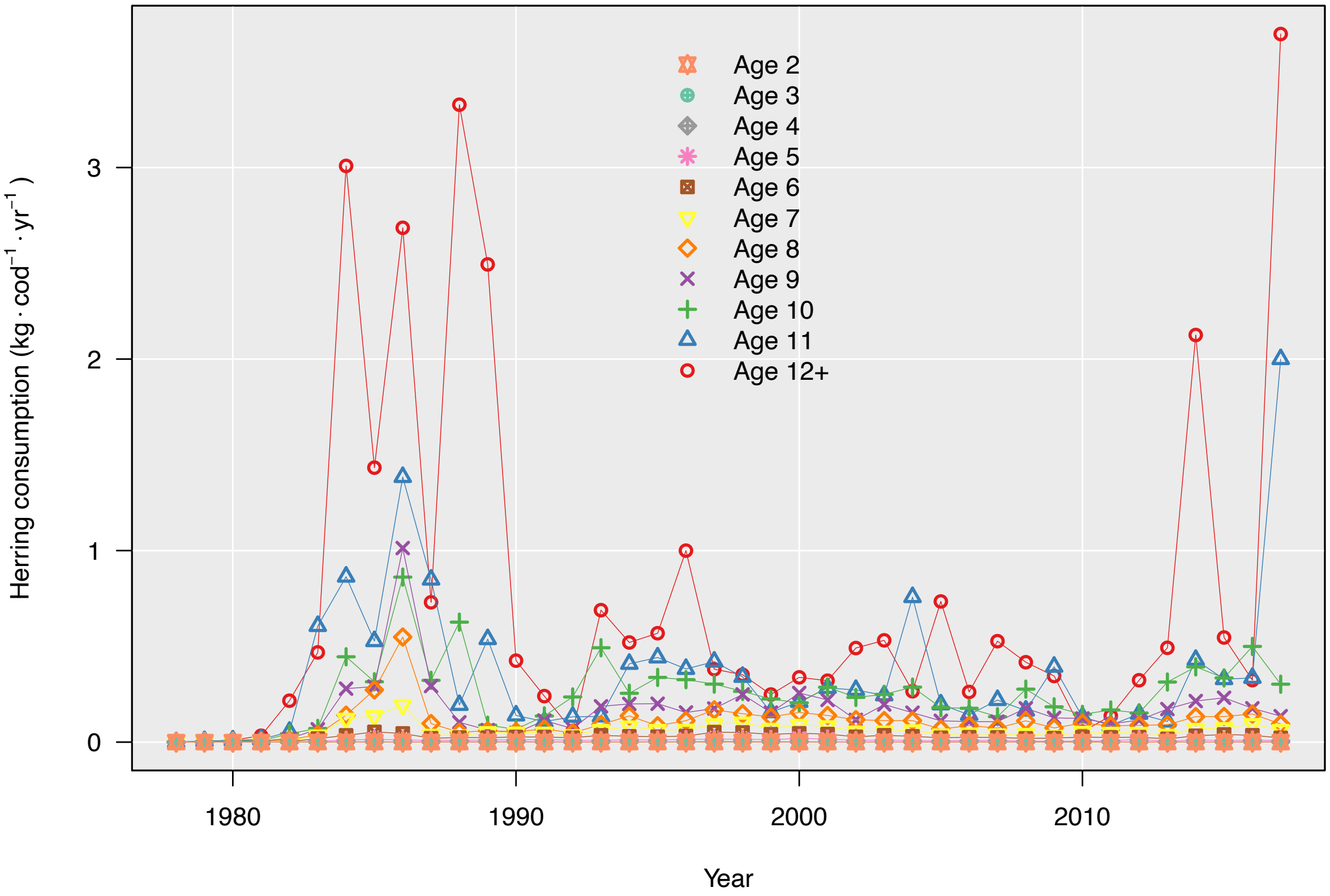

Figure B4. Herring consumption by individual sGSL Cod aged 2-12+ yr, 1978-2017.

## Appendix C: Stock-recruitment in Herring projections

In our analysis, we fit a Model of Intermediate Complexity for Ecosystem assessments (MICE) to observed data for Grey Seals, Atlantic Cod, and Atlantic Herring, then projected the model forward in time based on the optimized model parameters as well as stock-recruitment and functional response curves fitted to model outputs. In this appendix, we detail the (post-)fitting of stock-recruitment models to MICE estimates and summarize MICE projections using these post-fitted models.

### MICE estimates of herring recruitment and spawning biomass

The Spring, Fall-Middle, and Fall-South subpopulations exhibited generally positive relationships between spawning biomass and resulting recruitment, while there was no discernable pattern for Fall-North (Figure C1). Density-dependence was somewhat apparent for Spring herring, as the largest stock sizes generated only moderate recruitment (Figure C1). In contrast, large stock sizes for Fall-Middle and Fall-South generated large recruitment (Figure C1).

### Fitting spawner-recruitment relationships to MICE estimates

The shape of the stock-recruitment relationship for herring in the southern Gulf of St. Lawrence is unknown. Herring egg mortality often arises from suffocation at high densities (Haegele and Schweigert, 1985), which suggests that a dome-shaped function such as the Ricker may be preferable; however, overcompensation is not always evident in herring stock-recruitment analyses (Zheng 1996). Both Ricker and Beverton-Holt functions were previously fit to recruitment and spawning biomass estimates for herring in the southern Gulf of St. Lawrence, but neither curve was clearly preferable in that analysis (DFO, 2005). Given the *a priori* uncertainty around the shape of the stock-recruitment and given the lack of clear evidence for overcompensation in our MICE estimates, we tested both Ricker (1) and Beverton-Holt (2) functions in our analysis, i.e.,

$$\hat{R}_{x,t} = \Psi_{0,x}^{(\mathrm{R})} S_{x,t-2} \exp\left(-\Psi_{1,x}^{(\mathrm{R})} S_{x,t-2}\right) \quad (1)$$

$$\hat{R}_{x,t} = \Psi_{0,x}^{(\mathrm{BH})} S_{x,t-2} / \left(1 + \Psi_{1,x}^{(\mathrm{BH})} S_{x,t-2}\right) \quad (2)$$

where $R_{x,t}$ and $S_{x,t}$ to represent herring recruitment and spawning biomass, respectively, for subpopulation $x$ in year $t$, $\Psi_{0,x}$ represents density-independent fecundity and $\Psi_{1,x}$ represents the strength of density-dependence. We use the notation $R_{x,t}$ and $S_{x,t}$ for simplicity in this appendix; in MICE notation, these variables correspond to $N_{i=3,x,a=2,t}$ and $B_{i=3,x,t}$.

MICE-estimated recruitments were assumed to be lognormally-distributed around stock-recruitment function predictions, i.e.,

$$\log(R_{x,t}) \sim N(\log(\hat{R}_{x,t}), \sigma_x^2)$$

The standard deviation parameters ($\sigma_x^2$) were estimated. Each stock-recruitment model was fitted to each posterior MICE sample.

Ricker and Beverton-Holt models fits to the MICE estimates of spawning biomass and recruitment for the Fall subpopulations were nearly identical (Figure C2). Residual variance was particularly high for the Fall-North population, which was not surprising given the lack of a clear spawner-recruitment relationship in the MICE estimates. The North subpopulation had the highest productivity at low density and the strongest density-dependent effect (Table C1).

**Extirpation risk projections under Beverton-Holt recruitment**

Projections of the MICE assuming Beverton-Holt stock-recruitment dynamics for herring were nearly identical to the Ricker projections presented in the main article. The herring recruitment rate was insensitive to the choice of spawner-recruitment function, resulting in very similar levels of herring spawning biomass (Figure C3). Consequently, the choice of spawner-recruitment function had virtually no impact on the projected recovery of cod (Figure C4).

## Tables

Table C1. Post-fitted herring stock-recruitment parameter estimates. $\Psi_{0,x}$ represents fecundity for subpopulation x, $\Psi_{1,x}$ represents the effect of density dependence for subpopulation x, and $\sigma_x^2$ is the residual standard deviation for subpopulation x.

| **Parameter** | **Posterior quantiles** | | | | | |
|---|---|---|---|---|---|---|
| | **2.5%** | **50%** | **97.5%** | **2.5%** | **50%** | **97.5%** |
| | | *Ricker* | | | *Beverton-Holt* | |
| $\Psi_{0,x=1}$ | 13.47 | 15.41 | 17.26 | 16.36 | 21.72 | 29.42 |
| $\Psi_{0,x=2}$ | 8.12 | 13.25 | 23.92 | 8.29 | 19.04 | 109.85 |
| $\Psi_{0,x=3}$ | 8.64 | 9.24 | 9.97 | 8.64 | 9.26 | 10.00 |
| $\Psi_{0,x=4}$ | 10.45 | 12.79 | 14.45 | 11.09 | 14.32 | 17.25 |
| $\Psi_{1,x=1}$ | 8.10E-03 | 9.61E-03 | 1.10E-02 | 1.83E-02 | 2.87E-02 | 4.36E-02 |
| $\Psi_{1,x=2}$ | 1.37E-03 | 6.85E-03 | 1.44E-02 | 1.72E-03 | 1.85E-02 | 5.89E+04 |
| $\Psi_{1,x=3}$ | 3.06E-11 | 3.44E-05 | 2.14E-03 | 6.23E-15 | 1.20E-05 | 2.48E-03 |
| $\Psi_{1,x=4}$ | 2.19E-03 | 3.43E-03 | 4.50E-03 | 3.23E-03 | 6.11E-03 | 9.74E-03 |
| $\sigma_{x=1}^2$ | 0.492 | 0.575 | 0.730 | 0.486 | 0.568 | 0.728 |
| $\sigma_{x=2}^2$ | 0.763 | 1.081 | 1.430 | 0.763 | 1.081 | 1.428 |
| $\sigma_{x=3}^2$ | 0.419 | 0.477 | 0.566 | 0.419 | 0.477 | 0.566 |
| $\sigma_{x=4}^2$ | 0.465 | 0.528 | 0.599 | 0.462 | 0.524 | 0.596 |

**Figures**

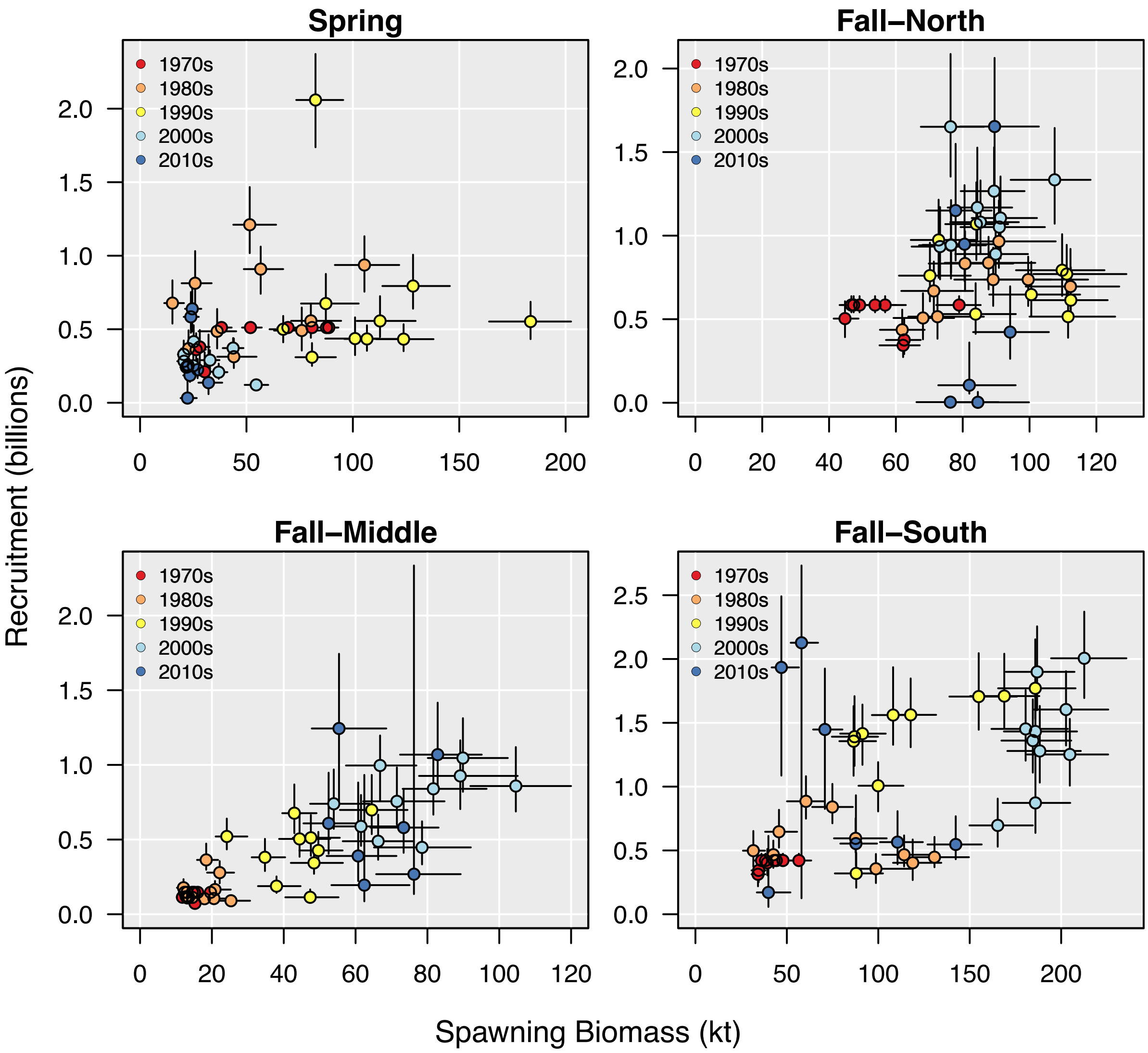

Figure C1. MICE estimates of spawning biomass and recruitment for each herring subpopulation. Circles represent posterior modes while lines indicate the central 95% posterior interval.

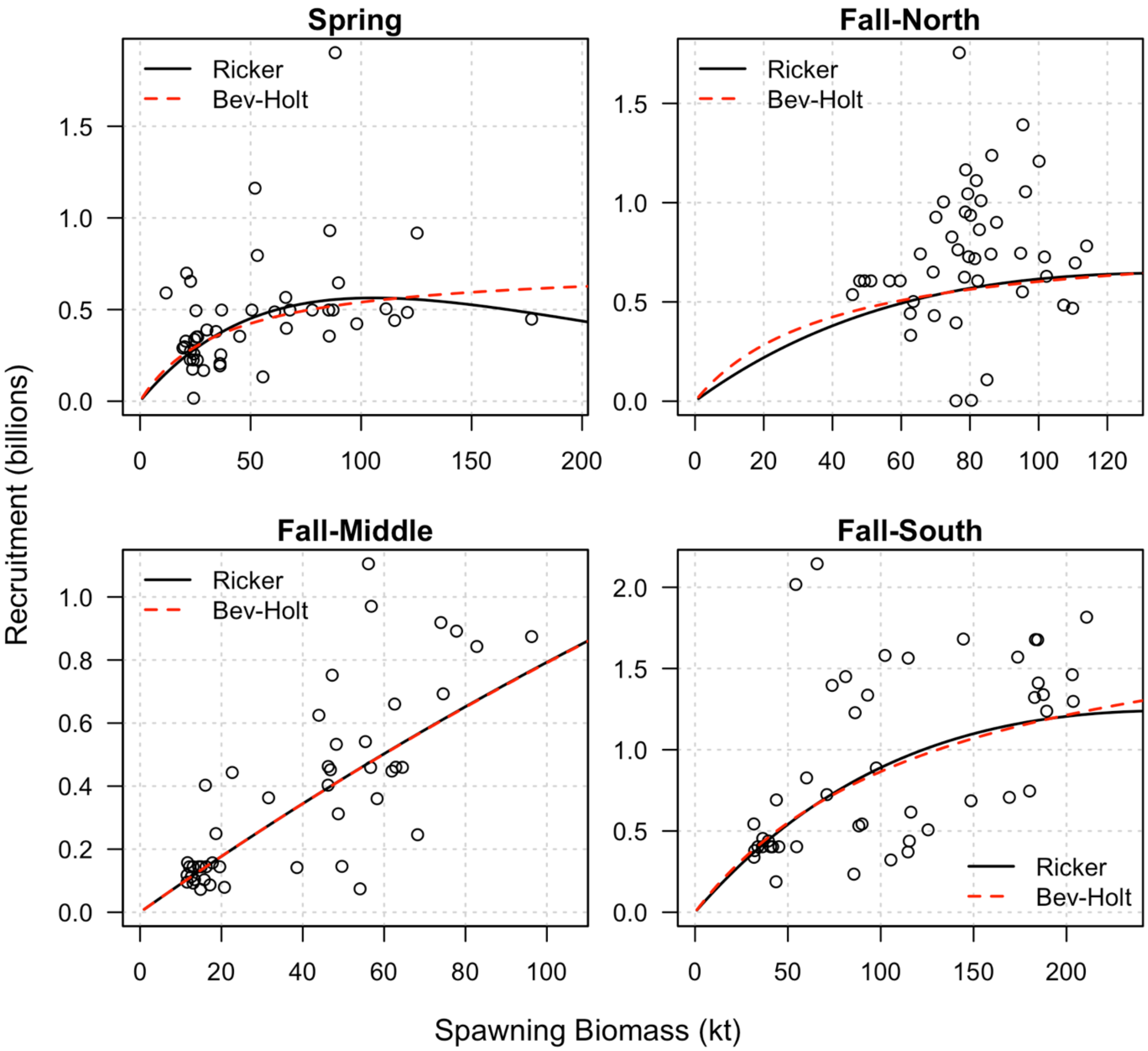

Figure C2. Post-fitted stock-recruitment relationships for a randomly selected posterior sample of spawning biomass and recruitment from the MICE.

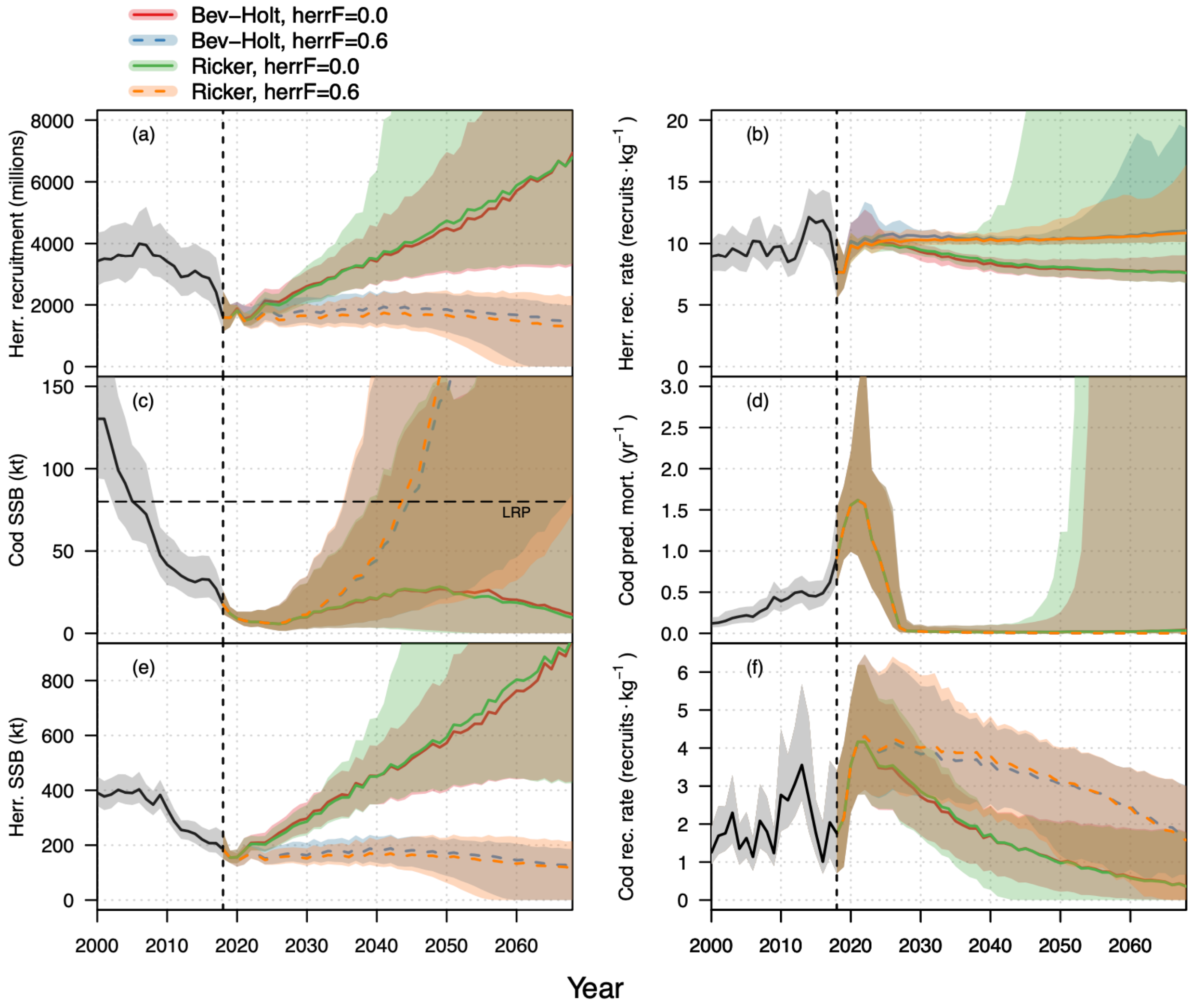

Figure C3. Model projections under four harvest plans with annual quotas of 12,000 seals targeting 50% YOY for 10 years. Lines represent posterior modes while shaded regions indicate the central 95% uncertainty interval. Black lines and grey shaded regions represent historical estimates while coloured lines and shaded regions represent projections.

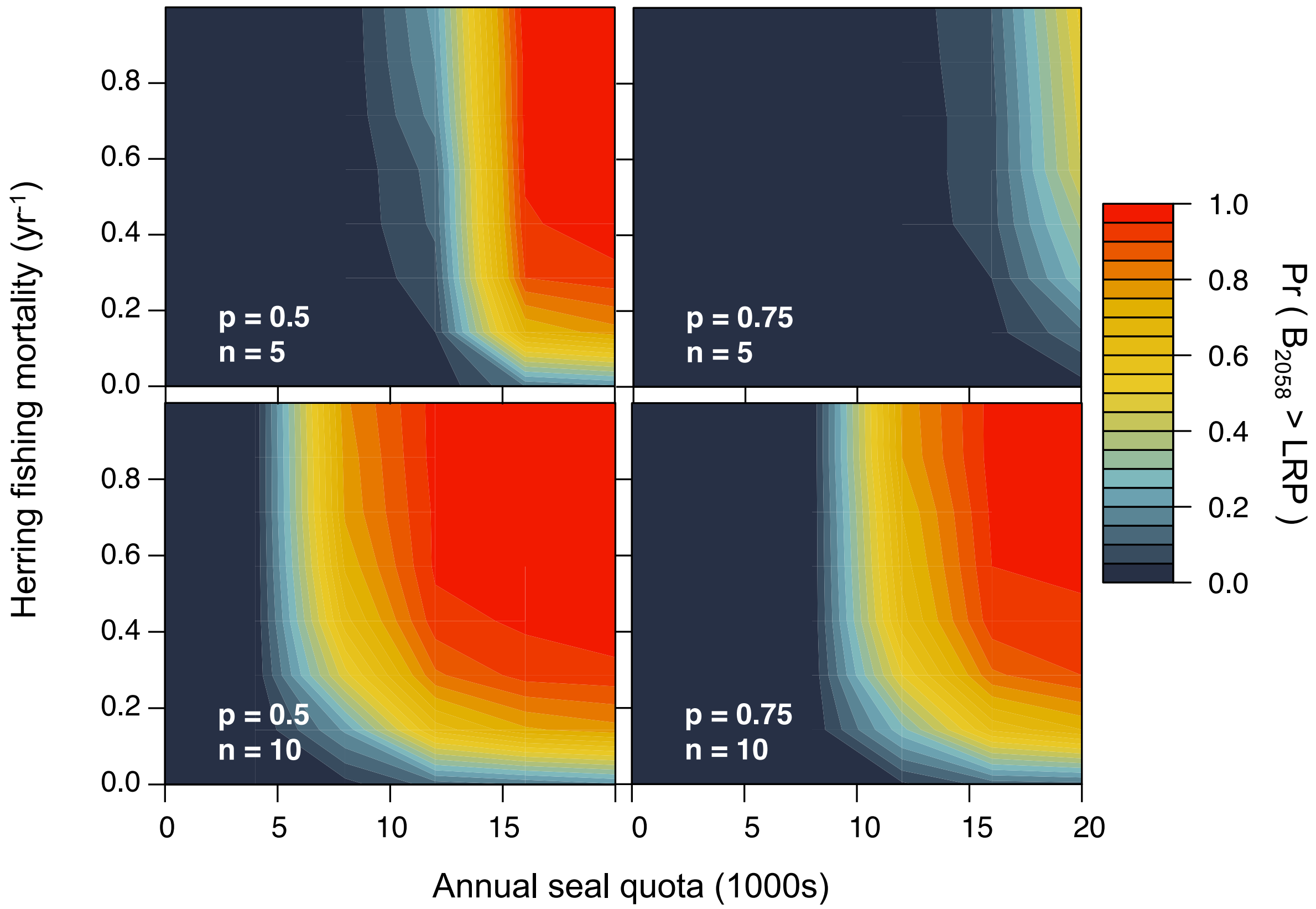

Figure C4. Probability of cod biomass in 2058 exceeding the limit reference point (LRP = 80 kt) given Beverton-Holt stock-recruitment functions for Herring, varying levels of seal quota (x-axis) and herring fishing mortality (y-axis), and four combinations of the proportion of YOY seals targeted for removal (p) and the length of the seal harvest period (n; number of years).

## Appendix D. Model fits and diagnostics

Table D1. Posterior distribution summaries and diagnostics. tESS and bESS represent tail and bulk effective sample sizes, respectively. For the grey seal component of the model, D and θ represent the half-saturation point and shape of the grey seal juvenile density-dependence relationship, $P_{max}$ is the reproductive rate among mature females, $a_{50\%}$ and $a_{step}$ are parameters of the reproductive rate-at-age.

| Parameter | Mean | SD | 2.5% | 25% | 50% | 75% | 97.5% | tESS | bESS | Rhat |
|---|---|---|---|---|---|---|---|---|---|---|
| *Grey seal population dynamics* | | | | | | | | | | |
| $\ln N_1^{(\text{init})}$ | 1.965 | 0.034 | 1.901 | 1.941 | 1.965 | 1.988 | 2.031 | 1710 | 1088 | 0.999 |
| $\ln N_2^{(\text{init})}$ | 2.395 | 0.060 | 2.291 | 2.353 | 2.389 | 2.431 | 2.525 | 1052 | 556 | 1.001 |
| $\ln D_{1,1}$ | 3.126 | 0.130 | 2.918 | 3.035 | 3.109 | 3.202 | 3.426 | 1765 | 864 | 1.000 |
| $\ln D_{1,2}$ | 1.777 | 0.061 | 1.648 | 1.738 | 1.778 | 1.818 | 1.892 | 945 | 904 | 1.003 |
| $\ln D_{2,1}$ | -0.403 | 0.134 | -0.687 | -0.487 | -0.395 | -0.306 | -0.161 | 1470 | 955 | 1.001 |
| $\ln D_{2,2}$ | 1.069 | 0.144 | 0.780 | 0.974 | 1.070 | 1.165 | 1.346 | 890 | 715 | 1.000 |
| $\text{logit}\bar{\gamma}$ | 2.053 | 0.055 | 1.948 | 2.015 | 2.053 | 2.092 | 2.164 | 2559 | 1122 | 1.003 |
| $\ln a^{(50\%)}$ | 1.575 | 0.009 | 1.558 | 1.569 | 1.575 | 1.581 | 1.592 | 3305 | 1147 | 1.006 |
| $\ln(a^{(95\%)}\text{-}a^{(50\%)})$ | 0.563 | 0.077 | 0.417 | 0.510 | 0.563 | 0.616 | 0.707 | 2616 | 1396 | 1.002 |
| *Grey seal relative consumption-at-age* | | | | | | | | | | |
| $\rho_{1,1}^{50\%}$ | 0.262 | 0.444 | -0.676 | -0.027 | 0.301 | 0.567 | 1.074 | 2343 | 1053 | 1.003 |
| $\rho_{1,2}^{50\%}$ | -0.383 | 0.619 | -1.710 | -0.767 | -0.335 | 0.051 | 0.678 | 2407 | 1147 | 1.001 |
| $\rho_{1,1}^{95\%}\text{-}\rho_{1,1}^{50\%}$ | 13.625 | 1.933 | 10.409 | 12.214 | 13.469 | 14.808 | 17.907 | 2158 | 1056 | 1.001 |
| $\rho_{1,2}^{95\%}\text{-}\rho_{1,2}^{50\%}$ | 18.623 | 3.170 | 13.709 | 16.472 | 18.212 | 20.276 | 26.680 | 1935 | 1024 | 1.001 |

| | | | | | | | | | |
|---|---|---|---|---|---|---|---|---|---|
| *Grey seal per-capita consumption* | | | | | | | | | |
| $\ln\varphi_{1,1,1,1}$ | -8.751 | 0.098 | -8.940 | -8.818 | -8.747 | -8.685 | -8.565 | 638 | 987 | 1.002 |
| $\ln\varphi_{1,1,2,1}$ | -8.388 | 0.122 | -8.621 | -8.467 | -8.389 | -8.304 | -8.148 | 395 | 681 | 1.006 |
| $\ln\varphi_{1,2,1,1}$ | -9.543 | 0.100 | -9.742 | -9.609 | -9.540 | -9.475 | -9.356 | 630 | 1011 | 1.002 |
| $\ln\varphi_{1,2,2,1}$ | -7.734 | 0.122 | -7.970 | -7.820 | -7.732 | -7.654 | -7.500 | 358 | 604 | 1.006 |
| $\ln\varphi_{1,3,1,1}$ | -5.121 | 0.097 | -5.315 | -5.188 | -5.120 | -5.055 | -4.944 | 711 | 1196 | 1.002 |
| $\ln\varphi_{1,3,2,1}$ | -5.021 | 0.122 | -5.254 | -5.099 | -5.019 | -4.937 | -4.780 | 402 | 634 | 1.006 |
| $\ln\varphi_{1,4,1,1}$ | -6.970 | 0.099 | -7.171 | -7.035 | -6.967 | -6.906 | -6.783 | 631 | 1222 | 1.000 |
| $\ln\varphi_{1,4,2,1}$ | -5.186 | 0.122 | -5.421 | -5.268 | -5.186 | -5.103 | -4.943 | 402 | 536 | 1.005 |
| $\varepsilon^{(\varphi)}_{1,2,2}$ | 0.055 | 0.120 | -0.181 | -0.024 | 0.056 | 0.130 | 0.308 | 1262 | 1054 | 0.999 |
| $\varepsilon^{(\varphi)}_{1,3,2}$ | 0.137 | 0.111 | -0.090 | 0.066 | 0.140 | 0.218 | 0.335 | 1355 | 961 | 1.002 |
| $\varepsilon^{(\varphi)}_{1,2,3}$ | 0.155 | 0.120 | -0.070 | 0.073 | 0.155 | 0.235 | 0.389 | 2284 | 1047 | 0.999 |
| $\varepsilon^{(\varphi)}_{1,3,3}$ | 0.033 | 0.108 | -0.183 | -0.039 | 0.033 | 0.105 | 0.239 | 3062 | 1020 | 1.005 |
| $\varepsilon^{(\varphi)}_{1,2,4}$ | 0.215 | 0.110 | -0.006 | 0.145 | 0.215 | 0.288 | 0.425 | 1971 | 1196 | 1.002 |
| $\varepsilon^{(\varphi)}_{1,3,4}$ | -0.001 | 0.112 | -0.222 | -0.080 | -0.002 | 0.078 | 0.215 | 2252 | 1125 | 0.999 |
| $\varepsilon^{(\varphi)}_{1,2,5}$ | 0.011 | 0.109 | -0.214 | -0.062 | 0.010 | 0.089 | 0.214 | 2079 | 1077 | 1.000 |
| $\varepsilon^{(\varphi)}_{1,3,5}$ | 0.121 | 0.107 | -0.088 | 0.046 | 0.121 | 0.194 | 0.335 | 2204 | 1217 | 1.001 |
| $\varepsilon^{(\varphi)}_{1,2,6}$ | -0.055 | 0.109 | -0.263 | -0.129 | -0.058 | 0.019 | 0.170 | 1915 | 1076 | 1.001 |
| $\varepsilon^{(\varphi)}_{1,3,6}$ | 0.034 | 0.107 | -0.185 | -0.038 | 0.038 | 0.105 | 0.233 | 2400 | 1074 | 1.000 |
| $\varepsilon_{1,2,7}^{(\varphi)}$ | -0.211 | 0.113 | -0.436 | -0.289 | -0.211 | -0.138 | 0.009 | 3248 | 1233 | 1.003 |
| $\varepsilon_{1,3,7}^{(\varphi)}$ | 0.014 | 0.104 | -0.178 | -0.057 | 0.014 | 0.081 | 0.215 | 2475 | 1084 | 1.002 |
| $\varepsilon_{1,2,8}^{(\varphi)}$ | -0.498 | 0.112 | -0.728 | -0.574 | -0.499 | -0.419 | -0.277 | 2947 | 1179 | 0.999 |
| $\varepsilon_{1,3,8}^{(\varphi)}$ | -0.158 | 0.100 | -0.350 | -0.226 | -0.161 | -0.092 | 0.037 | 2482 | 1273 | 1.002 |
| $\varepsilon_{1,2,9}^{(\varphi)}$ | -0.243 | 0.112 | -0.453 | -0.321 | -0.244 | -0.162 | -0.024 | 2357 | 1303 | 0.999 |
| $\varepsilon_{1,3,9}^{(\varphi)}$ | 0.052 | 0.107 | -0.151 | -0.021 | 0.050 | 0.122 | 0.273 | 2649 | 1169 | 1.001 |
| $\varepsilon_{1,2,10}^{(\varphi)}$ | -0.233 | 0.113 | -0.465 | -0.308 | -0.232 | -0.155 | -0.015 | 2766 | 1286 | 1.002 |
| $\varepsilon_{1,3,10}^{(\varphi)}$ | 0.103 | 0.104 | -0.103 | 0.035 | 0.099 | 0.175 | 0.304 | 2664 | 1077 | 1.002 |
| $\varepsilon_{1,2,11}^{(\varphi)}$ | -0.017 | 0.111 | -0.240 | -0.090 | -0.015 | 0.055 | 0.193 | 2214 | 954 | 1.001 |
| $\varepsilon_{1,3,11}^{(\varphi)}$ | -0.168 | 0.108 | -0.392 | -0.237 | -0.166 | -0.095 | 0.045 | 3066 | 1124 | 1.002 |
| $\varepsilon_{1,2,12}^{(\varphi)}$ | 0.020 | 0.110 | -0.189 | -0.052 | 0.015 | 0.090 | 0.243 | 2661 | 728 | 1.003 |
| $\varepsilon_{1,3,12}^{(\varphi)}$ | -0.117 | 0.108 | -0.326 | -0.191 | -0.117 | -0.044 | 0.088 | 2234 | 1076 | 0.999 |
| $\varepsilon_{1,2,13}^{(\varphi)}$ | 0.143 | 0.110 | -0.074 | 0.072 | 0.145 | 0.216 | 0.352 | 3130 | 1202 | 1.002 |
| $\varepsilon_{1,3,13}^{(\varphi)}$ | -0.177 | 0.104 | -0.379 | -0.248 | -0.178 | -0.108 | 0.027 | 1773 | 1344 | 1.001 |
| $\varepsilon_{1,2,14}^{(\varphi)}$ | -0.092 | 0.111 | -0.312 | -0.166 | -0.093 | -0.019 | 0.127 | 2422 | 1110 | 1.001 |
| $\varepsilon_{1,3,14}^{(\varphi)}$ | -0.200 | 0.109 | -0.422 | -0.273 | -0.200 | -0.125 | 0.017 | 1780 | 1120 | 1.000 |
| $\varepsilon_{1,2,15}^{(\varphi)}$ | -0.154 | 0.113 | -0.372 | -0.234 | -0.154 | -0.081 | 0.060 | 2812 | 1144 | 1.001 |
| $\varepsilon_{1,3,15}^{(\varphi)}$ | -0.144 | 0.108 | -0.359 | -0.215 | -0.140 | -0.071 | 0.056 | 1884 | 1160 | 0.999 |
| $\varepsilon^{(\varphi)}_{1,2,16}$ | 0.116 | 0.104 | -0.083 | 0.045 | 0.114 | 0.186 | 0.318 | 2552 | 993 | 1.000 |
| $\varepsilon^{(\varphi)}_{1,3,16}$ | -0.057 | 0.108 | -0.259 | -0.132 | -0.054 | 0.015 | 0.157 | 2428 | 1226 | 1.001 |
| $\varepsilon^{(\varphi)}_{1,2,17}$ | 0.187 | 0.112 | -0.036 | 0.110 | 0.189 | 0.263 | 0.405 | 1978 | 1131 | 1.003 |
| $\varepsilon^{(\varphi)}_{1,3,17}$ | 0.017 | 0.104 | -0.188 | -0.053 | 0.019 | 0.081 | 0.230 | 1987 | 1091 | 1.004 |
| $\varepsilon^{(\varphi)}_{1,2,18}$ | 0.105 | 0.112 | -0.115 | 0.029 | 0.106 | 0.180 | 0.321 | 1767 | 806 | 1.001 |
| $\varepsilon^{(\varphi)}_{1,3,18}$ | 0.008 | 0.109 | -0.199 | -0.066 | 0.006 | 0.081 | 0.225 | 2963 | 1164 | 1.005 |
| $\varepsilon^{(\varphi)}_{1,2,19}$ | 0.215 | 0.111 | -0.002 | 0.143 | 0.217 | 0.293 | 0.432 | 3393 | 1251 | 1.001 |
| $\varepsilon^{(\varphi)}_{1,3,19}$ | -0.055 | 0.111 | -0.273 | -0.130 | -0.060 | 0.019 | 0.171 | 2732 | 1168 | 1.001 |
| $\varepsilon^{(\varphi)}_{1,2,20}$ | 0.221 | 0.114 | 0.014 | 0.144 | 0.218 | 0.297 | 0.448 | 2393 | 1041 | 1.000 |
| $\varepsilon^{(\varphi)}_{1,3,20}$ | -0.073 | 0.107 | -0.279 | -0.144 | -0.075 | -0.008 | 0.146 | 2606 | 985 | 1.000 |
| $\varepsilon^{(\varphi)}_{1,2,21}$ | 0.255 | 0.109 | 0.050 | 0.180 | 0.256 | 0.330 | 0.462 | 2354 | 1038 | 1.000 |
| $\varepsilon^{(\varphi)}_{1,3,21}$ | 0.025 | 0.107 | -0.183 | -0.047 | 0.024 | 0.096 | 0.238 | 2825 | 1360 | 1.000 |
| $\varepsilon^{(\varphi)}_{1,2,22}$ | 0.094 | 0.118 | -0.141 | 0.014 | 0.097 | 0.174 | 0.320 | 2256 | 1144 | 1.001 |
| $\varepsilon^{(\varphi)}_{1,3,22}$ | -0.009 | 0.106 | -0.207 | -0.080 | -0.012 | 0.064 | 0.202 | 2982 | 1249 | 1.003 |
| $\varepsilon^{(\varphi)}_{1,2,23}$ | 0.003 | 0.111 | -0.216 | -0.068 | 0.001 | 0.077 | 0.223 | 2619 | 1044 | 1.001 |
| $\varepsilon^{(\varphi)}_{1,3,23}$ | -0.114 | 0.108 | -0.324 | -0.185 | -0.115 | -0.043 | 0.090 | 2884 | 1270 | 1.001 |
| $\varepsilon^{(\varphi)}_{1,2,24}$ | -0.007 | 0.109 | -0.221 | -0.083 | -0.009 | 0.067 | 0.203 | 2610 | 1176 | 1.000 |
| $\varepsilon^{(\varphi)}_{1,3,24}$ | -0.027 | 0.098 | -0.222 | -0.096 | -0.029 | 0.038 | 0.171 | 2773 | 1074 | 1.002 |

| | | | | | | | | | |
|---|---|---|---|---|---|---|---|---|---|
| $\varepsilon_{1,2,25}^{(\varphi)}$ | -0.029 | 0.111 | -0.238 | -0.107 | -0.031 | 0.042 | 0.191 | 2558 | 1207 | 1.001 |
| $\varepsilon_{1,3,25}^{(\varphi)}$ | 0.047 | 0.103 | -0.152 | -0.021 | 0.046 | 0.113 | 0.257 | 2164 | 1258 | 1.002 |
| $\varepsilon_{1,2,26}^{(\varphi)}$ | -0.058 | 0.113 | -0.278 | -0.132 | -0.061 | 0.017 | 0.174 | 3168 | 1124 | 0.999 |
| $\varepsilon_{1,3,26}^{(\varphi)}$ | 0.001 | 0.104 | -0.217 | -0.065 | -0.001 | 0.069 | 0.198 | 2234 | 981 | 1.000 |
| $\varepsilon_{1,2,27}^{(\varphi)}$ | -0.050 | 0.109 | -0.261 | -0.121 | -0.051 | 0.026 | 0.167 | 3574 | 1232 | 1.003 |
| $\varepsilon_{1,3,27}^{(\varphi)}$ | 0.020 | 0.117 | -0.203 | -0.058 | 0.019 | 0.093 | 0.268 | 3267 | 877 | 1.005 |
| $\varepsilon_{1,2,28}^{(\varphi)}$ | 0.071 | 0.109 | -0.143 | -0.001 | 0.071 | 0.141 | 0.286 | 3177 | 973 | 1.001 |
| $\varepsilon_{1,3,28}^{(\varphi)}$ | 0.023 | 0.114 | -0.194 | -0.057 | 0.021 | 0.099 | 0.244 | 2328 | 1192 | 1.001 |
| $\varepsilon_{1,2,29}^{(\varphi)}$ | 0.026 | 0.109 | -0.186 | -0.046 | 0.026 | 0.101 | 0.249 | 3021 | 1070 | 1.005 |
| $\varepsilon_{1,3,29}^{(\varphi)}$ | -0.041 | 0.109 | -0.247 | -0.116 | -0.044 | 0.035 | 0.173 | 2596 | 1237 | 1.003 |
| $\varepsilon_{1,2,30}^{(\varphi)}$ | -0.017 | 0.106 | -0.223 | -0.085 | -0.020 | 0.051 | 0.197 | 2613 | 1040 | 1.005 |
| $\varepsilon_{1,3,30}^{(\varphi)}$ | -0.013 | 0.101 | -0.216 | -0.079 | -0.015 | 0.054 | 0.189 | 2690 | 1107 | 1.002 |
| $\varepsilon_{1,2,31}^{(\varphi)}$ | 0.048 | 0.108 | -0.162 | -0.021 | 0.045 | 0.120 | 0.262 | 3182 | 1043 | 1.010 |
| $\varepsilon_{1,3,31}^{(\varphi)}$ | 0.016 | 0.105 | -0.192 | -0.057 | 0.019 | 0.086 | 0.224 | 3175 | 1406 | 1.001 |
| $\varepsilon_{1,2,32}^{(\varphi)}$ | 0.195 | 0.105 | -0.006 | 0.125 | 0.192 | 0.268 | 0.396 | 2689 | 1137 | 0.999 |
| $\varepsilon_{1,3,32}^{(\varphi)}$ | -0.046 | 0.107 | -0.254 | -0.121 | -0.048 | 0.026 | 0.159 | 2807 | 1314 | 1.000 |
| $\varepsilon_{1,2,33}^{(\varphi)}$ | 0.094 | 0.108 | -0.120 | 0.023 | 0.092 | 0.168 | 0.300 | 2333 | 1305 | 1.004 |
| $\varepsilon_{1,3,33}^{(\varphi)}$ | 0.029 | 0.110 | -0.191 | -0.045 | 0.030 | 0.104 | 0.250 | 2299 | 1333 | 1.001 |
| $\varepsilon_{1,2,34}^{(\varphi)}$ | 0.103 | 0.110 | -0.108 | 0.030 | 0.100 | 0.178 | 0.321 | 2689 | 1134 | 1.003 |
| $\varepsilon_{1,3,34}^{(\varphi)}$ | -0.014 | 0.110 | -0.237 | -0.090 | -0.016 | 0.061 | 0.192 | 2399 | 1250 | 1.004 |
| $\varepsilon_{1,2,35}^{(\varphi)}$ | 0.119 | 0.110 | -0.109 | 0.046 | 0.119 | 0.199 | 0.322 | 2786 | 1200 | 1.001 |
| $\varepsilon_{1,3,35}^{(\varphi)}$ | -0.027 | 0.109 | -0.236 | -0.100 | -0.025 | 0.045 | 0.189 | 2818 | 1311 | 1.000 |
| $\varepsilon_{1,2,36}^{(\varphi)}$ | 0.002 | 0.113 | -0.210 | -0.074 | 0.001 | 0.077 | 0.218 | 2402 | 994 | 1.004 |
| $\varepsilon_{1,3,36}^{(\varphi)}$ | 0.029 | 0.106 | -0.178 | -0.042 | 0.032 | 0.100 | 0.239 | 2218 | 1023 | 1.001 |
| $\varepsilon_{1,2,37}^{(\varphi)}$ | 0.197 | 0.108 | -0.005 | 0.121 | 0.198 | 0.270 | 0.410 | 2885 | 1172 | 1.001 |
| $\varepsilon_{1,3,37}^{(\varphi)}$ | 0.067 | 0.108 | -0.144 | -0.004 | 0.068 | 0.141 | 0.283 | 2489 | 1198 | 1.003 |
| $\varepsilon_{1,2,38}^{(\varphi)}$ | 0.129 | 0.107 | -0.077 | 0.054 | 0.129 | 0.204 | 0.337 | 2909 | 1313 | 1.003 |
| $\varepsilon_{1,3,38}^{(\varphi)}$ | 0.053 | 0.107 | -0.155 | -0.020 | 0.052 | 0.124 | 0.257 | 2466 | 1143 | 1.002 |
| $\varepsilon_{1,2,39}^{(\varphi)}$ | 0.332 | 0.110 | 0.114 | 0.257 | 0.330 | 0.407 | 0.550 | 3093 | 1287 | 1.001 |
| $\varepsilon_{1,3,39}^{(\varphi)}$ | -0.008 | 0.101 | -0.200 | -0.078 | -0.005 | 0.060 | 0.192 | 2276 | 1069 | 1.002 |
| $\varepsilon_{1,2,40}^{(\varphi)}$ | -0.069 | 0.107 | -0.275 | -0.140 | -0.070 | 0.002 | 0.141 | 2890 | 1458 | 1.000 |
| $\varepsilon_{1,3,40}^{(\varphi)}$ | 0.083 | 0.103 | -0.116 | 0.013 | 0.084 | 0.154 | 0.277 | 2677 | 1349 | 1.000 |
| $\varepsilon_{1,2,41}^{(\varphi)}$ | 0.085 | 0.099 | -0.114 | 0.019 | 0.084 | 0.153 | 0.274 | 2231 | 1233 | 1.002 |
| $\varepsilon_{1,3,41}^{(\varphi)}$ | 0.119 | 0.105 | -0.085 | 0.049 | 0.119 | 0.188 | 0.324 | 2448 | 1277 | 1.000 |
| $\varepsilon_{1,2,42}^{(\varphi)}$ | 0.167 | 0.102 | -0.039 | 0.102 | 0.166 | 0.238 | 0.363 | 2053 | 1160 | 1.001 |
| $\varepsilon_{1,3,42}^{(\varphi)}$ | 0.056 | 0.107 | -0.158 | -0.017 | 0.056 | 0.128 | 0.269 | 3133 | 1280 | 1.003 |
| $\varepsilon_{1,2,43}^{(\varphi)}$ | 0.071 | 0.103 | -0.122 | -0.001 | 0.068 | 0.141 | 0.278 | 2258 | 1307 | 1.001 |
| $\varepsilon_{1,3,43}^{(\varphi)}$ | 0.000 | 0.113 | -0.213 | -0.073 | 0.002 | 0.072 | 0.236 | 2177 | 1094 | 1.000 |
| $\varepsilon_{1,2,44}^{(\varphi)}$ | -0.091 | 0.100 | -0.294 | -0.154 | -0.094 | -0.023 | 0.106 | 2641 | 1138 | 1.002 |
| $\varepsilon_{1,3,44}^{(\varphi)}$ | 0.090 | 0.109 | -0.131 | 0.021 | 0.090 | 0.165 | 0.302 | 1590 | 1242 | 1.004 |
| $\varepsilon_{1,2,45}^{(\varphi)}$ | 0.027 | 0.102 | -0.169 | -0.045 | 0.029 | 0.097 | 0.214 | 2637 | 1277 | 1.001 |
| $\varepsilon_{1,3,45}^{(\varphi)}$ | 0.113 | 0.109 | -0.096 | 0.042 | 0.112 | 0.188 | 0.317 | 2224 | 1252 | 1.004 |
| $\varepsilon_{1,2,46}^{(\varphi)}$ | 0.110 | 0.104 | -0.094 | 0.040 | 0.109 | 0.179 | 0.315 | 2379 | 987 | 1.002 |
| $\varepsilon_{1,3,46}^{(\varphi)}$ | 0.078 | 0.113 | -0.141 | 0.000 | 0.077 | 0.152 | 0.300 | 2627 | 1004 | 0.999 |
| $\varepsilon_{1,2,47}^{(\varphi)}$ | 0.245 | 0.110 | 0.027 | 0.175 | 0.241 | 0.320 | 0.462 | 2152 | 1012 | 1.003 |
| $\varepsilon_{1,3,47}^{(\varphi)}$ | 0.092 | 0.110 | -0.136 | 0.020 | 0.090 | 0.169 | 0.300 | 2339 | 1125 | 1.006 |
| $\varepsilon_{1,2,48}^{(\varphi)}$ | 0.425 | 0.116 | 0.189 | 0.349 | 0.425 | 0.496 | 0.660 | 2655 | 1054 | 1.005 |
| $\varepsilon_{1,3,48}^{(\varphi)}$ | 0.156 | 0.113 | -0.070 | 0.082 | 0.158 | 0.232 | 0.379 | 2205 | 1306 | 1.000 |
| *Cod relative consumption-at-age* | | | | | | | | | | |
| $\rho_{2,1}^{50\%}$ | 6.398 | 0.011 | 6.376 | 6.391 | 6.398 | 6.405 | 6.419 | 2875 | 948 | 1.005 |
| $\rho_{2,1}^{95\%}$-$\rho_{2,1}^{50\%}$ | 1.316 | 0.004 | 1.308 | 1.313 | 1.316 | 1.319 | 1.324 | 2605 | 999 | 1.004 |
| *Cod per-capita consumption* | | | | | | | | | | |
| $\ln\varphi_{2,1,3,1}$ | -8.030 | 0.030 | -8.088 | -8.052 | -8.029 | -8.010 | -7.969 | 2331 | 1190 | 1.001 |
| $\varepsilon_{2,3,2}^{(\varphi)}$ | 0.109 | 0.032 | 0.048 | 0.087 | 0.109 | 0.129 | 0.169 | 2345 | 1160 | 1.002 |
| $\varepsilon_{2,3,3}^{(\varphi)}$ | -0.110 | 0.030 | -0.165 | -0.130 | -0.110 | -0.089 | -0.049 | 2356 | 1268 | 1.000 |
| $\varepsilon_{2,3,4}^{(\varphi)}$ | -0.061 | 0.029 | -0.117 | -0.081 | -0.062 | -0.042 | -0.005 | 1934 | 902 | 1.000 |
| $\varepsilon_{2,3,5}^{(\varphi)}$ | -0.303 | 0.031 | -0.366 | -0.324 | -0.303 | -0.282 | -0.243 | 2420 | 1077 | 1.006 |
| $\varepsilon_{2,3,6}^{(\varphi)}$ | 0.286 | 0.031 | 0.227 | 0.263 | 0.286 | 0.307 | 0.347 | 2521 | 1101 | 1.002 |
| $\varepsilon_{2,3,7}^{(\varphi)}$ | -0.255 | 0.031 | -0.318 | -0.276 | -0.255 | -0.234 | -0.197 | 2862 | 1313 | 1.000 |
| $\varepsilon_{2,3,8}^{(\varphi)}$ | -0.221 | 0.031 | -0.280 | -0.241 | -0.221 | -0.201 | -0.161 | 2692 | 1200 | 1.003 |
| $\varepsilon_{2,3,9}^{(\varphi)}$ | -0.638 | 0.032 | -0.702 | -0.660 | -0.638 | -0.617 | -0.575 | 3732 | 1006 | 1.004 |
| $\varepsilon_{2,3,10}^{(\varphi)}$ | 0.622 | 0.031 | 0.559 | 0.601 | 0.622 | 0.644 | 0.681 | 2657 | 1090 | 1.002 |
| $\varepsilon_{2,3,11}^{(\varphi)}$ | 0.140 | 0.032 | 0.075 | 0.118 | 0.140 | 0.162 | 0.202 | 2547 | 946 | 1.000 |
| $\varepsilon_{2,3,12}^{(\varphi)}$ | 0.125 | 0.033 | 0.060 | 0.104 | 0.127 | 0.147 | 0.188 | 3381 | 1174 | 0.999 |
| $\varepsilon_{2,3,13}^{(\varphi)}$ | 0.500 | 0.031 | 0.437 | 0.479 | 0.500 | 0.521 | 0.558 | 3673 | 1304 | 1.004 |
| $\varepsilon_{2,3,14}^{(\varphi)}$ | 0.768 | 0.031 | 0.708 | 0.747 | 0.769 | 0.789 | 0.825 | 2791 | 1054 | 1.001 |
| $\varepsilon_{2,3,15}^{(\varphi)}$ | 0.182 | 0.030 | 0.125 | 0.162 | 0.182 | 0.202 | 0.242 | 2505 | 1078 | 1.001 |
| $\varepsilon_{2,3,16}^{(\varphi)}$ | 0.123 | 0.031 | 0.063 | 0.103 | 0.123 | 0.143 | 0.183 | 2799 | 1192 | 1.000 |
| $\varepsilon_{2,3,17}^{(\varphi)}$ | -0.483 | 0.032 | -0.544 | -0.506 | -0.483 | -0.462 | -0.422 | 3052 | 1065 | 1.001 |
| $\varepsilon_{2,3,18}^{(\varphi)}$ | -0.021 | 0.031 | -0.081 | -0.041 | -0.021 | -0.001 | 0.038 | 2627 | 1174 | 1.005 |
| $\varepsilon_{2,3,19}^{(\varphi)}$ | -0.199 | 0.030 | -0.258 | -0.220 | -0.199 | -0.178 | -0.139 | 2827 | 1087 | 1.012 |
| $\varepsilon_{2,3,20}^{(\varphi)}$ | -0.181 | 0.031 | -0.243 | -0.202 | -0.182 | -0.159 | -0.124 | 3039 | 1121 | 1.001 |
| $\varepsilon_{2,3,21}^{(\varphi)}$ | -0.078 | 0.030 | -0.137 | -0.099 | -0.078 | -0.058 | -0.021 | 2975 | 1374 | 1.004 |
| $\varepsilon_{2,3,22}^{(\varphi)}$ | -0.053 | 0.031 | -0.115 | -0.072 | -0.052 | -0.032 | 0.008 | 2760 | 998 | 1.001 |
| $\varepsilon_{2,3,23}^{(\varphi)}$ | 0.204 | 0.032 | 0.144 | 0.181 | 0.205 | 0.227 | 0.265 | 2898 | 1173 | 0.999 |
| $\varepsilon_{2,3,24}^{(\varphi)}$ | 0.165 | 0.030 | 0.109 | 0.144 | 0.165 | 0.185 | 0.223 | 2432 | 1006 | 1.001 |
| $\varepsilon_{2,3,25}^{(\varphi)}$ | 0.136 | 0.032 | 0.076 | 0.115 | 0.136 | 0.158 | 0.196 | 3125 | 1142 | 1.001 |
| $\varepsilon_{2,3,26}^{(\varphi)}$ | 0.061 | 0.031 | 0.003 | 0.040 | 0.060 | 0.081 | 0.122 | 2730 | 1073 | 1.005 |
| $\varepsilon_{2,3,27}^{(\varphi)}$ | -0.016 | 0.030 | -0.074 | -0.038 | -0.017 | 0.004 | 0.044 | 2766 | 1172 | 0.999 |
| $\varepsilon_{2,3,28}^{(\varphi)}$ | -0.016 | 0.031 | -0.079 | -0.035 | -0.016 | 0.004 | 0.045 | 3080 | 1164 | 1.000 |
| $\varepsilon_{2,3,29}^{(\varphi)}$ | -0.127 | 0.031 | -0.190 | -0.148 | -0.126 | -0.106 | -0.066 | 3308 | 1239 | 1.000 |
| $\varepsilon_{2,3,30}^{(\varphi)}$ | 0.092 | 0.031 | 0.034 | 0.071 | 0.091 | 0.112 | 0.153 | 3799 | 1333 | 1.002 |
| $\varepsilon_{2,3,31}^{(\varphi)}$ | -0.067 | 0.030 | -0.127 | -0.087 | -0.068 | -0.047 | -0.005 | 2659 | 1211 | 1.001 |
| $\varepsilon_{2,3,32}^{(\varphi)}$ | -0.350 | 0.031 | -0.410 | -0.371 | -0.350 | -0.329 | -0.289 | 2753 | 1015 | 1.001 |
| $\varepsilon_{2,3,33}^{(\varphi)}$ | 0.064 | 0.031 | 0.003 | 0.044 | 0.063 | 0.085 | 0.124 | 2458 | 1210 | 1.007 |
| $\varepsilon_{2,3,34}^{(\varphi)}$ | -0.044 | 0.031 | -0.105 | -0.064 | -0.045 | -0.024 | 0.016 | 2861 | 1384 | 0.999 |
| $\varepsilon_{2,3,35}^{(\varphi)}$ | -0.084 | 0.031 | -0.143 | -0.104 | -0.082 | -0.064 | -0.022 | 3045 | 927 | 1.000 |
| $\varepsilon_{2,3,36}^{(\varphi)}$ | -0.002 | 0.031 | -0.063 | -0.022 | -0.001 | 0.019 | 0.057 | 3618 | 1194 | 1.005 |
| $\varepsilon_{2,3,37}^{(\varphi)}$ | -0.040 | 0.032 | -0.102 | -0.062 | -0.040 | -0.019 | 0.020 | 3030 | 1122 | 0.999 |
| $\varepsilon_{2,3,38}^{(\varphi)}$ | -0.101 | 0.032 | -0.164 | -0.122 | -0.101 | -0.079 | -0.041 | 3526 | 1135 | 0.999 |
| $\varepsilon^{(\varphi)}_{2,3,39}$ | -0.025 | 0.031 | -0.084 | -0.048 | -0.026 | -0.003 | 0.034 | 3388 | 1099 | 1.000 |
| $\varepsilon^{(\varphi)}_{2,3,40}$ | -0.069 | 0.032 | -0.134 | -0.090 | -0.069 | -0.049 | -0.007 | 2080 | 816 | 1.005 |
| $\varepsilon^{(\varphi)}_{2,3,41}$ | 0.032 | 0.031 | -0.026 | 0.011 | 0.031 | 0.053 | 0.092 | 2711 | 1206 | 1.002 |
| $\varepsilon^{(\varphi)}_{2,3,42}$ | -0.185 | 0.033 | -0.254 | -0.207 | -0.186 | -0.163 | -0.123 | 2655 | 980 | 1.005 |
| $\varepsilon^{(\varphi)}_{2,3,43}$ | 0.428 | 0.031 | 0.367 | 0.407 | 0.427 | 0.448 | 0.489 | 2504 | 1238 | 0.999 |
| $\varepsilon^{(\varphi)}_{2,3,44}$ | 0.496 | 0.034 | 0.430 | 0.474 | 0.496 | 0.518 | 0.562 | 3249 | 1060 | 1.001 |
| $\varepsilon^{(\varphi)}_{2,3,45}$ | -0.516 | 0.031 | -0.575 | -0.538 | -0.517 | -0.496 | -0.458 | 2180 | 1054 | 1.003 |
| $\varepsilon^{(\varphi)}_{2,3,46}$ | 0.099 | 0.032 | 0.036 | 0.078 | 0.100 | 0.120 | 0.160 | 2938 | 1139 | 1.003 |
| $\varepsilon^{(\varphi)}_{2,3,47}$ | 0.527 | 0.033 | 0.463 | 0.505 | 0.528 | 0.549 | 0.594 | 2716 | 929 | 1.000 |
| $\varepsilon^{(\varphi)}_{2,3,48}$ | -0.735 | 0.037 | -0.809 | -0.759 | -0.734 | -0.709 | -0.664 | 433 | 658 | 1.003 |
| *Cod fishery and survey selectivity* | | | | | | | | | | |
| $\ln s^{50\%}_{1,2,1,1:11}$ | 1.475 | 0.014 | 1.448 | 1.465 | 1.474 | 1.484 | 1.503 | 1947 | 1173 | 0.999 |
| $\ln s^{50\%}_{1,2,1,12:24}$ | 1.638 | 0.012 | 1.615 | 1.629 | 1.638 | 1.646 | 1.662 | 1705 | 1151 | 1.000 |
| $\ln s^{50\%}_{1,2,1,25:48}$ | 1.868 | 0.014 | 1.839 | 1.859 | 1.868 | 1.878 | 1.897 | 1388 | 952 | 1.001 |
| $\ln s^{50\%}_{2,2,1,1:48}$ | 1.395 | 0.019 | 1.356 | 1.382 | 1.395 | 1.407 | 1.432 | 1680 | 1169 | 1.001 |
| $\ln s^{50\%}_{3,2,1,1:48}$ | 1.324 | 0.029 | 1.268 | 1.304 | 1.325 | 1.345 | 1.380 | 1938 | 1019 | 1.008 |
| $\ln s^{50\%}_{4,2,1,1:48}$ | 1.970 | 0.014 | 1.945 | 1.961 | 1.970 | 1.980 | 1.998 | 1547 | 1018 | 1.001 |
| $\ln(s^{95\%}_{1,2,1,1:11}-s^{50\%}_{1,2,1,1:11})$ | -0.223 | 0.043 | -0.304 | -0.253 | -0.224 | -0.194 | -0.135 | 2111 | 1144 | 1.000 |
| $\ln(s^{95\%}_{1,2,1,12:24}-s^{50\%}_{1,2,1,12:24})$ | -0.057 | 0.032 | -0.120 | -0.079 | -0.057 | -0.036 | 0.007 | 1185 | 1278 | 1.001 |
| $\ln(s^{95\%}_{1,2,1,25:48}-s^{50\%}_{1,2,1,25:48})$ | 0.611 | 0.034 | 0.546 | 0.589 | 0.610 | 0.633 | 0.678 | 922 | 957 | 1.000 |
| $\ln(s^{95\%}_{2,2,1,1:48}-s^{50\%}_{2,2,1,1:48})$ | 0.835 | 0.052 | 0.734 | 0.800 | 0.834 | 0.870 | 0.941 | 1007 | 1036 | 1.002 |
| $\ln(s^{95\%}_{3,2,1,1:48}-s^{50\%}_{3,2,1,1:48})$ | 0.623 | 0.067 | 0.495 | 0.578 | 0.620 | 0.668 | 0.762 | 1888 | 1208 | 1.005 |
| $\ln(s^{95\%}_{4,2,1,1:48}-s^{50\%}_{4,2,1,1:48})$ | 0.976 | 0.038 | 0.901 | 0.951 | 0.976 | 1.001 | 1.048 | 2168 | 997 | 1.001 |
| *Recruitment deviations* | | | | | | | | | | |
| $\varepsilon^{(R)}_{2,1,2}$ | 0.421 | 0.100 | 0.222 | 0.352 | 0.416 | 0.492 | 0.615 | 2431 | 1038 | 1.004 |
| $\varepsilon^{(R)}_{2,1,3}$ | 0.071 | 0.096 | -0.111 | 0.005 | 0.068 | 0.138 | 0.260 | 2019 | 1055 | 1.007 |
| $\varepsilon^{(R)}_{2,1,4}$ | -0.035 | 0.101 | -0.231 | -0.104 | -0.035 | 0.036 | 0.153 | 2898 | 1132 | 1.000 |
| $\varepsilon^{(R)}_{2,1,5}$ | 1.056 | 0.096 | 0.869 | 0.993 | 1.057 | 1.123 | 1.235 | 2388 | 1146 | 1.003 |
| $\varepsilon^{(R)}_{2,1,6}$ | 0.273 | 0.097 | 0.083 | 0.205 | 0.272 | 0.338 | 0.467 | 2119 | 1199 | 1.000 |
| $\varepsilon^{(R)}_{2,1,7}$ | 0.278 | 0.092 | 0.090 | 0.214 | 0.277 | 0.340 | 0.455 | 2011 | 1323 | 0.999 |
| $\varepsilon^{(R)}_{2,1,8}$ | -0.311 | 0.097 | -0.496 | -0.381 | -0.314 | -0.245 | -0.123 | 2510 | 1022 | 1.004 |
| $\varepsilon^{(R)}_{2,1,9}$ | -0.152 | 0.096 | -0.336 | -0.215 | -0.152 | -0.087 | 0.039 | 3021 | 1270 | 1.000 |
| $\varepsilon^{(R)}_{2,1,10}$ | -0.161 | 0.101 | -0.346 | -0.231 | -0.160 | -0.094 | 0.028 | 2538 | 1053 | 1.003 |
| $\varepsilon^{(R)}_{2,1,11}$ | 0.793 | 0.102 | 0.588 | 0.724 | 0.791 | 0.864 | 0.992 | 2241 | 961 | 1.000 |
| $\varepsilon^{(R)}_{2,1,12}$ | 0.121 | 0.102 | -0.078 | 0.057 | 0.120 | 0.186 | 0.322 | 2169 | 1222 | 1.001 |
| $\varepsilon^{(R)}_{2,1,13}$ | -0.697 | 0.099 | -0.894 | -0.762 | -0.696 | -0.633 | -0.500 | 2682 | 1114 | 1.002 |
| $\varepsilon^{(R)}_{2,1,14}$ | -0.009 | 0.101 | -0.208 | -0.078 | -0.009 | 0.062 | 0.184 | 2781 | 1109 | 1.003 |
| $\varepsilon_{2,1,15}^{(R)}$ | -0.045 | 0.099 | -0.242 | -0.111 | -0.041 | 0.020 | 0.151 | 2467 | 1146 | 1.001 |
| $\varepsilon_{2,1,16}^{(R)}$ | -0.172 | 0.100 | -0.368 | -0.238 | -0.172 | -0.103 | 0.026 | 2438 | 1091 | 1.001 |
| $\varepsilon_{2,1,17}^{(R)}$ | -0.132 | 0.089 | -0.305 | -0.193 | -0.134 | -0.070 | 0.047 | 2337 | 1194 | 0.999 |
| $\varepsilon_{2,1,18}^{(R)}$ | -0.015 | 0.084 | -0.180 | -0.070 | -0.015 | 0.042 | 0.143 | 2531 | 1139 | 1.002 |
| $\varepsilon_{2,1,19}^{(R)}$ | 0.089 | 0.087 | -0.078 | 0.029 | 0.090 | 0.146 | 0.260 | 2615 | 1273 | 1.001 |
| $\varepsilon_{2,1,20}^{(R)}$ | -0.163 | 0.081 | -0.319 | -0.218 | -0.164 | -0.110 | 0.002 | 2429 | 1024 | 1.000 |
| $\varepsilon_{2,1,21}^{(R)}$ | -0.366 | 0.079 | -0.522 | -0.420 | -0.364 | -0.313 | -0.208 | 3096 | 1127 | 0.999 |
| $\varepsilon_{2,1,22}^{(R)}$ | -0.094 | 0.075 | -0.233 | -0.149 | -0.095 | -0.041 | 0.046 | 2682 | 1297 | 1.001 |
| $\varepsilon_{2,1,23}^{(R)}$ | -0.072 | 0.078 | -0.230 | -0.127 | -0.071 | -0.020 | 0.084 | 2648 | 1450 | 1.001 |
| $\varepsilon_{2,1,24}^{(R)}$ | 0.058 | 0.076 | -0.090 | 0.007 | 0.057 | 0.109 | 0.209 | 1878 | 1051 | 1.003 |
| $\varepsilon_{2,1,25}^{(R)}$ | -0.370 | 0.077 | -0.522 | -0.422 | -0.369 | -0.318 | -0.219 | 2663 | 1014 | 1.005 |
| $\varepsilon_{2,1,26}^{(R)}$ | 0.451 | 0.072 | 0.313 | 0.403 | 0.449 | 0.500 | 0.595 | 2201 | 983 | 1.000 |
| $\varepsilon_{2,1,27}^{(R)}$ | -0.015 | 0.069 | -0.146 | -0.059 | -0.015 | 0.029 | 0.125 | 2681 | 1190 | 1.007 |
| $\varepsilon_{2,1,28}^{(R)}$ | -0.154 | 0.071 | -0.294 | -0.203 | -0.153 | -0.107 | -0.010 | 2771 | 1219 | 1.002 |
| $\varepsilon_{2,1,29}^{(R)}$ | -0.051 | 0.069 | -0.178 | -0.097 | -0.052 | -0.005 | 0.086 | 2402 | 1215 | 1.006 |
| $\varepsilon_{2,1,30}^{(R)}$ | -0.388 | 0.071 | -0.534 | -0.431 | -0.389 | -0.340 | -0.246 | 2419 | 1194 | 1.003 |
| $\varepsilon_{2,1,31}^{(R)}$ | 0.224 | 0.068 | 0.090 | 0.180 | 0.225 | 0.268 | 0.358 | 2462 | 1223 | 1.000 |
| $\varepsilon_{2,1,32}^{(R)}$ | 0.056 | 0.069 | -0.079 | 0.009 | 0.055 | 0.102 | 0.191 | 2192 | 1033 | 0.999 |
| $\varepsilon^{(R)}_{2,1,33}$ | 0.248 | 0.064 | 0.125 | 0.203 | 0.248 | 0.293 | 0.375 | 2218 | 1318 | 1.000 |
| $\varepsilon^{(R)}_{2,1,34}$ | -0.646 | 0.065 | -0.774 | -0.693 | -0.646 | -0.599 | -0.521 | 2468 | 982 | 1.001 |
| $\varepsilon^{(R)}_{2,1,35}$ | 0.026 | 0.066 | -0.104 | -0.020 | 0.024 | 0.069 | 0.154 | 2828 | 1253 | 1.005 |
| $\varepsilon^{(R)}_{2,1,36}$ | -0.446 | 0.068 | -0.581 | -0.491 | -0.445 | -0.399 | -0.316 | 2235 | 1058 | 0.999 |
| $\varepsilon^{(R)}_{2,1,37}$ | 0.499 | 0.066 | 0.366 | 0.453 | 0.500 | 0.543 | 0.628 | 1738 | 1057 | 1.002 |
| $\varepsilon^{(R)}_{2,1,38}$ | -0.214 | 0.066 | -0.346 | -0.261 | -0.212 | -0.169 | -0.088 | 1804 | 1078 | 1.001 |
| $\varepsilon^{(R)}_{2,1,39}$ | -0.422 | 0.062 | -0.539 | -0.467 | -0.421 | -0.380 | -0.303 | 2537 | 1345 | 1.001 |
| $\varepsilon^{(R)}_{2,1,40}$ | 0.605 | 0.067 | 0.476 | 0.560 | 0.605 | 0.651 | 0.739 | 2204 | 1218 | 1.001 |
| $\varepsilon^{(R)}_{2,1,41}$ | -0.291 | 0.074 | -0.434 | -0.341 | -0.291 | -0.240 | -0.146 | 2263 | 953 | 1.002 |
| $\varepsilon^{(R)}_{2,1,42}$ | 0.033 | 0.080 | -0.126 | -0.020 | 0.034 | 0.090 | 0.180 | 2210 | 1144 | 1.000 |
| $\varepsilon^{(R)}_{2,1,43}$ | 0.074 | 0.084 | -0.086 | 0.018 | 0.075 | 0.129 | 0.237 | 2373 | 1432 | 1.002 |
| $\varepsilon^{(R)}_{2,1,44}$ | -0.391 | 0.098 | -0.582 | -0.458 | -0.388 | -0.326 | -0.202 | 1950 | 1175 | 1.004 |
| $\varepsilon^{(R)}_{2,1,45}$ | -0.476 | 0.114 | -0.704 | -0.552 | -0.477 | -0.398 | -0.254 | 2115 | 1116 | 1.000 |
| $\varepsilon^{(R)}_{2,1,46}$ | -0.585 | 0.135 | -0.849 | -0.671 | -0.587 | -0.492 | -0.320 | 2330 | 1221 | 1.003 |
| $\varepsilon^{(R)}_{2,1,47}$ | 0.771 | 0.154 | 0.469 | 0.671 | 0.767 | 0.876 | 1.074 | 2489 | 1378 | 1.001 |
| $\varepsilon^{(R)}_{2,1,48}$ | -0.194 | 0.216 | -0.619 | -0.341 | -0.196 | -0.045 | 0.211 | 3407 | 1118 | 1.005 |
| *Initial cod abundance* | | | | | | | | | | |
| $\ln N_{2,1,2,2}$ | 4.951 | 0.162 | 4.634 | 4.842 | 4.953 | 5.062 | 5.262 | 783 | 1088 | 1.002 |
| $\ln N_{2,1,2,3}$ | 4.987 | 0.118 | 4.757 | 4.906 | 4.989 | 5.066 | 5.213 | 857 | 1069 | 1.005 |
| $\ln N_{2,1,2,4}$ | 4.413 | 0.089 | 4.243 | 4.352 | 4.413 | 4.474 | 4.584 | 1341 | 956 | 1.000 |
| $\ln N_{2,1,2,5}$ | 3.831 | 0.080 | 3.669 | 3.776 | 3.835 | 3.889 | 3.982 | 2197 | 1062 | 1.005 |
| $\ln N_{2,1,2,6}$ | 3.521 | 0.084 | 3.358 | 3.463 | 3.522 | 3.579 | 3.684 | 2056 | 854 | 1.000 |
| $\ln N_{2,1,2,7}$ | 3.017 | 0.093 | 2.834 | 2.955 | 3.017 | 3.077 | 3.208 | 2434 | 905 | 1.002 |
| $\ln N_{2,1,2,8}$ | 1.642 | 0.109 | 1.433 | 1.569 | 1.643 | 1.718 | 1.852 | 3708 | 1178 | 1.001 |
| $\ln N_{2,1,2,9}$ | 0.785 | 0.133 | 0.523 | 0.696 | 0.784 | 0.875 | 1.048 | 2731 | 1015 | 1.011 |
| $\ln N_{2,1,2,10}$ | 0.746 | 0.150 | 0.437 | 0.653 | 0.748 | 0.846 | 1.030 | 3676 | 1009 | 1.004 |
| $\ln N_{2,1,2,11}$ | -0.046 | 0.207 | -0.451 | -0.185 | -0.050 | 0.099 | 0.350 | 2863 | 1140 | 1.000 |
| $\ln N_{2,1,2,12}$ | 0.334 | 0.310 | -0.253 | 0.122 | 0.335 | 0.541 | 0.929 | 2859 | 1095 | 0.999 |
| *Cod natural mortality* | | | | | | | | | | |
| $\ln M_{2,1,2:4,1:48}$ | -0.592 | 0.087 | -0.771 | -0.648 | -0.586 | -0.532 | -0.433 | 786 | 740 | 1.000 |
| $\varepsilon_2^{(\mathrm{M,c})}$ | 0.048 | 0.051 | -0.049 | 0.013 | 0.047 | 0.081 | 0.147 | 2282 | 1119 | 1.004 |
| $\varepsilon_3^{(\mathrm{M,c})}$ | 0.051 | 0.048 | -0.044 | 0.023 | 0.051 | 0.083 | 0.146 | 3181 | 1054 | 1.000 |
| $\varepsilon_4^{(\mathrm{M,c})}$ | 0.063 | 0.047 | -0.032 | 0.032 | 0.064 | 0.095 | 0.151 | 2157 | 1083 | 1.004 |
| $\varepsilon_5^{(\mathrm{M,c})}$ | 0.076 | 0.046 | -0.012 | 0.044 | 0.076 | 0.108 | 0.166 | 3840 | 925 | 1.000 |
| $\varepsilon_6^{(\mathrm{M,c})}$ | 0.080 | 0.050 | -0.026 | 0.050 | 0.081 | 0.114 | 0.181 | 4631 | 834 | 1.000 |
| $\varepsilon_7^{(\mathrm{M,c})}$ | 0.062 | 0.045 | -0.031 | 0.033 | 0.063 | 0.089 | 0.153 | 2639 | 1088 | 1.003 |
| $\varepsilon_8^{(\mathrm{M,c})}$ | 0.044 | 0.048 | -0.050 | 0.012 | 0.044 | 0.076 | 0.142 | 3199 | 1091 | 1.000 |
| $\varepsilon_9^{(\mathrm{M,c})}$ | 0.034 | 0.048 | -0.062 | 0.003 | 0.034 | 0.065 | 0.126 | 3442 | 1118 | 1.003 |
| $\varepsilon_{10}^{(M,c)}$ | 0.034 | 0.049 | -0.062 | 0.002 | 0.033 | 0.065 | 0.132 | 4085 | 1051 | 0.999 |
| $\varepsilon_{11}^{(M,c)}$ | 0.032 | 0.048 | -0.061 | 0.000 | 0.032 | 0.064 | 0.125 | 4446 | 1277 | 1.000 |
| $\varepsilon_{12}^{(M,c)}$ | 0.043 | 0.045 | -0.041 | 0.012 | 0.044 | 0.074 | 0.130 | 2938 | 1222 | 1.003 |
| $\varepsilon_{13}^{(M,c)}$ | 0.024 | 0.047 | -0.069 | -0.007 | 0.023 | 0.054 | 0.120 | 2591 | 1079 | 1.001 |
| $\varepsilon_{14}^{(M,c)}$ | 0.007 | 0.048 | -0.092 | -0.026 | 0.006 | 0.039 | 0.101 | 3149 | 1102 | 1.005 |
| $\varepsilon_{15}^{(M,c)}$ | 0.013 | 0.049 | -0.082 | -0.020 | 0.013 | 0.046 | 0.112 | 3813 | 1049 | 1.003 |
| $\varepsilon_{16}^{(M,c)}$ | 0.029 | 0.048 | -0.065 | -0.003 | 0.030 | 0.062 | 0.125 | 4152 | 1182 | 1.001 |
| $\varepsilon_{17}^{(M,c)}$ | 0.045 | 0.048 | -0.049 | 0.012 | 0.044 | 0.078 | 0.139 | 4339 | 1142 | 1.002 |
| $\varepsilon_{18}^{(M,c)}$ | 0.037 | 0.048 | -0.056 | 0.006 | 0.035 | 0.066 | 0.134 | 3825 | 1003 | 1.002 |
| $\varepsilon_{19}^{(M,c)}$ | 0.036 | 0.048 | -0.061 | 0.003 | 0.035 | 0.068 | 0.129 | 3424 | 1104 | 1.003 |
| $\varepsilon_{20}^{(M,c)}$ | 0.026 | 0.048 | -0.064 | -0.006 | 0.026 | 0.057 | 0.121 | 2937 | 984 | 1.004 |
| $\varepsilon_{21}^{(M,c)}$ | 0.004 | 0.046 | -0.087 | -0.027 | 0.003 | 0.035 | 0.095 | 3272 | 1126 | 1.001 |
| $\varepsilon_{22}^{(M,c)}$ | -0.011 | 0.047 | -0.106 | -0.041 | -0.011 | 0.019 | 0.082 | 3642 | 1105 | 1.003 |
| $\varepsilon_{23}^{(M,c)}$ | -0.042 | 0.050 | -0.139 | -0.073 | -0.042 | -0.009 | 0.057 | 4059 | 890 | 1.002 |
| $\varepsilon_{24}^{(M,c)}$ | -0.029 | 0.050 | -0.128 | -0.060 | -0.030 | 0.004 | 0.065 | 4605 | 1067 | 1.003 |
| $\varepsilon_{25}^{(M,c)}$ | -0.024 | 0.046 | -0.114 | -0.055 | -0.026 | 0.007 | 0.066 | 2536 | 1035 | 1.000 |
| $\varepsilon_{26}^{(M,c)}$ | -0.012 | 0.047 | -0.108 | -0.044 | -0.012 | 0.021 | 0.081 | 4094 | 1010 | 1.004 |
| $\varepsilon_{27}^{(M,c)}$ | -0.005 | 0.047 | -0.100 | -0.037 | -0.005 | 0.027 | 0.088 | 4218 | 1130 | 1.002 |
| $\varepsilon_{28}^{(M,c)}$ | 0.000 | 0.051 | -0.104 | -0.035 | 0.001 | 0.035 | 0.096 | 3953 | 1057 | 0.999 |
| $\varepsilon_{29}^{(M,c)}$ | -0.004 | 0.047 | -0.100 | -0.036 | -0.005 | 0.029 | 0.088 | 3427 | 978 | 1.001 |
| $\varepsilon_{30}^{(M,c)}$ | 0.022 | 0.045 | -0.068 | -0.008 | 0.021 | 0.053 | 0.113 | 3712 | 979 | 1.003 |
| $\varepsilon_{31}^{(M,c)}$ | 0.033 | 0.047 | -0.059 | 0.000 | 0.033 | 0.064 | 0.122 | 3445 | 1065 | 1.003 |
| $\varepsilon_{32}^{(M,c)}$ | 0.026 | 0.045 | -0.061 | -0.004 | 0.027 | 0.057 | 0.115 | 3316 | 1311 | 1.000 |
| $\varepsilon_{33}^{(M,c)}$ | 0.008 | 0.045 | -0.079 | -0.022 | 0.008 | 0.038 | 0.096 | 3658 | 1197 | 1.001 |
| $\varepsilon_{34}^{(M,c)}$ | -0.018 | 0.048 | -0.111 | -0.050 | -0.017 | 0.014 | 0.077 | 3900 | 832 | 1.004 |
| $\varepsilon_{35}^{(M,c)}$ | -0.019 | 0.046 | -0.112 | -0.050 | -0.017 | 0.012 | 0.071 | 3907 | 982 | 1.001 |
| $\varepsilon_{36}^{(M,c)}$ | -0.013 | 0.046 | -0.099 | -0.045 | -0.013 | 0.017 | 0.077 | 3276 | 970 | 1.005 |
| $\varepsilon_{37}^{(M,c)}$ | -0.018 | 0.050 | -0.112 | -0.053 | -0.019 | 0.017 | 0.078 | 3260 | 946 | 1.000 |
| $\varepsilon_{38}^{(M,c)}$ | -0.060 | 0.047 | -0.152 | -0.092 | -0.060 | -0.027 | 0.032 | 3875 | 1226 | 1.002 |
| $\varepsilon_{39}^{(M,c)}$ | -0.048 | 0.048 | -0.140 | -0.081 | -0.048 | -0.016 | 0.047 | 3847 | 1221 | 1.005 |
| $\varepsilon_{40}^{(M,c)}$ | -0.062 | 0.048 | -0.151 | -0.095 | -0.063 | -0.030 | 0.033 | 3844 | 1152 | 1.004 |
| $\varepsilon_{41}^{(M,c)}$ | -0.038 | 0.048 | -0.128 | -0.071 | -0.038 | -0.004 | 0.056 | 3161 | 1307 | 0.999 |
| $\varepsilon_{42}^{(M,c)}$ | -0.026 | 0.049 | -0.116 | -0.060 | -0.026 | 0.008 | 0.069 | 3477 | 1162 | 1.003 |
| $\varepsilon_{43}^{(M,c)}$ | -0.026 | 0.048 | -0.119 | -0.057 | -0.025 | 0.006 | 0.070 | 4214 | 1031 | 1.007 |
| $\varepsilon_{44}^{(M,c)}$ | -0.020 | 0.049 | -0.112 | -0.055 | -0.022 | 0.013 | 0.071 | 2999 | 815 | 1.006 |
| $\varepsilon_{45}^{(M,c)}$ | -0.004 | 0.047 | -0.095 | -0.037 | -0.003 | 0.029 | 0.088 | 2974 | 1005 | 1.000 |
| $\varepsilon_{46}^{(\mathrm{M,c})}$ | 0.004 | 0.048 | -0.094 | -0.027 | 0.002 | 0.036 | 0.100 | 3742 | 1167 | 1.009 |
| $\varepsilon_{47}^{(\mathrm{M,c})}$ | -0.012 | 0.049 | -0.105 | -0.045 | -0.013 | 0.020 | 0.090 | 3711 | 1035 | 1.001 |
| $\varepsilon_{48}^{(\mathrm{M,c})}$ | -0.010 | 0.050 | -0.106 | -0.045 | -0.009 | 0.024 | 0.082 | 4764 | 1282 | 1.000 |
| *Predator selectivity of herring* | | | | | | | | | | |
| $b^{50\%}$ | 7.684 | 0.169 | 7.334 | 7.577 | 7.690 | 7.799 | 8.002 | 314 | 573 | 1.002 |
| $\ln(b^{95\%}$-$b^{50\%})$ | 1.647 | 0.015 | 1.619 | 1.636 | 1.647 | 1.658 | 1.679 | 512 | 760 | 1.001 |
| $\ln k$ | 1.501 | 0.015 | 1.470 | 1.491 | 1.501 | 1.511 | 1.530 | 727 | 979 | 1.002 |
| $\ln\theta$ | 0.137 | 0.022 | 0.092 | 0.122 | 0.139 | 0.151 | 0.179 | 349 | 661 | 1.001 |
| *Herring recruitment* | | | | | | | | | | |
| $\ln\bar{R}_1$ | 6.238 | 0.025 | 6.192 | 6.219 | 6.237 | 6.255 | 6.289 | 895 | 972 | 1.001 |
| $\ln\bar{R}_2$ | 6.377 | 0.042 | 6.293 | 6.348 | 6.377 | 6.405 | 6.458 | 1189 | 945 | 1.003 |
| $\ln\bar{R}_3$ | 4.983 | 0.054 | 4.879 | 4.944 | 4.983 | 5.021 | 5.093 | 1530 | 1371 | 1.003 |
| $\ln\bar{R}_4$ | 6.056 | 0.056 | 5.945 | 6.018 | 6.056 | 6.095 | 6.162 | 1222 | 1115 | 1.000 |
| $\varepsilon_{3,1,9}^{(\mathrm{R})}$ | -0.874 | 0.248 | -1.398 | -1.026 | -0.867 | -0.708 | -0.379 | 1207 | 1226 | 1.000 |
| $\varepsilon_{3,2,9}^{(\mathrm{R})}$ | 0.519 | 0.349 | -0.194 | 0.301 | 0.523 | 0.743 | 1.204 | 1222 | 1102 | 1.001 |
| $\varepsilon_{3,3,9}^{(\mathrm{R})}$ | -0.040 | 0.261 | -0.524 | -0.212 | -0.046 | 0.124 | 0.490 | 1272 | 1062 | 1.000 |
| $\varepsilon_{3,4,9}^{(\mathrm{R})}$ | 0.090 | 0.221 | -0.345 | -0.061 | 0.085 | 0.230 | 0.543 | 1753 | 1133 | 1.005 |
| $\varepsilon_{3,1,10}^{(\mathrm{R})}$ | 0.577 | 0.178 | 0.207 | 0.460 | 0.580 | 0.698 | 0.924 | 1486 | 1094 | 1.000 |
| $\varepsilon_{3,2,10}^{(\mathrm{R})}$ | 0.221 | 0.149 | -0.079 | 0.122 | 0.224 | 0.325 | 0.501 | 1566 | 1277 | 1.003 |
| $\varepsilon_{3,3,10}^{(\mathrm{R})}$ | -0.538 | 0.169 | -0.877 | -0.650 | -0.536 | -0.422 | -0.211 | 1995 | 1192 | 1.000 |

| | | | | | | | | | |
|---|---|---|---|---|---|---|---|---|---|
| $\varepsilon^{(R)}_{3,4,10}$ | -0.419 | 0.205 | -0.814 | -0.554 | -0.417 | -0.283 | -0.023 | 1831 | 1300 | 1.000 |
| $\varepsilon^{(R)}_{3,1,11}$ | 0.413 | 0.217 | -0.022 | 0.275 | 0.414 | 0.555 | 0.868 | 1960 | 1139 | 1.001 |
| $\varepsilon^{(R)}_{3,2,11}$ | 0.651 | 0.192 | 0.273 | 0.521 | 0.652 | 0.782 | 1.033 | 1527 | 1063 | 1.001 |
| $\varepsilon^{(R)}_{3,3,11}$ | -0.532 | 0.148 | -0.824 | -0.632 | -0.541 | -0.430 | -0.244 | 1322 | 1056 | 1.000 |
| $\varepsilon^{(R)}_{3,4,11}$ | 0.484 | 0.120 | 0.267 | 0.401 | 0.477 | 0.563 | 0.719 | 1441 | 1168 | 1.001 |
| $\varepsilon^{(R)}_{3,1,12}$ | 0.322 | 0.099 | 0.128 | 0.256 | 0.322 | 0.387 | 0.526 | 1522 | 1352 | 1.002 |
| $\varepsilon^{(R)}_{3,2,12}$ | -0.563 | 0.136 | -0.827 | -0.651 | -0.561 | -0.471 | -0.291 | 1485 | 1244 | 1.001 |
| $\varepsilon^{(R)}_{3,3,12}$ | 1.071 | 0.117 | 0.843 | 0.990 | 1.069 | 1.149 | 1.310 | 1472 | 944 | 1.000 |
| $\varepsilon^{(R)}_{3,4,12}$ | -1.535 | 0.157 | -1.850 | -1.638 | -1.534 | -1.427 | -1.230 | 1939 | 1143 | 1.000 |
| $\varepsilon^{(R)}_{3,1,13}$ | 0.605 | 0.179 | 0.266 | 0.487 | 0.599 | 0.729 | 0.950 | 1835 | 971 | 0.999 |
| $\varepsilon^{(R)}_{3,2,13}$ | -0.342 | 0.158 | -0.653 | -0.452 | -0.343 | -0.233 | -0.031 | 1829 | 1095 | 1.000 |
| $\varepsilon^{(R)}_{3,3,13}$ | -0.036 | 0.164 | -0.365 | -0.150 | -0.030 | 0.078 | 0.276 | 1926 | 1157 | 1.001 |
| $\varepsilon^{(R)}_{3,4,13}$ | -0.235 | 0.157 | -0.549 | -0.338 | -0.234 | -0.123 | 0.070 | 2180 | 1317 | 1.000 |
| $\varepsilon^{(R)}_{3,1,14}$ | -0.003 | 0.146 | -0.281 | -0.105 | -0.006 | 0.103 | 0.290 | 2162 | 1220 | 1.001 |
| $\varepsilon^{(R)}_{3,2,14}$ | -0.303 | 0.160 | -0.602 | -0.414 | -0.301 | -0.197 | 0.019 | 1702 | 1016 | 1.001 |
| $\varepsilon^{(R)}_{3,3,14}$ | 0.441 | 0.159 | 0.128 | 0.333 | 0.441 | 0.545 | 0.741 | 1681 | 1049 | 1.001 |
| $\varepsilon^{(R)}_{3,4,14}$ | -1.399 | 0.149 | -1.703 | -1.495 | -1.397 | -1.297 | -1.122 | 1424 | 1365 | 1.001 |
| $\varepsilon^{(R)}_{3,1,15}$ | 1.089 | 0.148 | 0.773 | 0.996 | 1.087 | 1.186 | 1.386 | 1601 | 1192 | 1.001 |
| $\varepsilon^{(R)}_{3,2,15}$ | -0.243 | 0.150 | -0.530 | -0.348 | -0.245 | -0.140 | 0.056 | 1800 | 1216 | 1.000 |
| $\varepsilon^{(R)}_{3,3,15}$ | -0.329 | 0.162 | -0.639 | -0.437 | -0.330 | -0.218 | -0.015 | 1837 | 1072 | 1.002 |
| $\varepsilon^{(R)}_{3,4,15}$ | 0.160 | 0.147 | -0.128 | 0.059 | 0.162 | 0.260 | 0.431 | 1837 | 1191 | 1.002 |
| $\varepsilon^{(R)}_{3,1,16}$ | 0.084 | 0.141 | -0.189 | -0.006 | 0.080 | 0.175 | 0.363 | 1890 | 1143 | 1.001 |
| $\varepsilon^{(R)}_{3,2,16}$ | 0.471 | 0.144 | 0.183 | 0.373 | 0.474 | 0.574 | 0.741 | 1907 | 1255 | 1.000 |
| $\varepsilon^{(R)}_{3,3,16}$ | -0.464 | 0.148 | -0.760 | -0.562 | -0.465 | -0.366 | -0.173 | 2093 | 1246 | 1.002 |
| $\varepsilon^{(R)}_{3,4,16}$ | 0.219 | 0.154 | -0.084 | 0.114 | 0.227 | 0.317 | 0.516 | 1762 | 1156 | 1.003 |
| $\varepsilon^{(R)}_{3,1,17}$ | -0.121 | 0.151 | -0.400 | -0.229 | -0.114 | -0.012 | 0.169 | 1977 | 1140 | 0.999 |
| $\varepsilon^{(R)}_{3,2,17}$ | -0.264 | 0.166 | -0.575 | -0.379 | -0.263 | -0.153 | 0.060 | 1929 | 1225 | 1.000 |
| $\varepsilon^{(R)}_{3,3,17}$ | 0.981 | 0.160 | 0.676 | 0.872 | 0.980 | 1.092 | 1.287 | 2056 | 1312 | 1.000 |
| $\varepsilon^{(R)}_{3,4,17}$ | -1.125 | 0.179 | -1.463 | -1.254 | -1.124 | -1.000 | -0.767 | 1827 | 1345 | 1.000 |
| $\varepsilon^{(R)}_{3,1,18}$ | 1.174 | 0.201 | 0.772 | 1.046 | 1.173 | 1.305 | 1.556 | 2110 | 1077 | 1.000 |
| $\varepsilon^{(R)}_{3,2,18}$ | -0.907 | 0.270 | -1.427 | -1.086 | -0.905 | -0.715 | -0.404 | 2657 | 1162 | 1.000 |
| $\varepsilon^{(R)}_{3,3,18}$ | -0.436 | 0.522 | -1.469 | -0.779 | -0.422 | -0.087 | 0.560 | 2467 | 1379 | 1.000 |
| $\varepsilon^{(R)}_{3,4,18}$ | -1.177 | 0.746 | -2.756 | -1.677 | -1.130 | -0.657 | 0.164 | 2559 | 912 | 1.004 |
| $\varepsilon^{(R)}_{3,1,19}$ | -0.161 | 0.122 | -0.419 | -0.240 | -0.155 | -0.073 | 0.056 | 1171 | 1096 | 1.000 |
| $\varepsilon^{(R)}_{3,2,19}$ | -0.251 | 0.173 | -0.591 | -0.363 | -0.246 | -0.133 | 0.076 | 1445 | 1316 | 1.002 |
| $\varepsilon^{(R)}_{3,3,19}$ | -0.098 | 0.186 | -0.451 | -0.219 | -0.104 | 0.031 | 0.254 | 1649 | 1262 | 1.002 |
| $\varepsilon^{(R)}_{3,4,19}$ | 0.217 | 0.173 | -0.108 | 0.099 | 0.216 | 0.335 | 0.569 | 1918 | 1330 | 1.006 |
| $\varepsilon^{(R)}_{3,1,20}$ | 0.142 | 0.183 | -0.214 | 0.018 | 0.137 | 0.259 | 0.511 | 1814 | 1142 | 1.002 |
| $\varepsilon^{(R)}_{3,2,20}$ | 0.000 | 0.194 | -0.386 | -0.130 | -0.002 | 0.138 | 0.364 | 1635 | 1125 | 1.000 |
| $\varepsilon^{(R)}_{3,3,20}$ | 0.282 | 0.183 | -0.061 | 0.156 | 0.281 | 0.404 | 0.644 | 1774 | 978 | 1.001 |
| $\varepsilon^{(R)}_{3,4,20}$ | 0.218 | 0.166 | -0.094 | 0.100 | 0.218 | 0.332 | 0.544 | 1260 | 1140 | 1.002 |
| $\varepsilon^{(R)}_{3,1,21}$ | 0.166 | 0.156 | -0.134 | 0.053 | 0.166 | 0.273 | 0.472 | 1722 | 1296 | 1.001 |
| $\varepsilon^{(R)}_{3,2,21}$ | -0.289 | 0.164 | -0.612 | -0.402 | -0.288 | -0.176 | 0.026 | 1775 | 1249 | 1.003 |
| $\varepsilon^{(R)}_{3,3,21}$ | 0.113 | 0.162 | -0.202 | 0.004 | 0.112 | 0.222 | 0.432 | 1659 | 1145 | 1.000 |
| $\varepsilon^{(R)}_{3,4,21}$ | -0.148 | 0.130 | -0.407 | -0.233 | -0.143 | -0.056 | 0.104 | 1987 | 1286 | 1.005 |
| $\varepsilon^{(R)}_{3,1,22}$ | 0.010 | 0.158 | -0.315 | -0.096 | 0.010 | 0.112 | 0.307 | 1586 | 1215 | 1.002 |
| $\varepsilon^{(R)}_{3,2,22}$ | 0.064 | 0.181 | -0.301 | -0.056 | 0.068 | 0.185 | 0.411 | 1564 | 1208 | 1.002 |
| $\varepsilon^{(R)}_{3,3,22}$ | -0.407 | 0.195 | -0.778 | -0.535 | -0.407 | -0.277 | -0.022 | 1450 | 1342 | 1.003 |
| $\varepsilon^{(R)}_{3,4,22}$ | 0.457 | 0.201 | 0.049 | 0.330 | 0.456 | 0.591 | 0.831 | 1842 | 1139 | 1.003 |
| $\varepsilon^{(R)}_{3,1,23}$ | -0.189 | 0.176 | -0.529 | -0.303 | -0.190 | -0.067 | 0.149 | 1882 | 1358 | 1.002 |
| $\varepsilon^{(R)}_{3,2,23}$ | -0.047 | 0.173 | -0.382 | -0.163 | -0.051 | 0.071 | 0.285 | 1514 | 1302 | 1.000 |
| $\varepsilon^{(R)}_{3,3,23}$ | 0.520 | 0.166 | 0.203 | 0.411 | 0.521 | 0.632 | 0.839 | 1679 | 1283 | 1.006 |
| $\varepsilon^{(R)}_{3,4,23}$ | -0.122 | 0.165 | -0.454 | -0.235 | -0.116 | -0.009 | 0.180 | 1630 | 1295 | 1.001 |
| $\varepsilon^{(R)}_{3,1,24}$ | 0.016 | 0.170 | -0.308 | -0.096 | 0.017 | 0.133 | 0.345 | 1751 | 1160 | 1.004 |
| $\varepsilon^{(R)}_{3,2,24}$ | -0.229 | 0.178 | -0.573 | -0.348 | -0.234 | -0.105 | 0.127 | 2056 | 1263 | 1.003 |
| $\varepsilon^{(R)}_{3,3,24}$ | -0.343 | 0.195 | -0.724 | -0.473 | -0.341 | -0.214 | 0.058 | 1607 | 1208 | 1.000 |
| $\varepsilon^{(R)}_{3,4,24}$ | 0.520 | 0.180 | 0.170 | 0.403 | 0.519 | 0.639 | 0.866 | 1760 | 1379 | 1.001 |
| $\varepsilon^{(R)}_{3,1,25}$ | 0.140 | 0.143 | -0.152 | 0.050 | 0.142 | 0.239 | 0.406 | 1703 | 1187 | 1.004 |
| $\varepsilon^{(R)}_{3,2,25}$ | 0.040 | 0.141 | -0.230 | -0.052 | 0.037 | 0.135 | 0.331 | 1578 | 1151 | 1.000 |
| $\varepsilon^{(R)}_{3,3,25}$ | -0.161 | 0.160 | -0.483 | -0.264 | -0.156 | -0.054 | 0.145 | 1637 | 1250 | 0.999 |
| $\varepsilon^{(R)}_{3,4,25}$ | 0.051 | 0.169 | -0.277 | -0.063 | 0.050 | 0.164 | 0.388 | 1655 | 1305 | 1.000 |
| $\varepsilon^{(R)}_{3,1,26}$ | 0.256 | 0.153 | -0.040 | 0.152 | 0.257 | 0.363 | 0.561 | 1712 | 1079 | 1.000 |
| $\varepsilon^{(R)}_{3,2,26}$ | -0.149 | 0.151 | -0.443 | -0.250 | -0.148 | -0.046 | 0.151 | 1409 | 960 | 1.003 |
| $\varepsilon^{(R)}_{3,3,26}$ | 0.118 | 0.162 | -0.188 | 0.005 | 0.115 | 0.230 | 0.426 | 1372 | 1050 | 1.001 |
| $\varepsilon^{(R)}_{3,4,26}$ | 0.315 | 0.153 | 0.023 | 0.207 | 0.315 | 0.419 | 0.634 | 1221 | 1148 | 1.002 |
| $\varepsilon^{(R)}_{3,1,27}$ | -0.235 | 0.149 | -0.534 | -0.337 | -0.232 | -0.140 | 0.051 | 1248 | 1059 | 1.003 |
| $\varepsilon^{(R)}_{3,2,27}$ | -0.306 | 0.188 | -0.680 | -0.432 | -0.304 | -0.178 | 0.061 | 1470 | 1142 | 1.001 |
| $\varepsilon^{(R)}_{3,3,27}$ | 0.163 | 0.216 | -0.236 | 0.011 | 0.154 | 0.312 | 0.592 | 1233 | 1144 | 1.001 |
| $\varepsilon^{(R)}_{3,4,27}$ | 0.352 | 0.205 | -0.052 | 0.220 | 0.349 | 0.486 | 0.753 | 1230 | 1002 | 1.000 |
| $\varepsilon^{(R)}_{3,1,28}$ | -1.336 | 0.278 | -1.875 | -1.524 | -1.339 | -1.150 | -0.777 | 1705 | 1173 | 1.000 |
| $\varepsilon^{(R)}_{3,2,28}$ | -1.153 | 0.467 | -2.086 | -1.470 | -1.152 | -0.849 | -0.218 | 2470 | 1213 | 1.002 |
| $\varepsilon^{(R)}_{3,3,28}$ | -3.116 | 0.736 | -4.549 | -3.620 | -3.106 | -2.616 | -1.676 | 3066 | 1014 | 1.000 |
| $\varepsilon^{(R)}_{3,4,28}$ | -0.060 | 0.980 | -2.001 | -0.713 | -0.054 | 0.607 | 1.838 | 3867 | 1206 | 1.002 |
| $\varepsilon^{(R)}_{3,1,29}$ | -0.207 | 0.186 | -0.568 | -0.326 | -0.206 | -0.081 | 0.151 | 1925 | 1397 | 1.003 |
| $\varepsilon^{(R)}_{3,2,29}$ | -0.464 | 0.259 | -0.965 | -0.638 | -0.459 | -0.285 | 0.031 | 2052 | 1079 | 1.000 |
| $\varepsilon^{(R)}_{3,3,29}$ | 0.442 | 0.257 | -0.051 | 0.264 | 0.443 | 0.613 | 0.928 | 2185 | 1318 | 1.000 |
| $\varepsilon^{(R)}_{3,4,29}$ | 0.444 | 0.226 | 0.020 | 0.283 | 0.441 | 0.606 | 0.894 | 2100 | 1107 | 1.009 |
| $\varepsilon^{(R)}_{3,1,30}$ | -0.498 | 0.230 | -0.950 | -0.654 | -0.495 | -0.344 | -0.041 | 2169 | 1072 | 1.000 |
| $\varepsilon^{(R)}_{3,2,30}$ | 0.006 | 0.214 | -0.394 | -0.146 | 0.006 | 0.151 | 0.424 | 2174 | 974 | 1.001 |
| $\varepsilon^{(R)}_{3,3,30}$ | 0.058 | 0.178 | -0.285 | -0.059 | 0.064 | 0.176 | 0.411 | 2798 | 1210 | 1.001 |
| $\varepsilon^{(R)}_{3,4,30}$ | -0.073 | 0.181 | -0.424 | -0.197 | -0.073 | 0.048 | 0.286 | 2835 | 1203 | 1.004 |
| $\varepsilon^{(R)}_{3,1,31}$ | -0.170 | 0.197 | -0.538 | -0.303 | -0.167 | -0.045 | 0.235 | 3090 | 1230 | 1.003 |
| $\varepsilon^{(R)}_{3,2,31}$ | 0.613 | 0.195 | 0.256 | 0.483 | 0.606 | 0.743 | 1.002 | 1883 | 933 | 1.003 |
| $\varepsilon^{(R)}_{3,3,31}$ | 0.796 | 0.178 | 0.447 | 0.675 | 0.797 | 0.918 | 1.142 | 1481 | 1010 | 1.002 |
| $\varepsilon^{(R)}_{3,4,31}$ | -0.318 | 0.157 | -0.629 | -0.422 | -0.317 | -0.211 | -0.018 | 2134 | 1041 | 1.007 |
| $\varepsilon^{(R)}_{3,1,32}$ | -0.981 | 0.200 | -1.371 | -1.117 | -0.982 | -0.848 | -0.596 | 2469 | 1111 | 1.005 |
| $\varepsilon^{(R)}_{3,2,32}$ | 1.633 | 0.204 | 1.256 | 1.488 | 1.632 | 1.770 | 2.036 | 2268 | 1058 | 1.001 |
| $\varepsilon^{(R)}_{3,3,32}$ | -1.506 | 0.221 | -1.942 | -1.659 | -1.504 | -1.355 | -1.072 | 2193 | 1186 | 1.001 |
| $\varepsilon^{(R)}_{3,4,32}$ | 1.094 | 0.235 | 0.631 | 0.947 | 1.089 | 1.243 | 1.570 | 1990 | 1156 | 1.001 |
| $\varepsilon^{(R)}_{3,1,33}$ | -0.603 | 0.198 | -0.968 | -0.734 | -0.604 | -0.472 | -0.204 | 2125 | 996 | 1.004 |
| $\varepsilon^{(R)}_{3,2,33}$ | 0.809 | 0.189 | 0.434 | 0.690 | 0.808 | 0.933 | 1.175 | 2011 | 1051 | 1.003 |
| $\varepsilon^{(R)}_{3,3,33}$ | 0.481 | 0.176 | 0.131 | 0.365 | 0.482 | 0.597 | 0.820 | 1697 | 1144 | 1.002 |
| $\varepsilon^{(R)}_{3,4,33}$ | -0.301 | 0.169 | -0.637 | -0.417 | -0.300 | -0.186 | 0.022 | 1692 | 1347 | 1.002 |
| $\varepsilon^{(R)}_{3,1,34}$ | -0.288 | 0.173 | -0.618 | -0.396 | -0.289 | -0.175 | 0.066 | 1774 | 1266 | 1.003 |
| $\varepsilon^{(R)}_{3,2,34}$ | 0.289 | 0.183 | -0.069 | 0.169 | 0.289 | 0.406 | 0.665 | 1339 | 1296 | 1.002 |
| $\varepsilon^{(R)}_{3,3,34}$ | 0.330 | 0.193 | -0.048 | 0.210 | 0.325 | 0.463 | 0.715 | 1152 | 1064 | 1.003 |
| $\varepsilon^{(R)}_{3,4,34}$ | 0.301 | 0.183 | -0.068 | 0.178 | 0.296 | 0.427 | 0.661 | 1081 | 1166 | 1.003 |
| $\varepsilon^{(R)}_{3,1,35}$ | -0.243 | 0.177 | -0.580 | -0.369 | -0.243 | -0.119 | 0.103 | 1334 | 1168 | 1.001 |
| $\varepsilon^{(R)}_{3,2,35}$ | -0.221 | 0.183 | -0.577 | -0.346 | -0.219 | -0.096 | 0.130 | 1617 | 1111 | 1.003 |
| $\varepsilon^{(R)}_{3,3,35}$ | -0.288 | 0.190 | -0.642 | -0.424 | -0.290 | -0.153 | 0.085 | 1795 | 1333 | 1.002 |
| $\varepsilon^{(R)}_{3,4,35}$ | 0.659 | 0.176 | 0.305 | 0.542 | 0.663 | 0.776 | 1.014 | 1400 | 1222 | 1.003 |
| $\varepsilon^{(R)}_{3,1,36}$ | 0.173 | 0.145 | -0.115 | 0.076 | 0.175 | 0.270 | 0.449 | 1382 | 1096 | 1.000 |
| $\varepsilon^{(R)}_{3,2,36}$ | -0.211 | 0.150 | -0.496 | -0.315 | -0.211 | -0.108 | 0.087 | 1194 | 1196 | 0.999 |
| $\varepsilon^{(R)}_{3,3,36}$ | -0.090 | 0.161 | -0.406 | -0.196 | -0.089 | 0.021 | 0.222 | 1242 | 1197 | 1.002 |
| $\varepsilon^{(R)}_{3,4,36}$ | -0.437 | 0.188 | -0.800 | -0.563 | -0.443 | -0.307 | -0.084 | 1241 | 1197 | 1.000 |
| $\varepsilon^{(R)}_{3,1,37}$ | 0.612 | 0.188 | 0.241 | 0.486 | 0.609 | 0.743 | 0.980 | 1266 | 981 | 1.003 |
| $\varepsilon^{(R)}_{3,2,37}$ | 0.185 | 0.173 | -0.152 | 0.066 | 0.190 | 0.298 | 0.532 | 1354 | 1223 | 1.001 |
| $\varepsilon^{(R)}_{3,3,37}$ | -0.624 | 0.215 | -1.043 | -0.763 | -0.629 | -0.474 | -0.198 | 1194 | 1140 | 1.000 |
| $\varepsilon^{(R)}_{3,4,37}$ | 0.060 | 0.305 | -0.536 | -0.144 | 0.062 | 0.268 | 0.659 | 1628 | 1285 | 1.000 |
| $\varepsilon^{(R)}_{3,1,38}$ | -0.370 | 0.403 | -1.102 | -0.651 | -0.388 | -0.099 | 0.444 | 324 | 543 | 1.006 |
| $\varepsilon^{(R)}_{3,2,38}$ | 0.046 | 0.706 | -1.086 | -0.472 | -0.074 | 0.514 | 1.475 | 115 | 277 | 1.019 |
| $\varepsilon^{(R)}_{3,3,38}$ | -0.458 | 0.771 | -2.041 | -0.949 | -0.402 | 0.083 | 0.942 | 145 | 340 | 1.015 |
| $\varepsilon^{(R)}_{3,4,38}$ | 1.472 | 0.604 | 0.248 | 1.078 | 1.482 | 1.884 | 2.665 | 363 | 549 | 1.007 |
| $\varepsilon^{(R)}_{3,1,39}$ | -0.321 | 0.182 | -0.706 | -0.433 | -0.311 | -0.199 | 0.025 | 1112 | 1011 | 1.001 |
| $\varepsilon^{(R)}_{3,2,39}$ | 0.112 | 0.224 | -0.306 | -0.046 | 0.105 | 0.266 | 0.550 | 1284 | 1225 | 1.000 |
| $\varepsilon^{(R)}_{3,3,39}$ | 0.144 | 0.201 | -0.248 | 0.009 | 0.137 | 0.277 | 0.547 | 1451 | 1222 | 1.000 |
| $\varepsilon^{(R)}_{3,4,39}$ | 0.136 | 0.174 | -0.192 | 0.018 | 0.138 | 0.255 | 0.482 | 1638 | 1377 | 1.000 |
| $\varepsilon^{(R)}_{3,1,40}$ | 0.089 | 0.179 | -0.264 | -0.032 | 0.091 | 0.220 | 0.423 | 1370 | 978 | 1.001 |
| $\varepsilon^{(R)}_{3,2,40}$ | 0.276 | 0.169 | -0.028 | 0.156 | 0.271 | 0.385 | 0.620 | 1300 | 1153 | 1.001 |
| $\varepsilon^{(R)}_{3,3,40}$ | 0.311 | 0.132 | 0.050 | 0.223 | 0.312 | 0.403 | 0.561 | 1435 | 1302 | 1.003 |
| $\varepsilon^{(R)}_{3,4,40}$ | -0.058 | 0.109 | -0.281 | -0.129 | -0.058 | 0.017 | 0.151 | 1175 | 978 | 1.006 |
| $\varepsilon^{(R)}_{3,1,41}$ | -0.339 | 0.128 | -0.603 | -0.424 | -0.334 | -0.252 | -0.096 | 1415 | 986 | 1.003 |
| $\varepsilon^{(R)}_{3,2,41}$ | -0.558 | 0.179 | -0.927 | -0.671 | -0.551 | -0.435 | -0.218 | 1360 | 1014 | 1.002 |
| $\varepsilon^{(R)}_{3,3,41}$ | 0.328 | 0.185 | -0.016 | 0.202 | 0.316 | 0.454 | 0.693 | 1369 | 1029 | 1.004 |
| $\varepsilon^{(R)}_{3,4,41}$ | -0.023 | 0.153 | -0.320 | -0.129 | -0.023 | 0.077 | 0.281 | 1607 | 1092 | 1.001 |
| $\varepsilon^{(R)}_{3,1,42}$ | -0.149 | 0.228 | -0.599 | -0.297 | -0.147 | 0.004 | 0.292 | 1143 | 1159 | 1.000 |
| $\varepsilon^{(R)}_{3,2,42}$ | 0.893 | 0.225 | 0.466 | 0.738 | 0.885 | 1.045 | 1.327 | 1172 | 1198 | 0.999 |
| $\varepsilon^{(R)}_{3,3,42}$ | -1.122 | 0.236 | -1.611 | -1.280 | -1.121 | -0.963 | -0.681 | 1610 | 1173 | 1.005 |
| $\varepsilon^{(R)}_{3,4,42}$ | 1.458 | 0.220 | 1.038 | 1.310 | 1.454 | 1.602 | 1.915 | 1767 | 1180 | 1.002 |
| $\varepsilon^{(R)}_{3,1,43}$ | 0.022 | 0.077 | -0.133 | -0.031 | 0.025 | 0.076 | 0.166 | 1726 | 1381 | 0.999 |
| $\varepsilon^{(R)}_{3,2,43}$ | 0.098 | 0.071 | -0.039 | 0.051 | 0.098 | 0.145 | 0.239 | 1424 | 1041 | 1.005 |
| $\varepsilon^{(R)}_{3,3,43}$ | -0.152 | 0.083 | -0.317 | -0.206 | -0.150 | -0.097 | 0.010 | 1389 | 1196 | 1.002 |
| $\varepsilon^{(R)}_{3,4,43}$ | 0.179 | 0.086 | 0.018 | 0.123 | 0.175 | 0.236 | 0.353 | 1343 | 1277 | 1.002 |
| $\varepsilon^{(R)}_{3,1,44}$ | 0.064 | 0.072 | -0.075 | 0.016 | 0.064 | 0.112 | 0.206 | 1560 | 942 | 1.002 |
| $\varepsilon^{(R)}_{3,2,44}$ | 0.058 | 0.069 | -0.076 | 0.013 | 0.059 | 0.102 | 0.189 | 1540 | 1147 | 1.001 |
| $\varepsilon^{(R)}_{3,3,44}$ | -0.050 | 0.074 | -0.204 | -0.099 | -0.048 | 0.000 | 0.089 | 1302 | 1277 | 1.003 |
| $\varepsilon^{(R)}_{3,4,44}$ | -0.166 | 0.093 | -0.360 | -0.229 | -0.162 | -0.102 | 0.007 | 1261 | 1279 | 1.000 |
| $\varepsilon^{(R)}_{3,1,45}$ | -0.017 | 0.106 | -0.226 | -0.087 | -0.016 | 0.052 | 0.191 | 1258 | 999 | 1.004 |
| $\varepsilon^{(R)}_{3,2,45}$ | 0.100 | 0.095 | -0.078 | 0.039 | 0.098 | 0.162 | 0.292 | 1525 | 1171 | 1.000 |
| $\varepsilon^{(R)}_{3,3,45}$ | 0.230 | 0.078 | 0.082 | 0.176 | 0.232 | 0.281 | 0.384 | 1344 | 1284 | 1.003 |
| $\varepsilon^{(R)}_{3,4,45}$ | -0.057 | 0.081 | -0.219 | -0.112 | -0.059 | -0.002 | 0.094 | 1282 | 1100 | 1.001 |
| $\varepsilon^{(R)}_{3,1,46}$ | -0.320 | 0.103 | -0.516 | -0.390 | -0.322 | -0.251 | -0.119 | 1107 | 924 | 1.001 |
| $\varepsilon^{(R)}_{3,2,46}$ | -0.094 | 0.125 | -0.339 | -0.178 | -0.091 | -0.009 | 0.138 | 1098 | 1100 | 1.000 |
| $\varepsilon^{(R)}_{3,3,46}$ | 0.039 | 0.145 | -0.250 | -0.056 | 0.041 | 0.138 | 0.326 | 1040 | 1012 | 1.003 |
| $\varepsilon^{(R)}_{3,4,46}$ | -0.402 | 0.197 | -0.777 | -0.534 | -0.407 | -0.273 | -0.024 | 986 | 1063 | 1.006 |
| $\varepsilon^{(R)}_{3,1,47}$ | -0.223 | 0.211 | -0.631 | -0.364 | -0.234 | -0.077 | 0.205 | 1224 | 904 | 1.003 |
| $\varepsilon^{(R)}_{3,2,47}$ | -0.184 | 0.202 | -0.587 | -0.323 | -0.187 | -0.041 | 0.200 | 1519 | 1163 | 1.000 |
| $\varepsilon^{(R)}_{3,3,47}$ | -0.010 | 0.243 | -0.493 | -0.173 | -0.007 | 0.161 | 0.441 | 1324 | 905 | 1.001 |
| $\varepsilon^{(R)}_{3,4,47}$ | 0.049 | 0.296 | -0.513 | -0.145 | 0.051 | 0.247 | 0.641 | 1425 | 1079 | 0.999 |
| $\varepsilon^{(R)}_{3,1,48}$ | 0.827 | 0.266 | 0.296 | 0.655 | 0.837 | 1.006 | 1.316 | 990 | 1057 | 1.002 |
| $\varepsilon^{(R)}_{3,2,48}$ | 0.005 | 0.809 | -2.155 | 0.114 | 0.337 | 0.462 | 0.698 | 105 | 317 | 1.020 |
| $\varepsilon^{(R)}_{3,3,48}$ | 0.286 | 0.824 | -0.409 | -0.171 | -0.053 | 0.188 | 2.525 | 96 | 362 | 1.021 |
| $\varepsilon^{(R)}_{3,4,48}$ | -2.181 | 0.614 | -3.469 | -2.576 | -2.148 | -1.733 | -1.076 | 2317 | 1104 | 1.001 |
| *Herring catchability* | | | | | | | | | | |
| $\ln q_{2,3,1,1:20}$ | -5.003 | 0.136 | -5.269 | -5.090 | -5.003 | -4.915 | -4.735 | 916 | 1061 | 1.002 |
| $\ln q_{3,3,1,1:48}$ | -7.864 | 0.055 | -7.971 | -7.903 | -7.864 | -7.826 | -7.758 | 1094 | 1284 | 1.001 |
| $\ln q_{4,3,2:4,1:48}$ | -3.494 | 0.184 | -3.867 | -3.619 | -3.496 | -3.364 | -3.134 | 767 | 1010 | 0.999 |
| $\ln q_{5,3,2:4,1:48}$ | -3.478 | 0.148 | -3.747 | -3.588 | -3.483 | -3.374 | -3.189 | 1138 | 1012 | 1.001 |
| $\ln q_{2,3,2,1:20}$ | -3.475 | 0.139 | -3.756 | -3.564 | -3.473 | -3.384 | -3.205 | 1160 | 1264 | 1.000 |
| $\ln q_{2,3,3,1:20}$ | 0.029 | 0.046 | -0.059 | -0.002 | 0.030 | 0.060 | 0.116 | 4483 | 1219 | 1.000 |
| $\ln q_{2,3,4,1:20}$ | 0.019 | 0.046 | -0.073 | -0.013 | 0.019 | 0.050 | 0.109 | 2783 | 1029 | 1.007 |
| $\ln q_{6,3,1:4,1:48}$ | -0.033 | 0.047 | -0.123 | -0.065 | -0.033 | 0.000 | 0.060 | 2886 | 1064 | 1.000 |
| $\varepsilon^{(q)}_{1,21}$ | -0.023 | 0.045 | -0.110 | -0.053 | -0.023 | 0.006 | 0.066 | 3553 | 1141 | 1.004 |
| $\varepsilon_{1,22}^{(q)}$ | -0.004 | 0.044 | -0.088 | -0.034 | -0.004 | 0.026 | 0.080 | 4069 | 1316 | 1.001 |
| $\varepsilon_{1,23}^{(q)}$ | 0.041 | 0.045 | -0.050 | 0.010 | 0.042 | 0.070 | 0.126 | 2703 | 862 | 1.000 |
| $\varepsilon_{1,24}^{(q)}$ | 0.046 | 0.046 | -0.038 | 0.014 | 0.046 | 0.078 | 0.135 | 3565 | 907 | 1.000 |
| $\varepsilon_{1,25}^{(q)}$ | 0.068 | 0.041 | -0.011 | 0.040 | 0.067 | 0.095 | 0.148 | 3421 | 1084 | 1.006 |
| $\varepsilon_{1,26}^{(q)}$ | 0.097 | 0.043 | 0.015 | 0.069 | 0.098 | 0.126 | 0.181 | 2967 | 1239 | 1.005 |
| $\varepsilon_{1,27}^{(q)}$ | 0.163 | 0.043 | 0.080 | 0.135 | 0.163 | 0.192 | 0.246 | 3263 | 960 | 1.002 |
| $\varepsilon_{1,28}^{(q)}$ | 0.157 | 0.045 | 0.070 | 0.127 | 0.157 | 0.186 | 0.247 | 2983 | 958 | 1.003 |
| $\varepsilon_{1,29}^{(q)}$ | 0.132 | 0.043 | 0.046 | 0.102 | 0.132 | 0.162 | 0.218 | 3420 | 805 | 1.004 |
| $\varepsilon_{1,30}^{(q)}$ | 0.103 | 0.044 | 0.016 | 0.073 | 0.104 | 0.133 | 0.189 | 3054 | 1062 | 1.000 |
| $\varepsilon_{1,31}^{(q)}$ | 0.053 | 0.047 | -0.038 | 0.018 | 0.052 | 0.088 | 0.138 | 3392 | 1212 | 1.004 |
| $\varepsilon_{1,32}^{(q)}$ | 0.042 | 0.050 | -0.053 | 0.006 | 0.043 | 0.077 | 0.134 | 3040 | 949 | 1.001 |
| $\varepsilon_{1,33}^{(q)}$ | 0.015 | 0.046 | -0.068 | -0.019 | 0.014 | 0.048 | 0.101 | 3075 | 1274 | 1.001 |
| $\varepsilon_{1,34}^{(q)}$ | -0.003 | 0.046 | -0.097 | -0.034 | -0.003 | 0.027 | 0.088 | 4190 | 1048 | 1.003 |
| $\varepsilon_{1,35}^{(q)}$ | -0.017 | 0.045 | -0.104 | -0.047 | -0.018 | 0.011 | 0.071 | 3219 | 1108 | 1.006 |
| $\varepsilon_{1,36}^{(q)}$ | -0.038 | 0.045 | -0.125 | -0.069 | -0.038 | -0.008 | 0.053 | 3182 | 976 | 1.003 |
| $\varepsilon_{1,37}^{(q)}$ | -0.151 | 0.047 | -0.244 | -0.181 | -0.151 | -0.121 | -0.061 | 3227 | 1195 | 1.004 |
| $\varepsilon_{1,38}^{(q)}$ | -0.059 | 0.043 | -0.143 | -0.087 | -0.058 | -0.031 | 0.026 | 4124 | 1011 | 1.001 |
| $\varepsilon_{1,39}^{(q)}$ | -0.006 | 0.043 | -0.088 | -0.037 | -0.006 | 0.022 | 0.081 | 3622 | 1153 | 1.000 |
| $\varepsilon_{1,40}^{(q)}$ | 0.060 | 0.049 | -0.032 | 0.024 | 0.061 | 0.096 | 0.155 | 3961 | 1106 | 1.001 |
| $\varepsilon_{1,41}^{(q)}$ | -0.023 | 0.045 | -0.111 | -0.054 | -0.023 | 0.008 | 0.063 | 2904 | 1121 | 1.000 |
| $\varepsilon_{1,42}^{(q)}$ | -0.055 | 0.044 | -0.142 | -0.084 | -0.054 | -0.027 | 0.033 | 3435 | 1012 | 1.005 |
| $\varepsilon_{1,43}^{(q)}$ | -0.046 | 0.045 | -0.133 | -0.077 | -0.048 | -0.016 | 0.040 | 3715 | 1069 | 1.001 |
| $\varepsilon_{1,44}^{(q)}$ | 0.018 | 0.043 | -0.066 | -0.012 | 0.018 | 0.047 | 0.104 | 3481 | 1076 | 1.004 |
| $\varepsilon_{1,45}^{(q)}$ | -0.053 | 0.048 | -0.146 | -0.085 | -0.056 | -0.023 | 0.042 | 4764 | 1053 | 1.002 |
| $\varepsilon_{1,46}^{(q)}$ | 0.015 | 0.044 | -0.074 | -0.016 | 0.016 | 0.045 | 0.101 | 4276 | 1069 | 1.000 |
| $\varepsilon_{1,47}^{(q)}$ | 0.001 | 0.046 | -0.087 | -0.029 | 0.001 | 0.033 | 0.091 | 3399 | 1335 | 1.003 |
| $\varepsilon_{1,48}^{(q)}$ | 0.052 | 0.044 | -0.033 | 0.022 | 0.052 | 0.082 | 0.139 | 3084 | 1214 | 1.003 |
| $\varepsilon_{2,17}^{(q)}$ | -0.053 | 0.047 | -0.143 | -0.087 | -0.053 | -0.020 | 0.039 | 3131 | 1185 | 1.006 |
| $\varepsilon_{2,18}^{(q)}$ | 0.067 | 0.045 | -0.021 | 0.037 | 0.067 | 0.098 | 0.157 | 3725 | 965 | 1.004 |
| $\varepsilon_{2,19}^{(q)}$ | 0.016 | 0.046 | -0.071 | -0.014 | 0.017 | 0.047 | 0.103 | 3939 | 931 | 1.010 |
| $\varepsilon_{2,20}^{(q)}$ | -0.007 | 0.046 | -0.100 | -0.037 | -0.007 | 0.025 | 0.081 | 3332 | 1193 | 1.002 |
| $\varepsilon_{1,21}^{(q)}$ | 0.029 | 0.045 | -0.058 | 0.000 | 0.028 | 0.060 | 0.117 | 4104 | 1118 | 1.005 |
| $\varepsilon_{1,22}^{(q)}$ | 0.067 | 0.045 | -0.023 | 0.036 | 0.067 | 0.098 | 0.154 | 3396 | 1110 | 1.001 |
| $\varepsilon_{1,23}^{(q)}$ | 0.056 | 0.045 | -0.033 | 0.026 | 0.057 | 0.086 | 0.145 | 4107 | 1098 | 1.000 |
| $\varepsilon_{1,24}^{(q)}$ | 0.054 | 0.045 | -0.036 | 0.025 | 0.054 | 0.085 | 0.143 | 3988 | 929 | 1.001 |
| $\varepsilon_{1,25}^{(q)}$ | -0.005 | 0.048 | -0.098 | -0.040 | -0.005 | 0.027 | 0.086 | 3712 | 959 | 1.003 |
| $\varepsilon_{1,26}^{(q)}$ | 0.031 | 0.044 | -0.055 | 0.000 | 0.033 | 0.061 | 0.118 | 3601 | 884 | 1.008 |
| $\varepsilon_{1,27}^{(q)}$ | 0.045 | 0.047 | -0.046 | 0.014 | 0.046 | 0.080 | 0.137 | 3592 | 900 | 1.004 |
| $\varepsilon_{1,28}^{(q)}$ | -0.035 | 0.046 | -0.123 | -0.065 | -0.036 | -0.004 | 0.057 | 4349 | 1099 | 1.013 |
| $\varepsilon_{1,29}^{(q)}$ | 0.008 | 0.045 | -0.077 | -0.021 | 0.007 | 0.040 | 0.094 | 3684 | 1170 | 1.001 |
| $\varepsilon_{1,30}^{(q)}$ | 0.040 | 0.047 | -0.054 | 0.008 | 0.040 | 0.073 | 0.133 | 3703 | 1024 | 1.001 |
| $\varepsilon_{1,31}^{(q)}$ | 0.030 | 0.048 | -0.059 | -0.002 | 0.031 | 0.063 | 0.123 | 4283 | 974 | 1.001 |
| $\varepsilon_{1,32}^{(q)}$ | 0.081 | 0.047 | -0.010 | 0.049 | 0.081 | 0.114 | 0.178 | 3919 | 1183 | 1.011 |
| $\varepsilon_{1,33}^{(q)}$ | -0.095 | 0.047 | -0.187 | -0.127 | -0.093 | -0.065 | -0.006 | 3948 | 989 | 1.001 |
| $\varepsilon_{1,34}^{(q)}$ | 0.015 | 0.048 | -0.075 | -0.019 | 0.014 | 0.046 | 0.111 | 4216 | 953 | 1.000 |
| $\varepsilon_{1,35}^{(q)}$ | -0.081 | 0.045 | -0.168 | -0.111 | -0.080 | -0.051 | 0.005 | 3307 | 980 | 1.005 |
| $\varepsilon_{1,36}^{(q)}$ | 0.001 | 0.047 | -0.090 | -0.031 | 0.001 | 0.033 | 0.092 | 3762 | 1017 | 1.002 |
| $\varepsilon_{1,37}^{(q)}$ | -0.031 | 0.044 | -0.118 | -0.061 | -0.032 | 0.000 | 0.057 | 3119 | 946 | 1.007 |
| $\varepsilon_{1,38}^{(q)}$ | -0.048 | 0.047 | -0.138 | -0.080 | -0.048 | -0.017 | 0.043 | 3456 | 1059 | 1.001 |
| $\varepsilon_{1,39}^{(q)}$ | 0.015 | 0.048 | -0.085 | -0.016 | 0.013 | 0.046 | 0.107 | 3755 | 895 | 1.005 |
| $\varepsilon_{1,40}^{(q)}$ | 0.032 | 0.047 | -0.064 | 0.003 | 0.034 | 0.063 | 0.121 | 3740 | 1017 | 1.008 |
| $\varepsilon_{1,41}^{(q)}$ | 0.066 | 0.046 | -0.026 | 0.036 | 0.066 | 0.096 | 0.157 | 3526 | 1024 | 1.006 |
| $\varepsilon_{1,42}^{(q)}$ | 0.054 | 0.048 | -0.044 | 0.021 | 0.053 | 0.086 | 0.149 | 3521 | 1037 | 1.001 |
| $\varepsilon_{1,43}^{(q)}$ | -0.031 | 0.048 | -0.122 | -0.065 | -0.033 | 0.003 | 0.062 | 3093 | 1224 | 1.000 |
| $\varepsilon^{(q)}_{1,44}$ | -0.006 | 0.047 | -0.099 | -0.034 | -0.006 | 0.025 | 0.087 | 3239 | 1122 | 1.000 |
| $\varepsilon^{(q)}_{1,45}$ | 0.037 | 0.049 | -0.063 | 0.004 | 0.037 | 0.069 | 0.133 | 3254 | 903 | 1.002 |
| $\varepsilon^{(q)}_{1,46}$ | -0.066 | 0.048 | -0.162 | -0.097 | -0.067 | -0.035 | 0.031 | 3515 | 992 | 1.002 |
| $\varepsilon^{(q)}_{1,47}$ | -0.204 | 0.047 | -0.296 | -0.236 | -0.205 | -0.172 | -0.114 | 3571 | 1071 | 1.003 |
| $\varepsilon^{(q)}_{1,48}$ | -0.163 | 0.044 | -0.250 | -0.193 | -0.161 | -0.132 | -0.078 | 3257 | 1223 | 1.001 |
| $\varepsilon^{(q)}_{2,17}$ | -0.047 | 0.047 | -0.140 | -0.079 | -0.047 | -0.015 | 0.046 | 3905 | 1013 | 1.001 |
| $\varepsilon^{(q)}_{2,18}$ | -0.009 | 0.047 | -0.103 | -0.040 | -0.009 | 0.024 | 0.080 | 3407 | 960 | 1.003 |
| $\varepsilon^{(q)}_{2,19}$ | -0.068 | 0.045 | -0.156 | -0.098 | -0.068 | -0.039 | 0.019 | 4521 | 1206 | 1.002 |
| $\varepsilon^{(q)}_{2,20}$ | -0.049 | 0.046 | -0.142 | -0.078 | -0.048 | -0.019 | 0.041 | 3889 | 875 | 1.005 |
| $\varepsilon^{(q)}_{1,21}$ | -0.045 | 0.044 | -0.128 | -0.074 | -0.047 | -0.016 | 0.042 | 4169 | 1016 | 1.005 |
| $\varepsilon^{(q)}_{1,22}$ | 0.063 | 0.045 | -0.025 | 0.034 | 0.063 | 0.093 | 0.155 | 2658 | 1045 | 1.001 |
| $\varepsilon^{(q)}_{1,23}$ | 0.073 | 0.045 | -0.017 | 0.040 | 0.074 | 0.105 | 0.161 | 3634 | 944 | 1.012 |
| $\varepsilon^{(q)}_{1,24}$ | 0.065 | 0.045 | -0.024 | 0.036 | 0.065 | 0.095 | 0.153 | 4073 | 1069 | 1.003 |
| $\varepsilon^{(q)}_{1,25}$ | -0.030 | 0.044 | -0.115 | -0.061 | -0.030 | 0.001 | 0.058 | 3450 | 1050 | 1.003 |
| $\varepsilon^{(q)}_{1,26}$ | 0.036 | 0.046 | -0.055 | 0.005 | 0.037 | 0.067 | 0.123 | 2697 | 997 | 1.002 |
| $\varepsilon^{(q)}_{1,27}$ | 0.036 | 0.045 | -0.049 | 0.004 | 0.036 | 0.067 | 0.124 | 3518 | 1101 | 1.004 |
| $\varepsilon^{(q)}_{1,28}$ | -0.054 | 0.046 | -0.141 | -0.086 | -0.054 | -0.022 | 0.037 | 2533 | 1069 | 1.001 |
| $\varepsilon^{(q)}_{1,29}$ | -0.062 | 0.046 | -0.153 | -0.092 | -0.064 | -0.034 | 0.030 | 3112 | 980 | 1.002 |
| $\varepsilon_{1,30}^{(q)}$ | -0.019 | 0.045 | -0.103 | -0.050 | -0.021 | 0.012 | 0.067 | 2893 | 982 | 1.000 |
| $\varepsilon_{1,31}^{(q)}$ | 0.077 | 0.046 | -0.013 | 0.048 | 0.077 | 0.107 | 0.168 | 3013 | 732 | 1.003 |
| $\varepsilon_{1,32}^{(q)}$ | 0.115 | 0.047 | 0.026 | 0.083 | 0.113 | 0.146 | 0.205 | 2648 | 899 | 1.002 |
| $\varepsilon_{1,33}^{(q)}$ | 0.049 | 0.047 | -0.045 | 0.020 | 0.047 | 0.082 | 0.142 | 3157 | 1076 | 1.002 |
| $\varepsilon_{1,34}^{(q)}$ | 0.016 | 0.046 | -0.076 | -0.016 | 0.016 | 0.047 | 0.110 | 3092 | 1199 | 1.001 |
| $\varepsilon_{1,35}^{(q)}$ | -0.059 | 0.045 | -0.148 | -0.090 | -0.059 | -0.030 | 0.028 | 4103 | 1335 | 0.999 |
| $\varepsilon_{1,36}^{(q)}$ | -0.085 | 0.046 | -0.176 | -0.115 | -0.084 | -0.052 | 0.004 | 4764 | 844 | 1.005 |
| $\varepsilon_{1,37}^{(q)}$ | -0.111 | 0.045 | -0.195 | -0.140 | -0.111 | -0.079 | -0.026 | 3241 | 1272 | 1.001 |
| $\varepsilon_{1,38}^{(q)}$ | -0.009 | 0.045 | -0.100 | -0.038 | -0.008 | 0.020 | 0.077 | 2998 | 1113 | 1.001 |
| $\varepsilon_{1,39}^{(q)}$ | 0.059 | 0.045 | -0.030 | 0.030 | 0.059 | 0.088 | 0.147 | 2989 | 1048 | 1.000 |
| $\varepsilon_{1,40}^{(q)}$ | 0.098 | 0.045 | 0.009 | 0.069 | 0.097 | 0.129 | 0.183 | 2560 | 1122 | 1.005 |
| $\varepsilon_{1,41}^{(q)}$ | 0.020 | 0.047 | -0.071 | -0.012 | 0.019 | 0.052 | 0.111 | 3212 | 908 | 1.010 |
| $\varepsilon_{1,42}^{(q)}$ | 0.165 | 0.046 | 0.079 | 0.134 | 0.164 | 0.196 | 0.257 | 3061 | 1005 | 1.007 |
| $\varepsilon_{1,43}^{(q)}$ | 0.065 | 0.047 | -0.027 | 0.033 | 0.065 | 0.096 | 0.156 | 2779 | 924 | 1.001 |
| $\varepsilon_{1,44}^{(q)}$ | -0.040 | 0.047 | -0.129 | -0.072 | -0.042 | -0.008 | 0.055 | 2681 | 799 | 0.999 |
| $\varepsilon_{1,45}^{(q)}$ | 0.102 | 0.049 | 0.003 | 0.069 | 0.101 | 0.135 | 0.201 | 2383 | 1083 | 1.008 |
| $\varepsilon_{1,46}^{(q)}$ | -0.049 | 0.046 | -0.144 | -0.080 | -0.050 | -0.021 | 0.045 | 4764 | 1172 | 1.005 |
| $\varepsilon_{1,47}^{(q)}$ | -0.009 | 0.043 | -0.095 | -0.037 | -0.008 | 0.020 | 0.078 | 2310 | 813 | 1.004 |
| $\varepsilon_{1,48}^{(q)}$ | 0.070 | 0.047 | -0.024 | 0.037 | 0.069 | 0.100 | 0.163 | 3081 | 984 | 1.004 |
| $\varepsilon_{2,17}^{(q)}$ | -0.014 | 0.043 | -0.101 | -0.044 | -0.014 | 0.015 | 0.068 | 3735 | 1152 | 1.000 |
| $\varepsilon_{2,18}^{(q)}$ | -0.075 | 0.044 | -0.162 | -0.107 | -0.074 | -0.044 | 0.007 | 3710 | 907 | 1.000 |
| $\varepsilon_{2,19}^{(q)}$ | -0.090 | 0.043 | -0.180 | -0.118 | -0.090 | -0.064 | -0.004 | 2808 | 1092 | 1.008 |
| $\varepsilon_{2,20}^{(q)}$ | -0.126 | 0.044 | -0.214 | -0.156 | -0.126 | -0.096 | -0.044 | 3667 | 1196 | 1.003 |
| $\varepsilon_{1,21}^{(q)}$ | -0.086 | 0.045 | -0.172 | -0.117 | -0.087 | -0.056 | -0.002 | 3258 | 1015 | 1.002 |
| $\varepsilon_{1,22}^{(q)}$ | -0.081 | 0.043 | -0.160 | -0.110 | -0.082 | -0.053 | 0.003 | 4304 | 1237 | 1.000 |
| $\varepsilon_{1,23}^{(q)}$ | 0.015 | 0.043 | -0.068 | -0.015 | 0.015 | 0.043 | 0.098 | 2736 | 1163 | 1.000 |
| $\varepsilon_{1,24}^{(q)}$ | 0.021 | 0.044 | -0.065 | -0.009 | 0.021 | 0.051 | 0.108 | 3255 | 1071 | 1.007 |
| $\varepsilon_{1,25}^{(q)}$ | -0.123 | 0.044 | -0.209 | -0.153 | -0.122 | -0.094 | -0.041 | 4137 | 1113 | 1.001 |
| $\varepsilon_{1,26}^{(q)}$ | -0.034 | 0.046 | -0.124 | -0.069 | -0.033 | -0.002 | 0.052 | 3347 | 1009 | 1.006 |
| $\varepsilon_{1,27}^{(q)}$ | -0.036 | 0.043 | -0.124 | -0.065 | -0.035 | -0.007 | 0.049 | 3575 | 1370 | 1.005 |
| $\varepsilon_{1,28}^{(q)}$ | 0.016 | 0.042 | -0.066 | -0.013 | 0.016 | 0.044 | 0.100 | 3244 | 1109 | 1.007 |
| $\varepsilon_{1,29}^{(q)}$ | 0.001 | 0.045 | -0.086 | -0.029 | 0.002 | 0.032 | 0.086 | 2892 | 1080 | 1.003 |
| $\varepsilon_{1,30}^{(q)}$ | 0.011 | 0.044 | -0.075 | -0.020 | 0.012 | 0.041 | 0.094 | 3341 | 899 | 1.000 |
| $\varepsilon_{1,31}^{(q)}$ | -0.009 | 0.046 | -0.098 | -0.041 | -0.006 | 0.024 | 0.079 | 3525 | 995 | 1.001 |
| $\varepsilon_{1,32}^{(q)}$ | 0.047 | 0.048 | -0.045 | 0.016 | 0.047 | 0.078 | 0.142 | 3773 | 1160 | 1.001 |
| $\varepsilon_{1,33}^{(q)}$ | 0.059 | 0.045 | -0.027 | 0.028 | 0.058 | 0.090 | 0.147 | 3695 | 1097 | 1.003 |

| | | | | | | | | | |
|---|---|---|---|---|---|---|---|---|---|
| $\varepsilon_{1,34}^{(q)}$ | 0.040 | 0.045 | -0.048 | 0.011 | 0.040 | 0.072 | 0.129 | 2942 | 1058 | 1.000 |
| $\varepsilon_{1,35}^{(q)}$ | -0.072 | 0.045 | -0.162 | -0.102 | -0.072 | -0.043 | 0.018 | 2808 | 1099 | 1.000 |
| $\varepsilon_{1,36}^{(q)}$ | -0.024 | 0.044 | -0.111 | -0.054 | -0.024 | 0.006 | 0.059 | 3733 | 1182 | 1.003 |
| $\varepsilon_{1,37}^{(q)}$ | -0.007 | 0.044 | -0.093 | -0.037 | -0.006 | 0.024 | 0.077 | 2766 | 1193 | 1.006 |
| $\varepsilon_{1,38}^{(q)}$ | -0.025 | 0.045 | -0.114 | -0.056 | -0.025 | 0.007 | 0.062 | 3750 | 1224 | 1.001 |
| $\varepsilon_{1,39}^{(q)}$ | -0.041 | 0.045 | -0.129 | -0.070 | -0.042 | -0.013 | 0.045 | 3260 | 807 | 1.007 |
| $\varepsilon_{1,40}^{(q)}$ | 0.125 | 0.044 | 0.044 | 0.095 | 0.124 | 0.156 | 0.209 | 2597 | 1272 | 1.000 |
| $\varepsilon_{1,41}^{(q)}$ | 0.106 | 0.047 | 0.021 | 0.073 | 0.106 | 0.136 | 0.200 | 2045 | 935 | 1.004 |
| $\varepsilon_{4,42}^{(q)}$ | 0.031 | 0.049 | -0.065 | -0.002 | 0.032 | 0.065 | 0.122 | 1412 | 1235 | 1.004 |
| $\varepsilon_{4,43}^{(q)}$ | -0.122 | 0.048 | -0.217 | -0.153 | -0.123 | -0.091 | -0.023 | 1970 | 857 | 1.006 |
| $\varepsilon_{4,44}^{(q)}$ | 0.015 | 0.049 | -0.083 | -0.018 | 0.015 | 0.048 | 0.110 | 2237 | 896 | 1.000 |
| $\varepsilon_{4,45}^{(q)}$ | -0.036 | 0.052 | -0.136 | -0.069 | -0.036 | -0.002 | 0.063 | 1655 | 906 | 1.003 |
| $\varepsilon_{4,46}^{(q)}$ | -5.003 | 0.136 | -5.269 | -5.090 | -5.003 | -4.915 | -4.735 | 916 | 1061 | 1.002 |
| $\varepsilon_{4,47}^{(q)}$ | -7.864 | 0.055 | -7.971 | -7.903 | -7.864 | -7.826 | -7.758 | 1094 | 1284 | 1.001 |
| $\varepsilon_{4,48}^{(q)}$ | -3.494 | 0.184 | -3.867 | -3.619 | -3.496 | -3.364 | -3.134 | 767 | 1010 | 0.999 |
| *Herring selectivity* | | | | | | | | | | |
| $\ln s_{1,3,1,8:19}^{(50\%)}$ | 1.389 | 0.043 | 1.311 | 1.360 | 1.388 | 1.417 | 1.479 | 449 | 623 | 1.006 |
| $\ln s_{1,3,2,8:19}^{(50\%)}$ | 2.110 | 0.131 | 1.883 | 2.019 | 2.099 | 2.189 | 2.390 | 1405 | 995 | 1.001 |

| | | | | | | | | | |
|---|---|---|---|---|---|---|---|---|---|
| $\ln s_{1,3,3,8:19}^{(50\%)}$ | 1.388 | 0.024 | 1.339 | 1.373 | 1.389 | 1.404 | 1.435 | 1464 | 1083 | 1.003 |
| $\ln s_{1,3,4,8:19}^{(50\%)}$ | 1.468 | 0.032 | 1.404 | 1.447 | 1.469 | 1.490 | 1.532 | 880 | 849 | 1.001 |
| $\ln s_{1,3,1,20:34}^{(50\%)}$ | 1.449 | 0.025 | 1.403 | 1.432 | 1.449 | 1.466 | 1.496 | 1366 | 1347 | 1.004 |
| $\ln s_{1,3,2,20:34}^{(50\%)}$ | 1.488 | 0.024 | 1.443 | 1.471 | 1.488 | 1.505 | 1.534 | 1208 | 1122 | 1.000 |
| $\ln s_{1,3,3,20:34}^{(50\%)}$ | 1.515 | 0.021 | 1.475 | 1.500 | 1.515 | 1.530 | 1.558 | 1890 | 1127 | 1.001 |
| $\ln s_{1,3,4,20:34}^{(50\%)}$ | 1.662 | 0.018 | 1.627 | 1.650 | 1.662 | 1.675 | 1.697 | 1460 | 1198 | 1.002 |
| $\ln s_{1,3,1,35:48}^{(50\%)}$ | 1.840 | 0.032 | 1.778 | 1.818 | 1.840 | 1.861 | 1.905 | 1454 | 1115 | 1.007 |
| $\ln s_{1,3,2,35:48}^{(50\%)}$ | 1.563 | 0.018 | 1.527 | 1.552 | 1.563 | 1.576 | 1.599 | 1274 | 1129 | 1.000 |
| $\ln s_{1,3,3,35:48}^{(50\%)}$ | 1.667 | 0.018 | 1.632 | 1.655 | 1.667 | 1.678 | 1.702 | 1700 | 1144 | 1.003 |
| $\ln s_{1,3,4,35:48}^{(50\%)}$ | 1.730 | 0.014 | 1.703 | 1.721 | 1.730 | 1.740 | 1.756 | 1252 | 1034 | 1.001 |
| $\ln s_{2,3,1,8:34}^{(50\%)}$ | 1.808 | 0.034 | 1.746 | 1.785 | 1.807 | 1.830 | 1.877 | 1317 | 1192 | 1.001 |
| $\ln s_{2,3,2,8:34}^{(50\%)}$ | 2.007 | 0.046 | 1.923 | 1.976 | 2.003 | 2.037 | 2.095 | 710 | 1155 | 1.000 |
| $\ln s_{2,3,3,8:34}^{(50\%)}$ | 1.777 | 0.030 | 1.723 | 1.757 | 1.776 | 1.795 | 1.840 | 1141 | 953 | 1.003 |
| $\ln s_{2,3,4,8:34}^{(50\%)}$ | 1.960 | 0.034 | 1.896 | 1.936 | 1.959 | 1.983 | 2.030 | 1271 | 1139 | 1.000 |
| $\ln s_{2,3,1,35:48}^{(50\%)}$ | 1.745 | 0.018 | 1.709 | 1.733 | 1.744 | 1.758 | 1.781 | 1557 | 1275 | 0.999 |
| $\ln s_{2,3,2,35:48}^{(50\%)}$ | 1.856 | 0.017 | 1.821 | 1.844 | 1.855 | 1.867 | 1.890 | 944 | 1147 | 1.004 |
| $\ln s_{2,3,3,35:48}^{(50\%)}$ | 1.806 | 0.016 | 1.775 | 1.795 | 1.806 | 1.816 | 1.837 | 997 | 1005 | 1.003 |
| $\ln s_{2,3,4,35:48}^{(50\%)}$ | 1.810 | 0.017 | 1.777 | 1.799 | 1.810 | 1.821 | 1.843 | 1201 | 1119 | 1.001 |
| $\ln s_{3,3,1:4,1:7}^{(50\%)}$ | 1.569 | 0.016 | 1.638 | 1.659 | 1.698 | 1.701 | 1.715 | 1419 | 1102 | 1.000 |
| $\ln(s_{1,3,1,8:19}^{(95\%)}\text{-}s_{1,3,1,8:19}^{(50\%)})$ | 1.204 | 0.162 | 0.873 | 1.103 | 1.204 | 1.317 | 1.508 | 422 | 670 | 1.006 |
| $\ln(s_{1,3,2,8:19}^{(95\%)}\text{-}s_{1,3,2,8:19}^{(50\%)})$ | 2.015 | 0.119 | 1.779 | 1.936 | 2.016 | 2.097 | 2.240 | 1394 | 1053 | 1.000 |
| $\ln(s_{1,3,3,8:19}^{(95\%)}\text{-}s_{1,3,3,8:19}^{(50\%)})$ | -0.366 | 0.074 | -0.516 | -0.414 | -0.368 | -0.320 | -0.218 | 1431 | 1056 | 1.003 |
| $\ln(s_{1,3,4,8:19}^{(95\%)}\text{-}s_{1,3,4,8:19}^{(50\%)})$ | 0.090 | 0.080 | -0.062 | 0.036 | 0.091 | 0.146 | 0.240 | 904 | 1271 | 1.002 |
| $\ln(s_{1,3,1,20:34}^{(95\%)}\text{-}s_{1,3,1,20:34}^{(50\%)})$ | 0.172 | 0.066 | 0.045 | 0.127 | 0.173 | 0.216 | 0.301 | 1455 | 1177 | 1.000 |
| $\ln(s_{1,3,2,20:34}^{(95\%)}\text{-}s_{1,3,2,20:34}^{(50\%)})$ | -0.014 | 0.061 | -0.131 | -0.058 | -0.016 | 0.029 | 0.103 | 1218 | 1197 | 1.001 |
| $\ln(s_{1,3,3,20:34}^{(95\%)}\text{-}s_{1,3,3,20:34}^{(50\%)})$ | -0.414 | 0.080 | -0.568 | -0.472 | -0.414 | -0.361 | -0.258 | 1797 | 1146 | 1.002 |
| $\ln(s_{1,3,4,20:34}^{(95\%)}\text{-}s_{1,3,4,20:34}^{(50\%)})$ | 0.183 | 0.049 | 0.084 | 0.152 | 0.184 | 0.216 | 0.278 | 1345 | 1422 | 1.004 |
| $\ln(s_{1,3,1,35:48}^{(95\%)}\text{-}s_{1,3,1,35:48}^{(50\%)})$ | 0.858 | 0.063 | 0.732 | 0.813 | 0.857 | 0.899 | 0.985 | 1339 | 914 | 1.006 |
| $\ln(s_{1,3,2,35:48}^{(95\%)}\text{-}s_{1,3,2,35:48}^{(50\%)})$ | 0.038 | 0.053 | -0.064 | -0.001 | 0.036 | 0.074 | 0.145 | 2258 | 1303 | 1.000 |
| $\ln(s_{1,3,3,35:48}^{(95\%)}\text{-}s_{1,3,3,35:48}^{(50\%)})$ | -0.098 | 0.052 | -0.196 | -0.135 | -0.101 | -0.063 | 0.005 | 1555 | 1089 | 1.000 |
| $\ln(s_{1,3,4,35:48}^{(95\%)}\text{-}s_{1,3,4,35:48}^{(50\%)})$ | -0.096 | 0.042 | -0.177 | -0.125 | -0.096 | -0.067 | -0.014 | 511 | 866 | 1.003 |
| $\ln(s_{2,3,1,8:34}^{(95\%)}\text{-}s_{2,3,1,8:34}^{(50\%)})$ | 0.911 | 0.097 | 0.732 | 0.848 | 0.910 | 0.972 | 1.108 | 1496 | 1197 | 0.999 |
| $\ln(s_{2,3,2,8:34}^{(95\%)}\text{-}s_{2,3,2,8:34}^{(50\%)})$ | 1.030 | 0.083 | 0.871 | 0.973 | 1.030 | 1.088 | 1.189 | 999 | 1203 | 1.001 |
| $\ln(s_{2,3,3,8:34}^{(95\%)}\text{-}s_{2,3,3,8:34}^{(50\%)})$ | 0.695 | 0.090 | 0.512 | 0.636 | 0.694 | 0.753 | 0.879 | 1541 | 1014 | 1.002 |
| $\ln(s_{2,3,4,8:34}^{(95\%)}\text{-}s_{2,3,4,8:34}^{(50\%)})$ | 0.961 | 0.083 | 0.806 | 0.901 | 0.959 | 1.020 | 1.119 | 1612 | 1112 | 1.000 |
| $\ln(s_{2,3,1,35:48}^{(95\%)}\text{-}s_{2,3,1,35:48}^{(50\%)})$ | 0.375 | 0.066 | 0.248 | 0.330 | 0.375 | 0.421 | 0.503 | 1988 | 1342 | 1.001 |
| $\ln(s^{(95\%)}_{2,3,2,35:48}\text{-}s^{(50\%)}_{2,3,2,35:48})$ | 0.119 | 0.045 | 0.029 | 0.090 | 0.118 | 0.148 | 0.208 | 1643 | 1220 | 1.002 |
| $\ln(s^{(95\%)}_{2,3,3,35:48}\text{-}s^{(50\%)}_{2,3,3,35:48})$ | 0.190 | 0.054 | 0.087 | 0.155 | 0.191 | 0.227 | 0.303 | 458 | 1019 | 1.010 |
| $\ln(s^{(95\%)}_{2,3,4,35:48}\text{-}s^{(50\%)}_{2,3,4,35:48})$ | 0.066 | 0.058 | -0.046 | 0.029 | 0.066 | 0.104 | 0.174 | 648 | 812 | 1.001 |
| $\ln(s^{(95\%)}_{3,3,1:4,1:7}\text{-}s^{(50\%)}_{6,3,1:4,1:7})$ | 0.319 | 0.107 | 1.106 | 1.263 | 1.307 | 1.421 | 1.663 | 1021 | 1125 | 1.000 |
| *Herring natural mortality* | | | | | | | | | | |
| $\varepsilon^{(\mathrm{M,h1})}_{2}$ | 0.269 | 0.018 | 0.236 | 0.257 | 0.269 | 0.281 | 0.304 | 1993 | 1069 | 1.002 |
| $\varepsilon^{(\mathrm{M,h1})}_{3}$ | 0.193 | 0.019 | 0.155 | 0.179 | 0.194 | 0.206 | 0.231 | 2568 | 1089 | 1.003 |
| $\varepsilon^{(\mathrm{M,h1})}_{4}$ | 0.114 | 0.019 | 0.075 | 0.102 | 0.114 | 0.127 | 0.151 | 3532 | 1113 | 1.003 |
| $\varepsilon^{(\mathrm{M,h1})}_{5}$ | -0.008 | 0.019 | -0.045 | -0.021 | -0.008 | 0.005 | 0.029 | 2921 | 982 | 1.001 |
| $\varepsilon^{(\mathrm{M,h1})}_{6}$ | -0.105 | 0.018 | -0.142 | -0.117 | -0.105 | -0.092 | -0.069 | 2672 | 1117 | 0.999 |
| $\varepsilon^{(\mathrm{M,h1})}_{7}$ | -0.060 | 0.018 | -0.095 | -0.072 | -0.061 | -0.048 | -0.022 | 2340 | 942 | 1.007 |
| $\varepsilon^{(\mathrm{M,h1})}_{8}$ | 0.101 | 0.020 | 0.063 | 0.087 | 0.101 | 0.116 | 0.139 | 2292 | 1085 | 1.004 |
| $\varepsilon^{(\mathrm{M,h1})}_{9}$ | -0.038 | 0.021 | -0.079 | -0.054 | -0.039 | -0.024 | 0.005 | 2890 | 1197 | 1.000 |
| $\varepsilon^{(\mathrm{M,h1})}_{10}$ | -0.094 | 0.021 | -0.134 | -0.108 | -0.093 | -0.080 | -0.053 | 3536 | 1176 | 1.004 |
| $\varepsilon^{(\mathrm{M,h1})}_{11}$ | -0.098 | 0.022 | -0.140 | -0.113 | -0.098 | -0.083 | -0.058 | 2999 | 1067 | 1.001 |
| $\varepsilon^{(\mathrm{M,h1})}_{12}$ | -0.114 | 0.023 | -0.156 | -0.129 | -0.113 | -0.097 | -0.070 | 3344 | 1291 | 1.001 |
| $\varepsilon^{(\mathrm{M,h1})}_{13}$ | -0.092 | 0.023 | -0.136 | -0.107 | -0.092 | -0.077 | -0.044 | 3741 | 998 | 1.000 |
| $\varepsilon^{(\mathrm{M,h1})}_{14}$ | -0.007 | 0.022 | -0.050 | -0.021 | -0.007 | 0.007 | 0.035 | 3330 | 1204 | 1.007 |
| $\varepsilon_{15}^{(\mathrm{M,h1})}$ | 0.035 | 0.022 | -0.008 | 0.020 | 0.033 | 0.049 | 0.079 | 2963 | 1016 | 1.000 |
| $\varepsilon_{16}^{(\mathrm{M,h1})}$ | 0.046 | 0.023 | 0.001 | 0.031 | 0.047 | 0.060 | 0.090 | 3515 | 1095 | 1.002 |
| $\varepsilon_{17}^{(\mathrm{M,h1})}$ | 0.009 | 0.022 | -0.034 | -0.005 | 0.010 | 0.023 | 0.052 | 3027 | 991 | 1.009 |
| $\varepsilon_{18}^{(\mathrm{M,h1})}$ | -0.012 | 0.024 | -0.059 | -0.027 | -0.011 | 0.004 | 0.034 | 3271 | 1011 | 1.005 |
| $\varepsilon_{19}^{(\mathrm{M,h1})}$ | -0.005 | 0.024 | -0.052 | -0.021 | -0.005 | 0.011 | 0.041 | 2768 | 1072 | 1.001 |
| $\varepsilon_{20}^{(\mathrm{M,h1})}$ | -0.003 | 0.023 | -0.048 | -0.018 | -0.004 | 0.012 | 0.040 | 2179 | 956 | 1.003 |
| $\varepsilon_{21}^{(\mathrm{M,h1})}$ | -0.007 | 0.022 | -0.051 | -0.023 | -0.007 | 0.007 | 0.035 | 3056 | 1226 | 1.002 |
| $\varepsilon_{22}^{(\mathrm{M,h1})}$ | -0.011 | 0.022 | -0.055 | -0.026 | -0.011 | 0.004 | 0.032 | 3194 | 1163 | 1.000 |
| $\varepsilon_{23}^{(\mathrm{M,h1})}$ | 0.010 | 0.023 | -0.036 | -0.005 | 0.010 | 0.025 | 0.055 | 4658 | 1240 | 1.002 |
| $\varepsilon_{24}^{(\mathrm{M,h1})}$ | 0.048 | 0.022 | 0.004 | 0.035 | 0.048 | 0.063 | 0.091 | 3501 | 1076 | 1.005 |
| $\varepsilon_{25}^{(\mathrm{M,h1})}$ | 0.061 | 0.021 | 0.018 | 0.046 | 0.061 | 0.075 | 0.101 | 3061 | 1162 | 1.002 |
| $\varepsilon_{26}^{(\mathrm{M,h1})}$ | 0.054 | 0.022 | 0.011 | 0.039 | 0.054 | 0.069 | 0.097 | 3462 | 1057 | 1.001 |
| $\varepsilon_{27}^{(\mathrm{M,h1})}$ | 0.038 | 0.022 | -0.004 | 0.022 | 0.038 | 0.053 | 0.079 | 2842 | 1140 | 1.000 |
| $\varepsilon_{28}^{(\mathrm{M,h1})}$ | 0.009 | 0.022 | -0.031 | -0.005 | 0.009 | 0.024 | 0.053 | 3599 | 878 | 1.001 |
| $\varepsilon_{29}^{(\mathrm{M,h1})}$ | -0.010 | 0.023 | -0.054 | -0.025 | -0.010 | 0.006 | 0.034 | 3439 | 952 | 1.000 |
| $\varepsilon_{30}^{(\mathrm{M,h1})}$ | 0.009 | 0.022 | -0.036 | -0.006 | 0.009 | 0.025 | 0.051 | 3406 | 1024 | 0.999 |
| $\varepsilon_{31}^{(\mathrm{M,h1})}$ | 0.009 | 0.022 | -0.032 | -0.006 | 0.009 | 0.024 | 0.053 | 2972 | 1097 | 1.005 |
| $\varepsilon_{32}^{(\mathrm{M,h1})}$ | -0.015 | 0.020 | -0.055 | -0.028 | -0.015 | -0.001 | 0.025 | 3004 | 1267 | 1.001 |
| $\varepsilon_{33}^{(M,h1)}$ | -0.006 | 0.022 | -0.049 | -0.021 | -0.006 | 0.009 | 0.036 | 3945 | 888 | 1.004 |
| $\varepsilon_{34}^{(M,h1)}$ | 0.024 | 0.021 | -0.018 | 0.009 | 0.024 | 0.038 | 0.066 | 3039 | 1313 | 1.001 |
| $\varepsilon_{35}^{(M,h1)}$ | 0.040 | 0.022 | -0.002 | 0.027 | 0.040 | 0.054 | 0.084 | 3220 | 906 | 1.000 |
| $\varepsilon_{36}^{(M,h1)}$ | 0.043 | 0.022 | 0.002 | 0.028 | 0.043 | 0.058 | 0.086 | 2671 | 1135 | 1.003 |
| $\varepsilon_{37}^{(M,h1)}$ | 0.030 | 0.022 | -0.012 | 0.016 | 0.030 | 0.043 | 0.073 | 2697 | 1037 | 1.001 |
| $\varepsilon_{38}^{(M,h1)}$ | 0.010 | 0.021 | -0.029 | -0.004 | 0.010 | 0.025 | 0.051 | 2592 | 1026 | 1.006 |
| $\varepsilon_{39}^{(M,h1)}$ | 0.012 | 0.020 | -0.028 | -0.001 | 0.012 | 0.025 | 0.052 | 2324 | 1136 | 0.999 |
| $\varepsilon_{40}^{(M,h1)}$ | 0.011 | 0.021 | -0.030 | -0.004 | 0.011 | 0.025 | 0.053 | 2560 | 1181 | 1.004 |
| $\varepsilon_{41}^{(M,h1)}$ | -0.002 | 0.021 | -0.042 | -0.017 | -0.002 | 0.012 | 0.037 | 2478 | 1134 | 1.007 |
| $\varepsilon_{42}^{(M,h1)}$ | -0.024 | 0.021 | -0.064 | -0.039 | -0.023 | -0.010 | 0.017 | 3304 | 1114 | 1.002 |
| $\varepsilon_{43}^{(M,h1)}$ | 0.001 | 0.021 | -0.039 | -0.014 | 0.001 | 0.016 | 0.042 | 3174 | 1095 | 1.004 |
| $\varepsilon_{44}^{(M,h1)}$ | 0.074 | 0.021 | 0.033 | 0.060 | 0.075 | 0.088 | 0.111 | 2249 | 947 | 1.000 |
| $\varepsilon_{45}^{(M,h1)}$ | 0.044 | 0.021 | 0.002 | 0.030 | 0.044 | 0.058 | 0.086 | 2690 | 979 | 1.000 |
| $\varepsilon_{46}^{(M,h1)}$ | -0.016 | 0.022 | -0.058 | -0.031 | -0.016 | -0.001 | 0.027 | 3934 | 1080 | 1.004 |
| $\varepsilon_{47}^{(M,h1)}$ | 0.046 | 0.022 | 0.002 | 0.031 | 0.046 | 0.061 | 0.090 | 3249 | 980 | 1.002 |
| $\varepsilon_{48}^{(M,h1)}$ | 0.038 | 0.025 | -0.010 | 0.021 | 0.039 | 0.055 | 0.088 | 3577 | 1309 | 1.010 |
| $\varepsilon_{2}^{(M,h2)}$ | -0.060 | 0.024 | -0.108 | -0.075 | -0.059 | -0.044 | -0.012 | 1460 | 786 | 1.001 |
| $\varepsilon_{3}^{(M,h2)}$ | -0.065 | 0.024 | -0.110 | -0.081 | -0.065 | -0.050 | -0.020 | 2126 | 1042 | 1.003 |
| $\varepsilon_{4}^{(M,h2)}$ | -0.096 | 0.024 | -0.141 | -0.113 | -0.096 | -0.080 | -0.049 | 3548 | 1057 | 1.002 |
| $\varepsilon_{5}^{(M,h2)}$ | -0.118 | 0.023 | -0.164 | -0.133 | -0.117 | -0.102 | -0.072 | 2741 | 960 | 1.000 |
| $\varepsilon_{6}^{(M,h2)}$ | -0.111 | 0.023 | -0.156 | -0.126 | -0.111 | -0.095 | -0.065 | 3576 | 832 | 1.004 |
| $\varepsilon_{7}^{(M,h2)}$ | -0.078 | 0.023 | -0.121 | -0.093 | -0.078 | -0.062 | -0.032 | 4045 | 1152 | 1.001 |
| $\varepsilon_{8}^{(M,h2)}$ | -0.024 | 0.023 | -0.068 | -0.040 | -0.024 | -0.009 | 0.022 | 3000 | 1108 | 1.004 |
| $\varepsilon_{9}^{(M,h2)}$ | 0.012 | 0.024 | -0.033 | -0.005 | 0.013 | 0.028 | 0.058 | 3617 | 1143 | 1.000 |
| $\varepsilon_{10}^{(M,h2)}$ | 0.031 | 0.025 | -0.018 | 0.014 | 0.031 | 0.047 | 0.081 | 4256 | 1054 | 0.999 |
| $\varepsilon_{11}^{(M,h2)}$ | 0.038 | 0.025 | -0.013 | 0.021 | 0.038 | 0.056 | 0.086 | 3079 | 906 | 1.005 |
| $\varepsilon_{12}^{(M,h2)}$ | 0.036 | 0.025 | -0.015 | 0.021 | 0.036 | 0.052 | 0.084 | 4311 | 1000 | 1.001 |
| $\varepsilon_{13}^{(M,h2)}$ | 0.032 | 0.025 | -0.018 | 0.015 | 0.033 | 0.049 | 0.079 | 4030 | 1023 | 1.002 |
| $\varepsilon_{14}^{(M,h2)}$ | 0.012 | 0.024 | -0.034 | -0.005 | 0.012 | 0.028 | 0.058 | 2703 | 1137 | 1.002 |
| $\varepsilon_{15}^{(M,h2)}$ | 0.001 | 0.024 | -0.045 | -0.015 | 0.001 | 0.018 | 0.050 | 2569 | 1032 | 1.002 |
| $\varepsilon_{16}^{(M,h2)}$ | -0.007 | 0.024 | -0.054 | -0.023 | -0.006 | 0.010 | 0.038 | 2746 | 934 | 1.001 |
| $\varepsilon_{17}^{(M,h2)}$ | -0.008 | 0.025 | -0.059 | -0.024 | -0.008 | 0.009 | 0.042 | 3619 | 994 | 1.003 |
| $\varepsilon_{18}^{(M,h2)}$ | -0.017 | 0.024 | -0.065 | -0.033 | -0.017 | 0.000 | 0.032 | 3465 | 999 | 1.003 |
| $\varepsilon_{19}^{(M,h2)}$ | -0.027 | 0.025 | -0.077 | -0.044 | -0.026 | -0.010 | 0.021 | 3402 | 872 | 1.009 |
| $\varepsilon_{20}^{(M,h2)}$ | -0.029 | 0.023 | -0.074 | -0.044 | -0.029 | -0.014 | 0.019 | 3253 | 984 | 1.005 |
| $\varepsilon_{21}^{(M,h2)}$ | -0.025 | 0.023 | -0.071 | -0.040 | -0.025 | -0.008 | 0.021 | 3002 | 875 | 1.001 |
| $\varepsilon_{22}^{(M,h2)}$ | -0.017 | 0.025 | -0.066 | -0.034 | -0.017 | 0.000 | 0.031 | 3684 | 884 | 1.012 |
| $\varepsilon_{23}^{(M,h2)}$ | -0.019 | 0.025 | -0.069 | -0.036 | -0.020 | -0.002 | 0.029 | 3917 | 1173 | 1.002 |
| $\varepsilon_{24}^{(M,h2)}$ | -0.025 | 0.025 | -0.075 | -0.043 | -0.025 | -0.008 | 0.024 | 3819 | 1167 | 1.008 |
| $\varepsilon_{25}^{(M,h2)}$ | -0.022 | 0.024 | -0.068 | -0.039 | -0.022 | -0.006 | 0.025 | 3322 | 865 | 1.000 |
| $\varepsilon_{26}^{(M,h2)}$ | -0.008 | 0.026 | -0.061 | -0.025 | -0.009 | 0.010 | 0.044 | 3650 | 748 | 1.004 |
| $\varepsilon_{27}^{(M,h2)}$ | -0.002 | 0.026 | -0.051 | -0.019 | -0.002 | 0.016 | 0.048 | 3046 | 1131 | 1.002 |
| $\varepsilon_{28}^{(M,h2)}$ | 0.001 | 0.024 | -0.044 | -0.015 | 0.001 | 0.018 | 0.047 | 3950 | 1111 | 1.001 |
| $\varepsilon_{29}^{(M,h2)}$ | -0.004 | 0.024 | -0.050 | -0.021 | -0.004 | 0.012 | 0.044 | 3388 | 1163 | 1.000 |
| $\varepsilon_{30}^{(M,h2)}$ | -0.005 | 0.022 | -0.049 | -0.020 | -0.006 | 0.010 | 0.039 | 4565 | 1140 | 0.999 |
| $\varepsilon_{31}^{(M,h2)}$ | -0.001 | 0.024 | -0.047 | -0.018 | -0.002 | 0.015 | 0.046 | 3864 | 1264 | 1.003 |
| $\varepsilon_{32}^{(M,h2)}$ | 0.004 | 0.026 | -0.047 | -0.013 | 0.003 | 0.022 | 0.051 | 3057 | 877 | 1.002 |
| $\varepsilon_{33}^{(M,h2)}$ | 0.006 | 0.024 | -0.041 | -0.011 | 0.005 | 0.022 | 0.051 | 2819 | 1082 | 1.014 |
| $\varepsilon_{34}^{(M,h2)}$ | 0.004 | 0.025 | -0.045 | -0.011 | 0.004 | 0.020 | 0.054 | 3834 | 1043 | 1.002 |
| $\varepsilon_{35}^{(M,h2)}$ | 0.006 | 0.026 | -0.047 | -0.012 | 0.006 | 0.024 | 0.056 | 3461 | 1102 | 1.002 |
| $\varepsilon_{36}^{(M,h2)}$ | 0.014 | 0.024 | -0.032 | -0.003 | 0.015 | 0.031 | 0.063 | 4535 | 1053 | 1.001 |
| $\varepsilon_{37}^{(M,h2)}$ | 0.018 | 0.026 | -0.032 | 0.001 | 0.018 | 0.036 | 0.067 | 3128 | 1073 | 1.000 |
| $\varepsilon_{38}^{(M,h2)}$ | 0.019 | 0.024 | -0.028 | 0.004 | 0.019 | 0.034 | 0.069 | 2813 | 843 | 1.003 |
| $\varepsilon_{39}^{(M,h2)}$ | 0.010 | 0.023 | -0.037 | -0.006 | 0.010 | 0.026 | 0.056 | 3688 | 1255 | 1.002 |
| $\varepsilon_{40}^{(M,h2)}$ | 0.008 | 0.025 | -0.038 | -0.009 | 0.009 | 0.026 | 0.054 | 3185 | 872 | 1.005 |
| $\varepsilon_{41}^{(M,h2)}$ | 0.011 | 0.025 | -0.039 | -0.006 | 0.011 | 0.027 | 0.059 | 3817 | 972 | 1.002 |
| $\varepsilon_{42}^{(M,h2)}$ | 0.017 | 0.024 | -0.029 | 0.001 | 0.017 | 0.034 | 0.064 | 4107 | 1062 | 1.001 |
| $\varepsilon_{43}^{(M,h2)}$ | 0.029 | 0.024 | -0.019 | 0.013 | 0.029 | 0.044 | 0.078 | 4764 | 681 | 1.001 |
| $\varepsilon_{44}^{(M,h2)}$ | 0.038 | 0.024 | -0.010 | 0.022 | 0.038 | 0.054 | 0.086 | 3508 | 783 | 1.000 |
| $\varepsilon_{45}^{(M,h2)}$ | 0.041 | 0.024 | -0.003 | 0.025 | 0.042 | 0.058 | 0.085 | 3186 | 1062 | 1.001 |
| $\varepsilon_{46}^{(M,h2)}$ | 0.040 | 0.025 | -0.008 | 0.023 | 0.041 | 0.057 | 0.087 | 2542 | 1202 | 1.001 |
| $\varepsilon_{47}^{(M,h2)}$ | 0.022 | 0.024 | -0.024 | 0.006 | 0.022 | 0.038 | 0.070 | 3499 | 1022 | 1.003 |
| $\varepsilon_{48}^{(M,h2)}$ | 0.006 | 0.024 | -0.040 | -0.012 | 0.005 | 0.022 | 0.051 | 4292 | 1071 | 1.002 |

Table D2. Post-fitted cod stock-recruitment parameter estimates. $\beta_0$ represents fecundity while $\beta_C$ and $\beta_H$ represent the effect of cod and herring biomass, respectively, on cod recruitment.

| **Parameter** | **Posterior quantiles** | | |
|---|---|---|---|
| | **2.5%** | **50%** | **97.5%** |
| $\beta_0$ | 4.78 | 6.79 | 9.42 |
| $\beta_C$ | 1.05E-03 | 1.57E-03 | 2.17E-03 |
| $\beta_H$ | 2.63E-03 | 3.22E-03 | 3.87E-03 |

Table D3. Post-fitted functional response parameter estimates. η, h, and λ represent the encounter rate, handling time, and shape of the functional response, respectively.

| Predator | Par. | Prey | Posterior quantiles | | |
|---|---|---|---|---|---|
| | | | **2.5%** | **50%** | **97.5%** |
| Seal, Shelf-M | $\eta$ | Cod | 3.65E+01 | 3.61E+02 | 1.23E+05 |
| | | Herring | 1.53E+01 | 1.58E+02 | 4.66E+04 |
| | $h$ | Cod | 5.05E+02 | 1.09E+04 | 4.61E+06 |
| | | Herring | 6.16E-06 | 1.41E+02 | 3.39E+04 |
| | $\lambda$ | Cod | 7.66E-05 | 3.61E-02 | 2.49E-01 |
| | | Herring | 3.22E-08 | 4.05E-03 | 1.67E-01 |
| Seal, Shelf-F | $\eta$ | Cod | 4.45E-01 | 2.79E+02 | 4.31E+09 |
| | | Herring | 5.50E-01 | 4.04E+02 | 5.27E+09 |
| | $h$ | Cod | 1.57E-06 | 6.11E+03 | 4.83E+09 |
| | | Herring | 1.22E-08 | 4.57E+02 | 5.25E+09 |
| | $\lambda$ | Cod | 1.90E-07 | 1.64E-02 | 3.75E-01 |
| | | Herring | 2.83E-09 | 9.36E-03 | 1.02E+00 |
| Seal, Gulf-M | $\eta$ | Cod | 2.37E+01 | 5.32E+02 | 1.21E+07 |
| | | Herring | 7.37E+00 | 1.63E+02 | 3.62E+06 |
| | $h$ | Cod | 1.51E-04 | 2.63E+02 | 1.04E+05 |
| | | Herring | 5.81E-01 | 7.11E+01 | 4.35E+06 |
| | $\lambda$ | Cod | 1.85E-07 | 1.68E-02 | 6.93E-02 |
| | | Herring | 2.25E-08 | 3.18E-03 | 1.46E-01 |
| Seal, Gulf-F | $\eta$ | Cod | 1.62E+01 | 9.61E+02 | 1.58E+08 |
| | | Herring | 2.10E+01 | 1.50E+03 | 2.36E+08 |
| | $h$ | Cod | 2.07E-05 | 2.69E+02 | 1.74E+05 |
| | | Herring | 1.69E+00 | 2.86E+03 | 6.44E+08 |
| | $\lambda$ | Cod | 1.85E-07 | 3.71E-04 | 6.09E-02 |
| | | Herring | 3.83E-15 | 6.47E-05 | 8.14E-02 |
| Cod | $\eta$ | Herring | 7.00E-12 | 3.54E-07 | 1.17E+00 |
| | $h$ | Herring | 1.92E-11 | 3.89E-08 | 6.19E+00 |
| | $\lambda$ | Herring | 2.76E-01 | 2.23E+00 | 4.22E+00 |

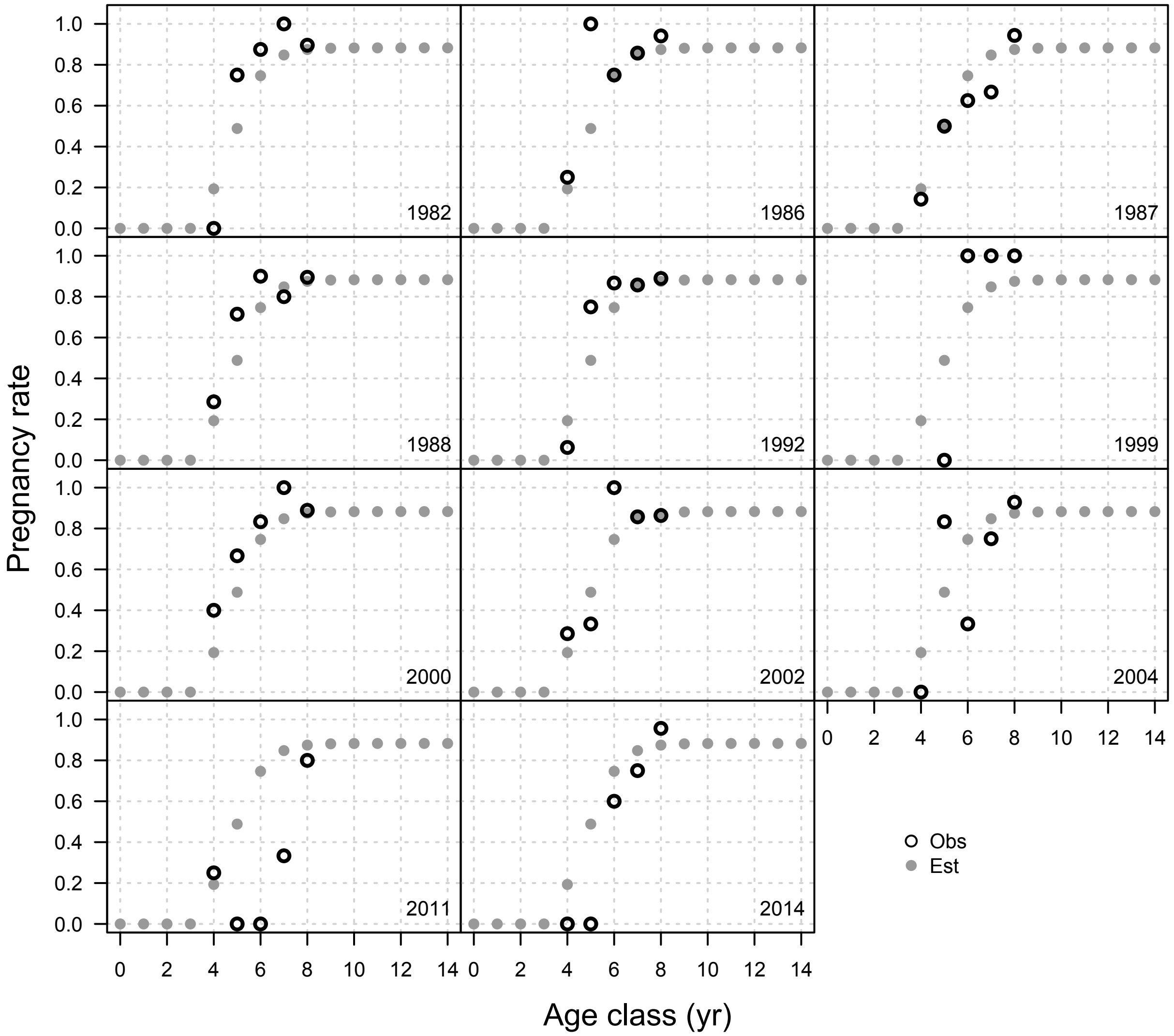

Figure D1. Model fits (grey points) to observed grey seal pregnancy rates (circles) for years with more than 15 observations.

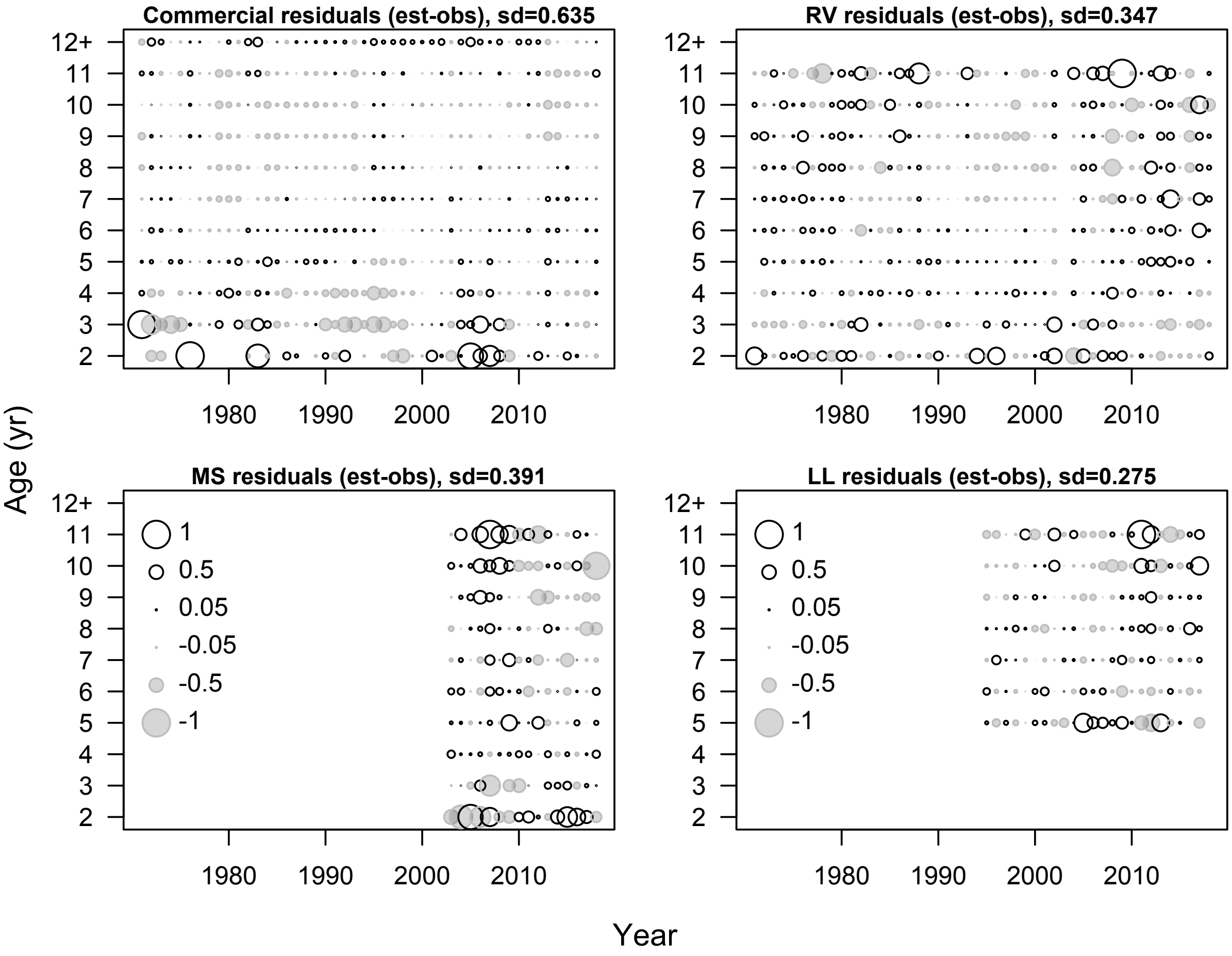

Figure D2. Residuals between the observed proportions-at-age in the cod abundance indices and the model-estimated proportions. Residuals (estimated - observed) are proportional to circle radii.

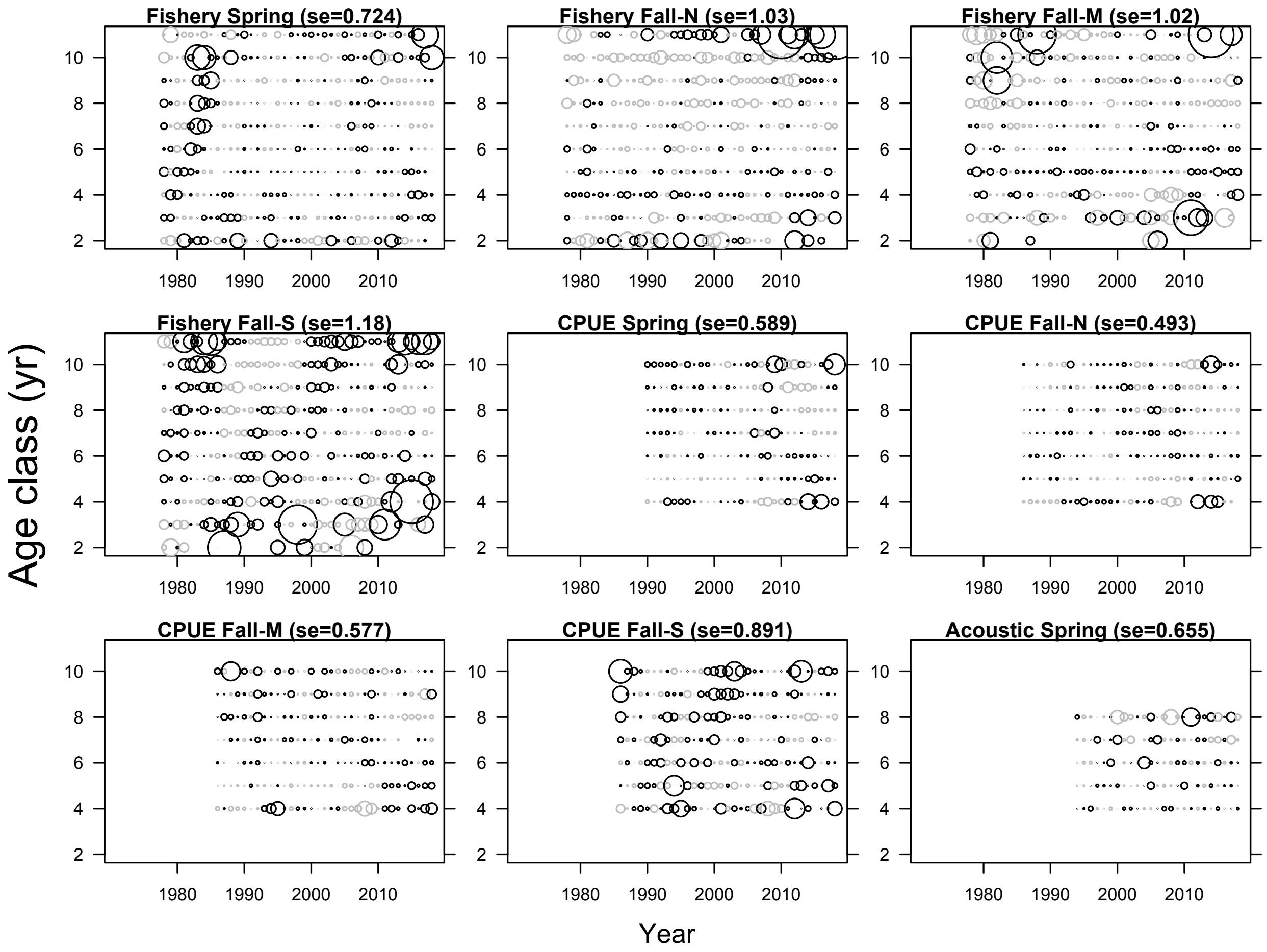

Figure D3. Residuals between the observed proportions-at-age in the herring abundance indices and the model-estimated proportions. Residuals (estimated - observed) are proportional to circle radii. Black circles denote positive residuals.

**Appendix E: Sensitivity analysis of consumption prior**

In the multispecies model, we calculated the per-capita consumption by predator *j* in subpopulation *y* at age *b* of prey *i* in year *t* ($\hat{c}_{j,y,b,i,t}$) and applied lognormal prior to this quantity to constrain prey consumption within biologically plausible ranges:

$$\ln \hat{c}_{j,y,b,i,t} \sim \mathrm{N}\left(\ln c_{j,y,b,i,t}\,, \sigma^2_{\mathrm{c},j}\right)$$

Prior means $c$ were calculated from the bioenergetic requirements of predators (grey seals and cod), predator diet composition data, and the spatiotemporal overlap between predator and prey species (Appendix A). Prior standard deviation ($\sigma$) was set to 1.0 for seal predation and 0.1 for cod predation. In this appendix, we summarize the sensitivity of the multispecies model to alternative choices of prior standard deviation.

**Seal predation**

We tested alternative values of $\sigma_{\mathrm{c},j=1}$ = 0.1, 0.5, and 1.5 for seal predation on cod and herring. Each of the alternative models fit the observed data approximately as well as the base model ($\sigma_{\mathrm{c},j=1}$ = 1.0); increasing $\sigma$ did not visually improve fits to abundance or biomass indices (Fig. E1). Seal consumption of cod and herring was more variable under less constraining priors but estimates were within biologically plausible ranges (Figs. E2-E4). Notably, per-capita seal consumption of cod declined by more than 50% over the last 15 years under $\sigma$=1.0 and $\sigma$=1.5 (Fig. E4). It is difficult to tell whether these patterns are representative of underlying dynamics or whether the model is overfitting to noise. Despite these differences, estimates of biomass and key population processes for all species are very similar across different values of $\sigma_{\mathrm{c},j=1}$ (Fig. E5).

**Cod predation**

We tested alternative values of $\sigma_{\mathrm{c},j=2}$ = 0.05, 0.20, and 1.00 for cod predation on herring. Each of the alternative models fit the observed data approximately as well as the base model ($\sigma_{\mathrm{c},j=2}$ = 0.1; Fig. E6). Increasing prior standard deviation resulted in less intense predation in the mid-1980s and the mid- to late-1990s, which was offset by higher levels of other natural mortality and reduced cod biomass (Fig. E7). The more modest consumption estimated by models with higher $\sigma_{c,j=2}$, particularly $\sigma_{c,j=2}$=1.00, was inconsistent with previous studies of cod predation on herring (Benoît and Rail, 2016).

## Figures

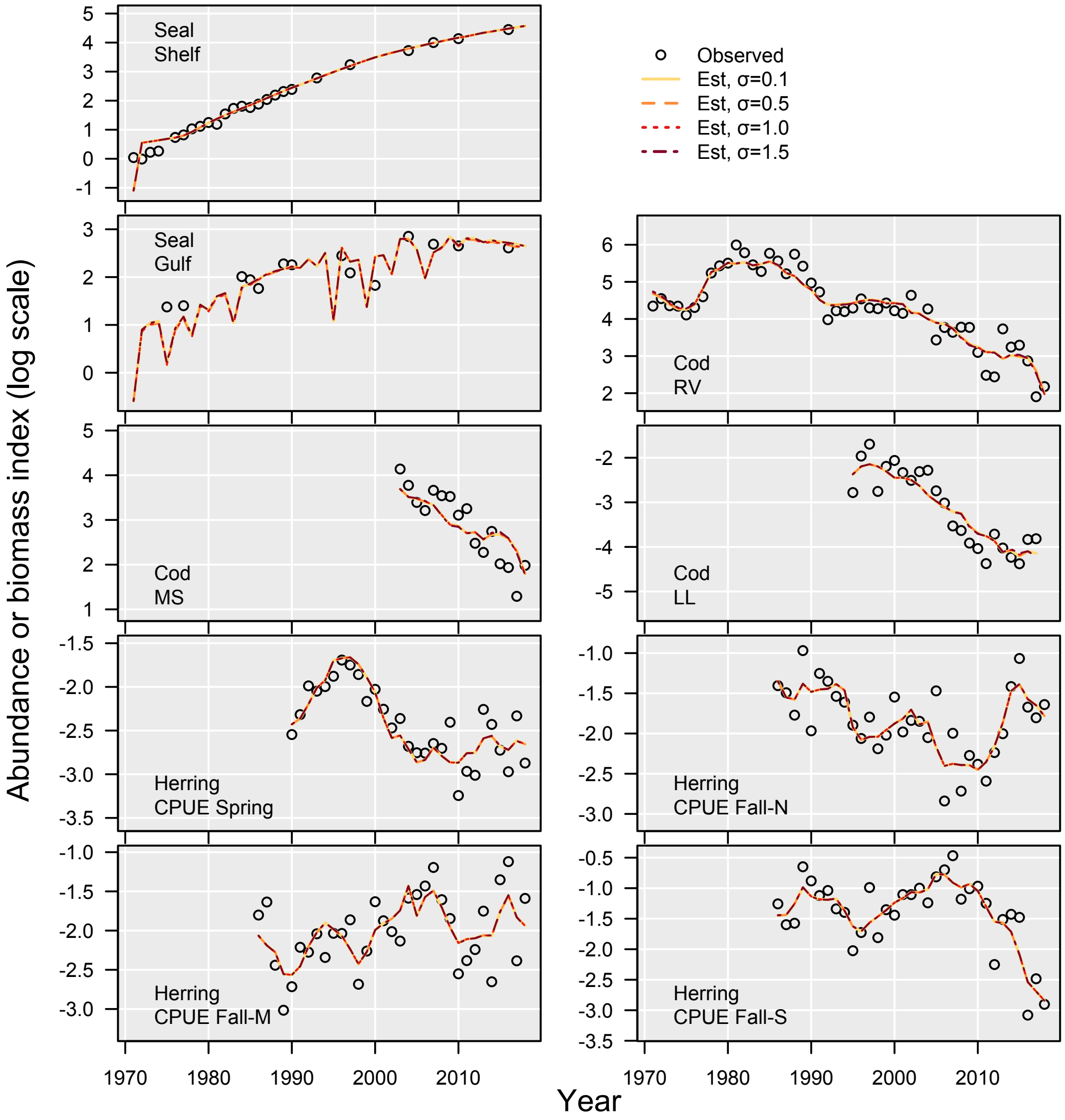

Figure E1. Model fits (posterior modes; lines) under four values of $\sigma_{c,j=1}$ to observed abundance or biomass indices (circles). The index represents pup production in numbers for grey seals and vulnerable biomass for cod and herring.

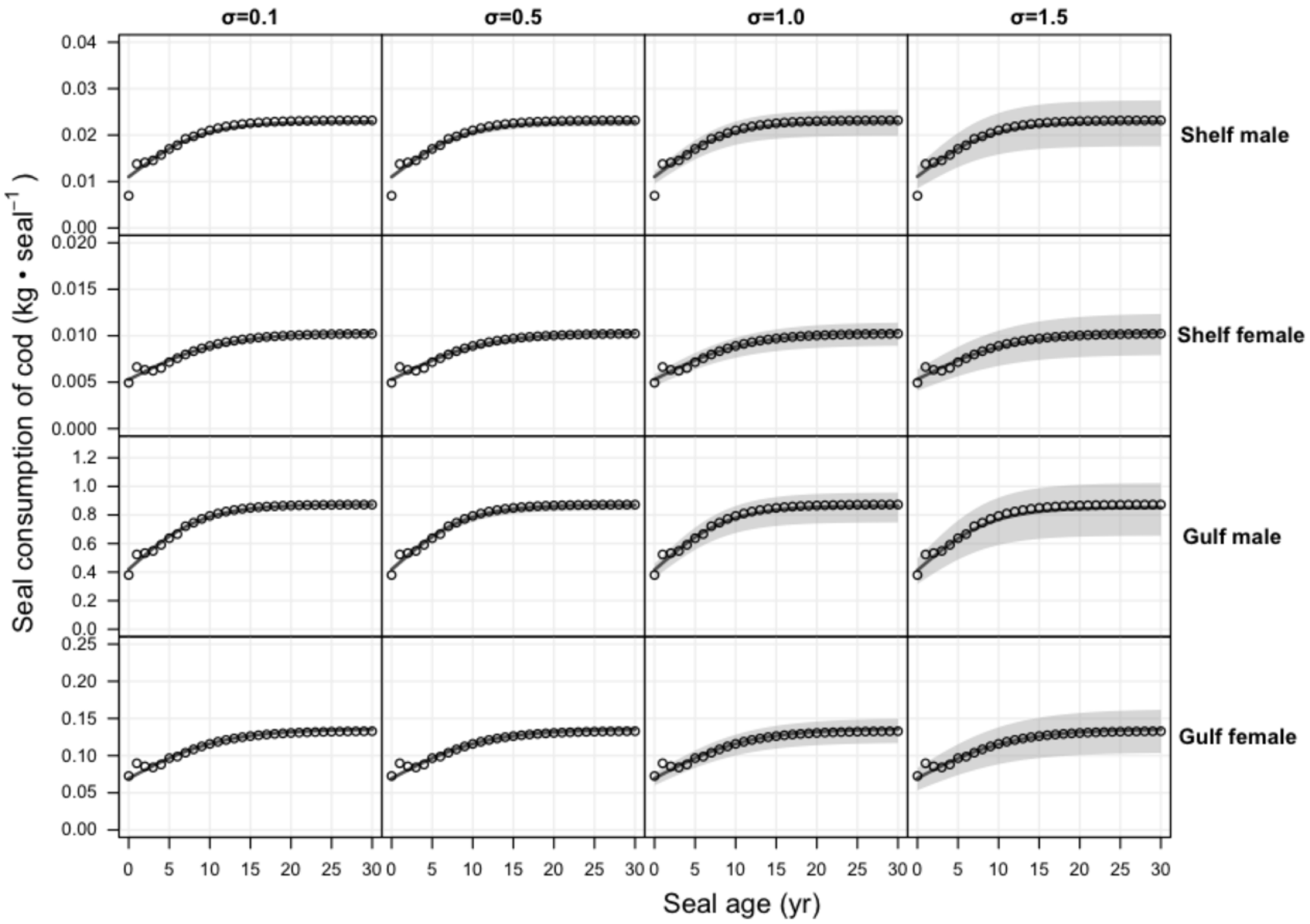

Figure E2. Prior mean (circles) and model-estimated (lines, shaded regions) cod consumption per seal using alternative values of $\sigma_{c,j=1}$ (columns). Each row represents a seal subpopulation. Lines represent posterior modes while shaded regions represent central 95% uncertainty intervals.

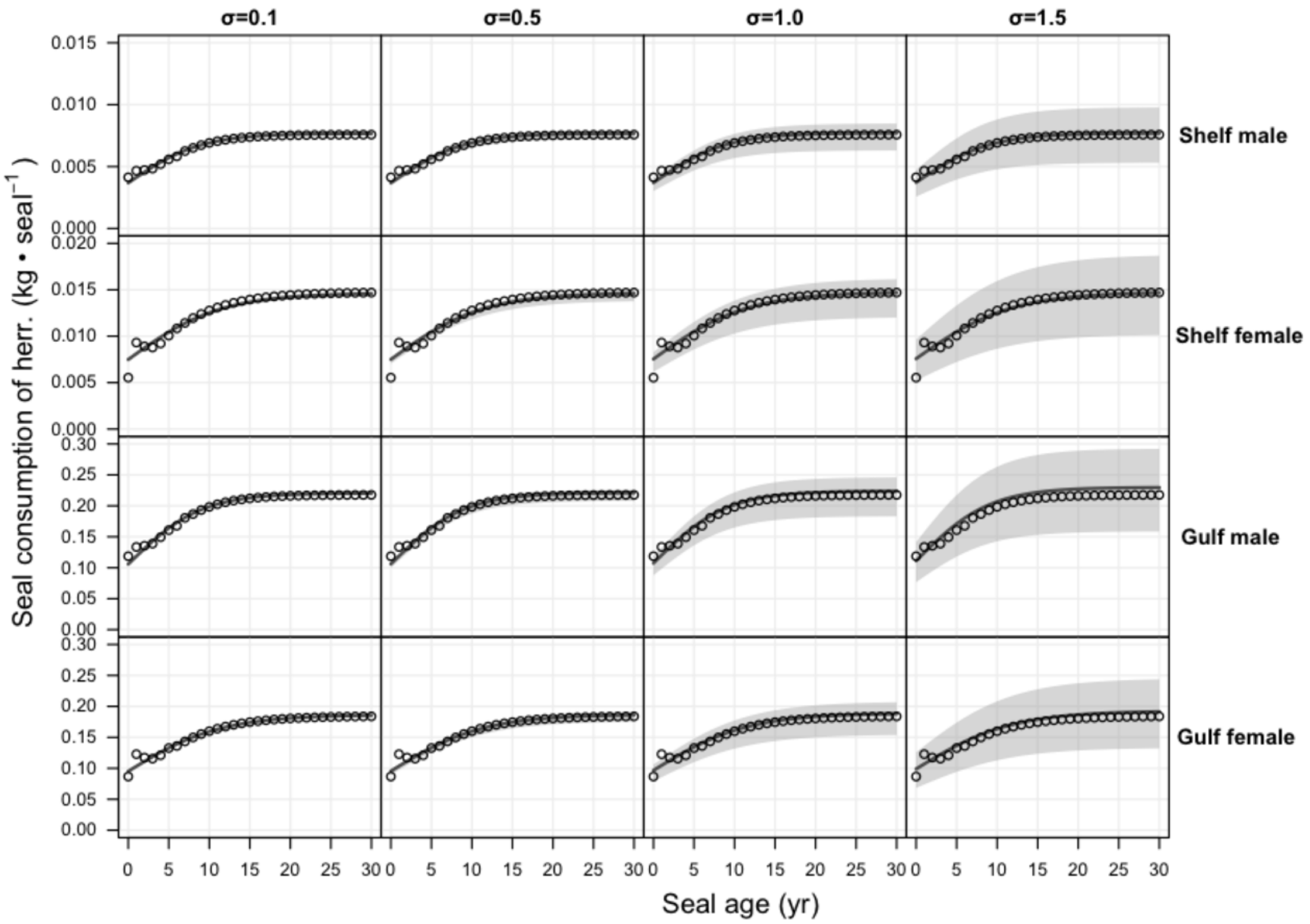

Figure E3. Prior mean (circles) and model-estimated (lines, shaded regions) herring consumption per seal using alternative values of $\sigma_{c,j=1}$ (columns). Each row represents a seal subpopulation. Lines represent posterior modes while shaded regions represent central 95% uncertainty intervals.

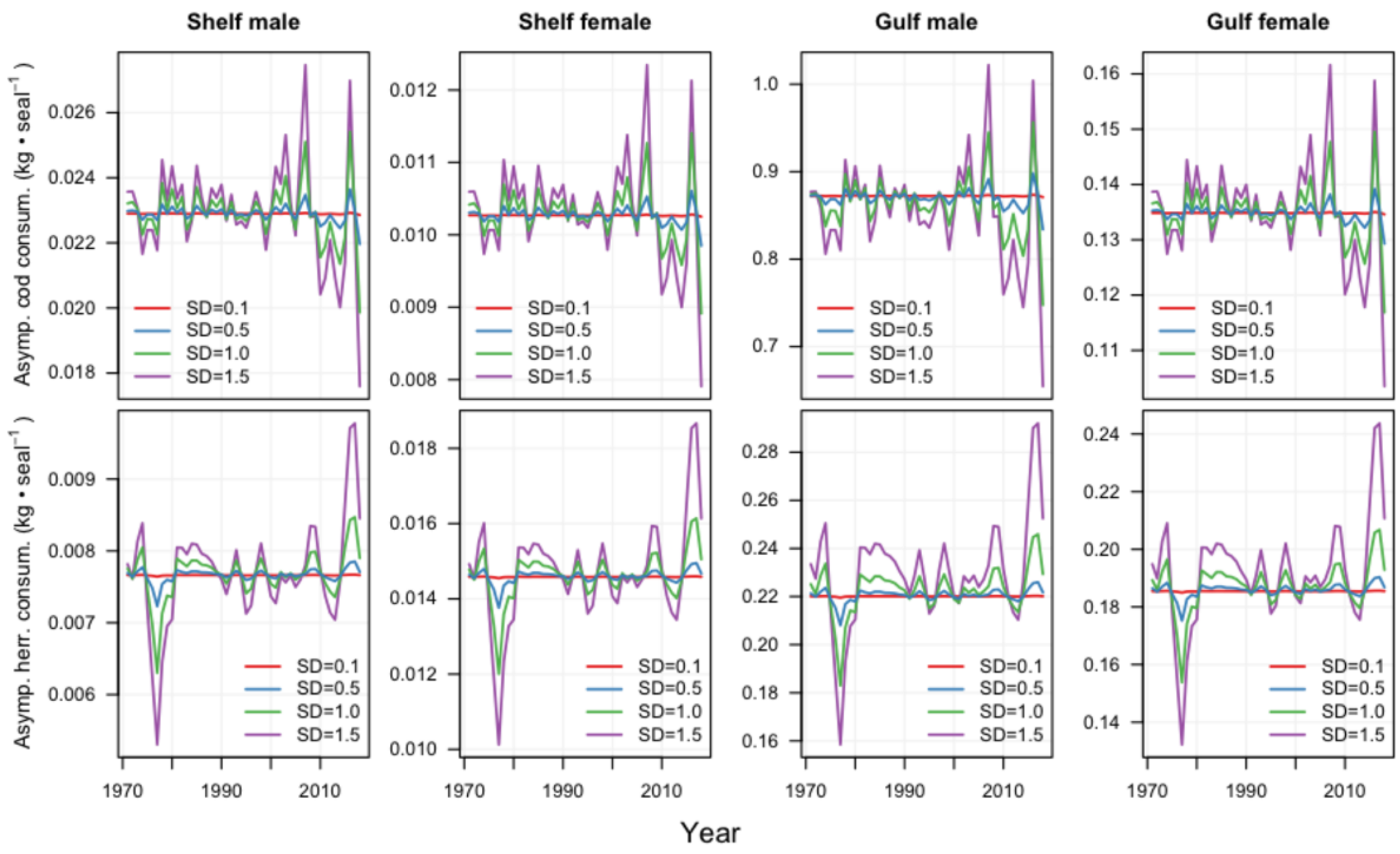

Figure E4. Model-estimated prey consumption per seal for the oldest seal age-class (30+) using alternative values of $\sigma_{c,j=1}$ (columns).

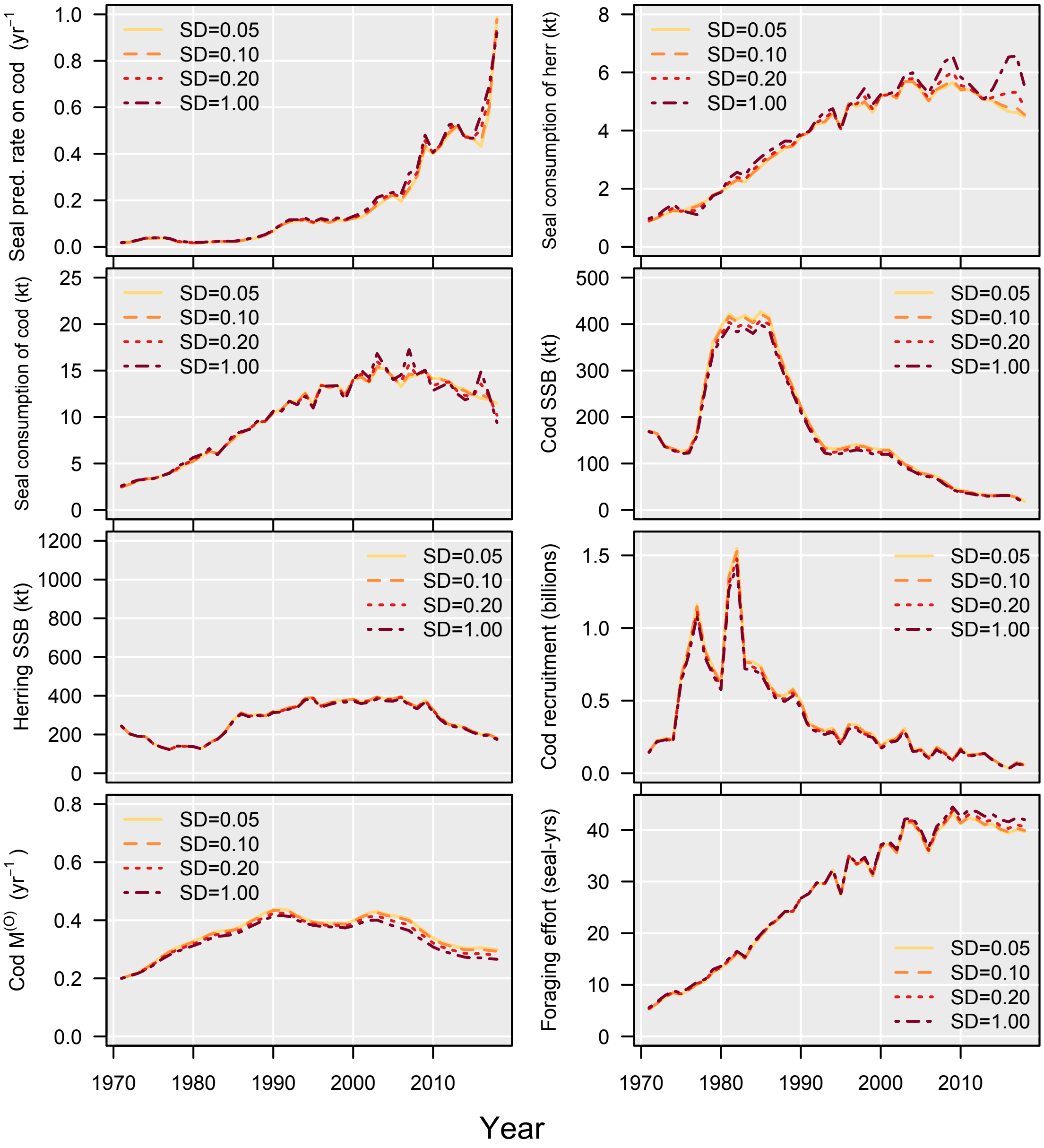

Figure E5. Estimated time-series of key model quantities using alternative values of $\sigma_{c,j=1}$.

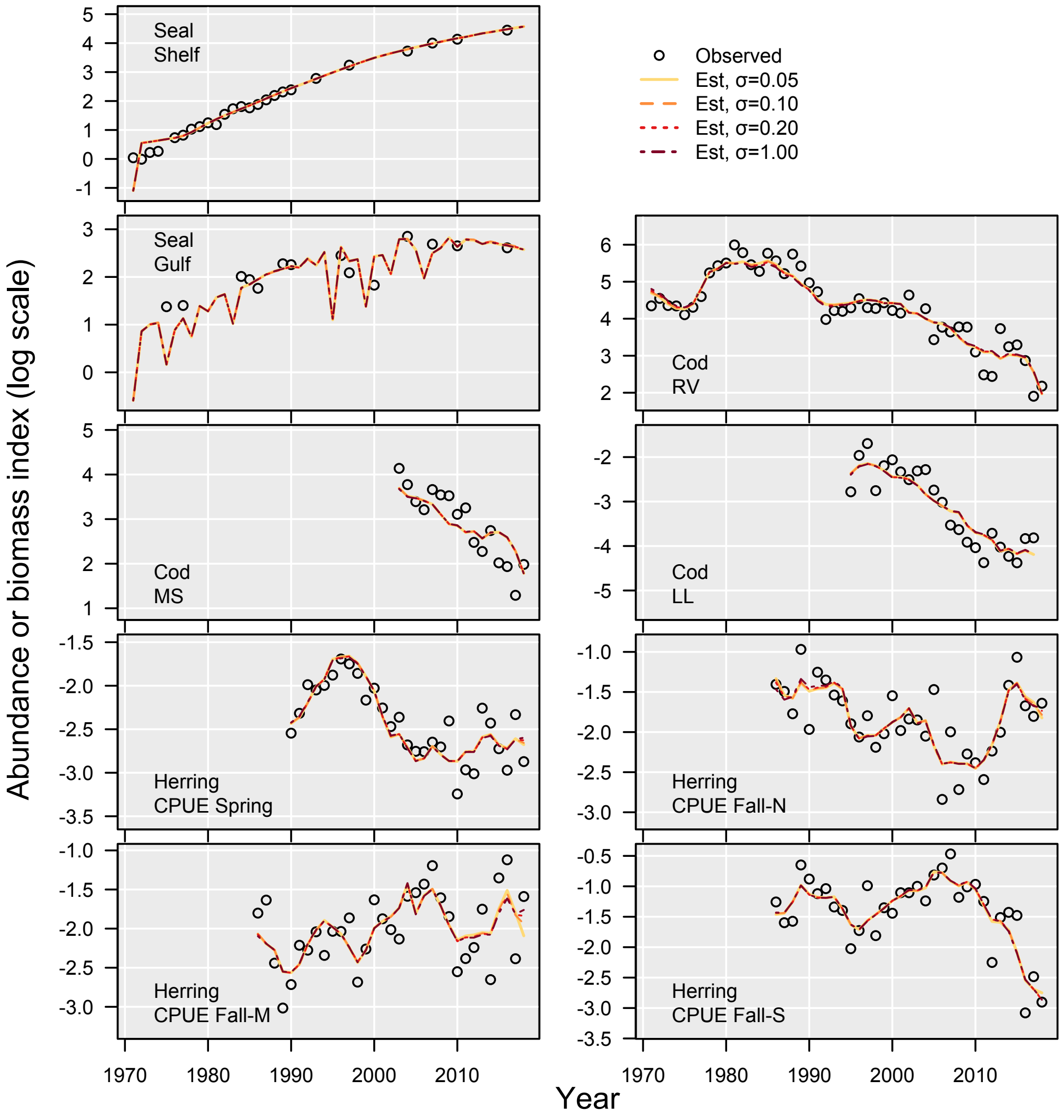

Figure E6. Model fits (posterior modes; lines) under four values of $\sigma_{c,j=2}$ to observed abundance or biomass indices (circles). The index represents pup production in numbers for grey seals and vulnerable biomass for cod and herring.

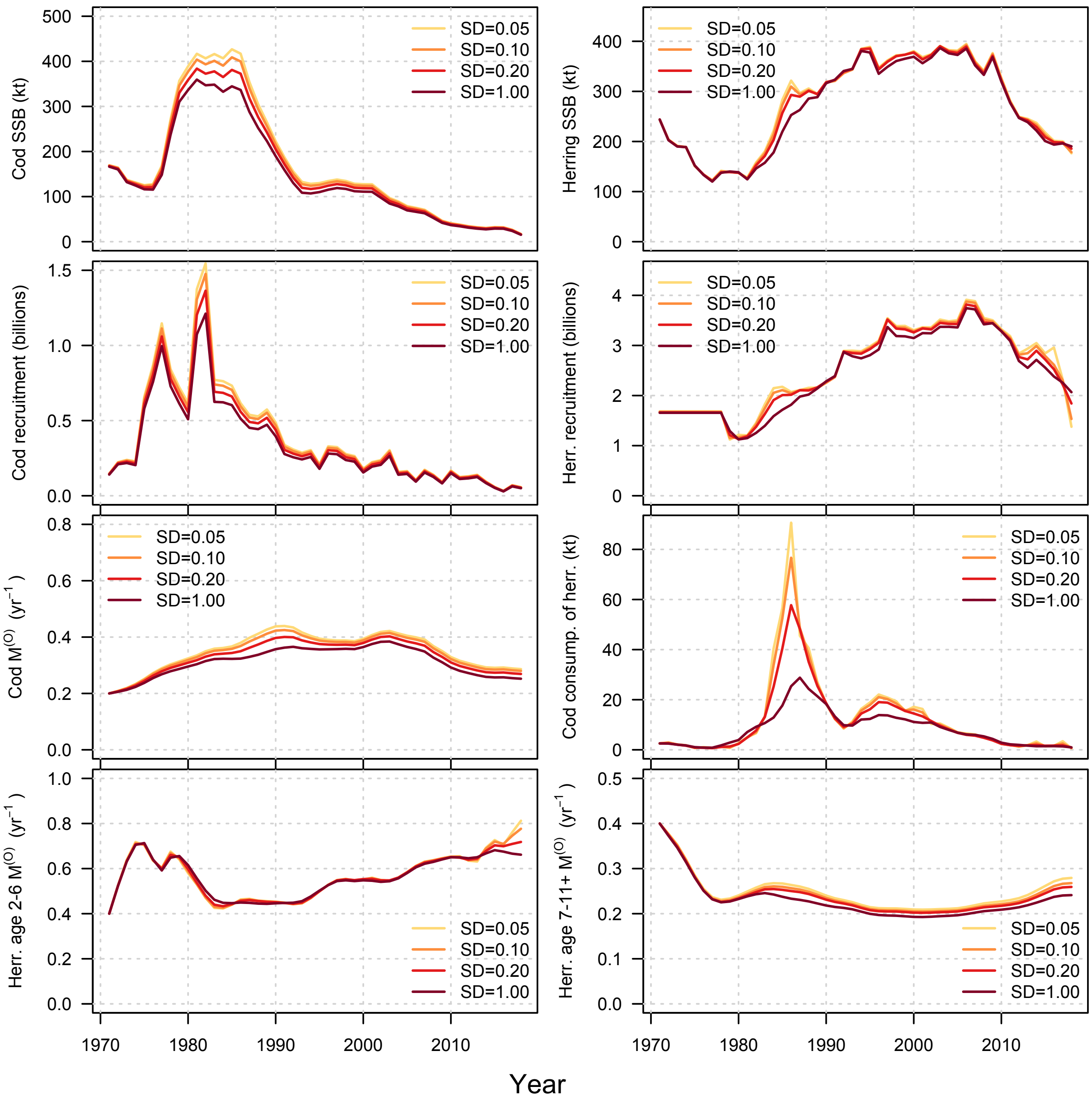

Figure E7. Estimated time-series of key model quantities using alternative values of $\sigma_{c,j=2}$.